\documentclass[a4paper,12pt]{article}

\usepackage{amsmath,amsthm,amssymb,amsfonts}
\makeatletter

\@addtoreset{equation}{section}
\makeatother

\newtheorem{theorem}{Theorem}[section]
\newtheorem{lemma}[theorem]{Lemma}
\newtheorem{proposition}[theorem]{Proposition}
\newtheorem{corollary}[theorem]{Corollary}

\newtheorem{remark}[theorem]{Remark}

\def\ep{\varepsilon}

\newcommand{\dsum}{\displaystyle\sum}
\newcommand{\dint}{\displaystyle\int}
\newcommand{\pdr}[2]{\dfrac{\partial{#1}}{\partial{#2}}}

\def\R{\mathbb R}
\def\S{\mathbb S}
\def\N{\mathbb N}
\def\pa{\partial}
\def\b{\backslash}

\def\diam{{\rm diam}(X)}
\newcommand{\cout}[1]{}

\begin{document}
\title{Time-dependent angularly averaged inverse transport\\
(extended version)}

\author{Guillaume Bal and Alexandre Jollivet
        \thanks{Department of Applied Physics and 
        Applied Mathematics, Columbia University, 
        New York NY, 10027; gb2030@columbia.edu and aj2315@columbia.edu}}

\maketitle

\begin{abstract}
  This paper concerns the reconstruction of the absorption and
  scattering parameters in a time-dependent linear transport equation
  from knowledge of angularly averaged measurements performed at the
  boundary of a domain of interest. We show that the absorption
  coefficient and the spatial component of the scattering coefficient
  are uniquely determined by such measurements. We obtain stability
  results on the reconstruction of the absorption and scattering
  parameters with respect to the measured albedo operator. The
  stability results are obtained by a precise decomposition of the
  measurements into components with different singular behavior in the
  time domain.
\end{abstract}

\section{Introduction}

Inverse transport theory has many applications in e.g. medical and
geophysical imaging. It consists of reconstructing optical parameters
in a domain of interest from measurements of the transport solution at
the boundary of that domain. The optical parameters are the total
absorption (extinction) parameter $\sigma(x)$ and the scattering
parameter $k(x,v',v)$, which measures the probability of a particle at
position $x\in X\subset\R^n$ to scatter from direction $v'\in\S^{n-1}$ to
direction $v\in\S^{n-1}$, where $\S^{n-1}$ is the unit sphere in
$\R^n$.

The domain of interest is probed as follows. A known flux of particles
enters the domain and the flux of outgoing particles is measured at
the domain's boundary. Several inverse theories may then be envisioned
based on available data. In this paper, we assume availability of time
dependent measurements that are angularly averaged. Also the source
term used to probe the domain is not resolved angularly in order to
e.g. save time in the acquisition of data.  More precisely, the
incoming density of particles $\phi(t,x,v)$ as a function of time $t$,
at position $x\in\partial X$ at the boundary of the domain of
interest, and for incoming directions $v$, is of the form
$\phi_S(t,x,v)=\phi(t,x)S(x,v)$, where $\phi(t,x)$ is arbitrary but
$S(x,v)$ is fixed.  This paper is concerned with the reconstruction of
the optical parameters from such measurements. We show that the
attenuation coefficient is uniquely determined and that the spatial
structure of the scattering coefficient can be reconstructed provided
that scattering vanishes in the vicinity of the domain's boundary
(except in dimension $n=2$ and when $X$ is a disc, where our theory
does not require $k$ to vanish in the vicinity of $\partial X$). For
instance, when $k(x,v',v)=k_0(x)g(v',v)$ with $g(v',v)$ known a
priori, then $k_0(x)$ is uniquely determined by the measurements.
Similar results were announced in \cite{B-PISA-08} when measurements
are available in the modulation frequency variable, which is the dual
(Fourier) variable to the time variable.

Several other regimes have been considered in the literature. The
uniqueness of the reconstruction of the optical parameters from
knowledge of angularly resolved measurements both in the
time-dependent and time-independent settings was proved in
\cite{CS-CPDE-96,CS-OSAKA-99}; see also \cite{S-IO-03} for a review.
Stability in the time-independent case has been analyzed in dimension
$n=2,3$ under smallness assumptions for the optical parameters in
\cite{Rom-JIIPP-97,R-MT-98} and in dimension $n=2$ in
\cite{SU-MAA-03}. Stability results in the presence of full, angularly
resolved, measurements have been obtained in
\cite{BJ-IPI-08,BJ-SIMA-09,W-AIHP-99}. The intermediate case of
angularly averaged measurements with angularly resolved sources was
considered in \cite{L-IP-08}. The lack of stability of the
reconstruction in the time independent setting with angularly averaged
measurements and isotropic sources is treated in \cite{BLM-IPI-08}.
See also \cite{B-IP-09} for a recent review of results in inverse
transport theory.

The rest of the paper is structured as follows. Section
\ref{sec:albedo} recalls known results on the transport equation and
the decomposition of the albedo operator.
In section \ref{sec:albedoaveraged} we define and decompose the averaged albedo operator  (Proposition \ref{prop:kernelA}) and we study its distributional kernel (Theorems \ref{theorem:normesimplescat}--\ref{theorem:limH2}).   
Our main results on
uniqueness and stability are presented in section \ref{sec:stab} (Theorems \ref{theorem:stabH1}--\ref{theorem:stabH2}, Theorems \ref{cor_ball}--\ref{cor} and Corollary \ref{cor2}).  We
show that the absorption coefficient and the spatial structure of the
scattering coefficient (the phase function describing scattering from
$v$ to $v'$ has to be known in advance) can be reconstructed stably
from angularly averaged time dependent data. The reconstruction of the
scattering coefficient requires invertion of a weighted Radon
transform in the general case. In the specific case of a spherical
geometry (measurements are performed at the boundary of a sphere),
then the scattering coefficient may be obtained by inverting a {\em
  classical} Radon transform. 
In section \ref{proof_thm_H1H2} we prove Theorems \ref{theorem:limH1}--\ref{theorem:limH2}.
In section \ref{proof_thm_stabH1H2_cor} we prove Theorems \ref{theorem:stabH1}--\ref{theorem:stabH2}, Theorem \ref{cor} and Theorem \ref{cor_ball} \eqref{eq:cor2H1}.
In section \ref{proof_thm_normesimplescat} we prove Theorem \ref{theorem:normesimplescat}.
In section \ref{proof_thm_normemultiplescat} we prove Theorem \ref{theorem:normemultiplescat}.
In section \ref{proof_N} we prove Lemmas \ref{lem:prep}--\ref{lem:n=3nulaubord} that are used in section \ref{proof_thm_normemultiplescat}.
In section \ref{proof_kernelA} we prove Proposition \ref{prop:kernelA}.

The derivation of the results is fairly technical and is based on a
careful analysis of the temporal behavior of the decomposition of the
albedo operator into components that are multi-linear in the
scattering coefficient. Our results are based on showing that the
ballistic and single scattering components may be separated from the
rest of the data. These two components are then used to obtain our
uniqueness and stability results.  It turns out that the structure of
single scattering is different depending on whether $k$ vanishes on
$\partial X$ or not. When $k$ does not vanish on $\partial X$, the
main singularities of the single scattering component do not allow us
to ``see inside'' the domain as they only depend on values of $k$ at
the domain's boundary in dimension $n\geq3$. The singular structure of
single scattering and the resulting stability estimates are presented
in detail when both $k$ vanishes and does not vanish on $\partial X$.

This is the extended version of a submitted paper \cite{BJ-IPShort-09}.

\section{The forward problem and albedo operator}
\label{sec:albedo}
\subsection{The linear Boltzmann transport equation}
We now introduce notation and recall some known results on the linear
transport equation.  Let $X$ be a bounded open subset of $\R^n$, $n\ge
2$, with a $C^1$ boundary $\pa X$.  Let $\nu(x)$ denote the outward
normal unit vector to $\pa X$ at $x\in \pa X$.  Let
$\Gamma_{\pm}=\{(x,v)\in \pa X\times \S^{n-1}\ |\ \pm\nu(x)\cdot
v>0\}$ be the sets of incoming and outgoing conditions.  For $(x,v)\in
\bar X\times \S^{n-1}$ we define $\tau_\pm (x,v)$ and $\tau(x,v)$ by
$\tau_{\pm}(x,v):=\inf\{s\in (0,+\infty)\ |\ x\pm sv \not \in X\}$ and
$\tau(x,v):=\tau_-(x,v)+\tau_+(x,v)$.
For $x\in \pa X$ we define $\S_{x,\pm}^{n-1}:=\{v\in \S^{n-1}\ |\ \pm \nu(x)\cdot v>0\}$.

Consider $\sigma:X\times\S^{n-1}\to \R$ and $k:X\times \S^{n-1}\times
\S^{n-1}\to \R$ two nonnegative measurable functions.  We assume that
$(\sigma,k)$ is admissible  when
\begin{equation} \label{eq:hyp1p}
  \begin{array}{l}
  0\le \sigma\in L^{\infty}(X\times\S^{n-1}),\\
  0\le k(x,v',.)\in L^1(\S^{n-1})\textrm{ for a.e. }
     (x,v')\in X\times \S^{n-1}, \\
  \sigma_p(x,v')=\dint_{\S^{n-1}}k(x,v',v)dv\textrm{ belongs to } 
  L^{\infty}(X\times \S^{n-1}).
\end{array}
\end{equation}
Let $T>\eta>0$. We consider the following linear Boltzmann transport
equation
\begin{equation}
  \label{eq:B1}
  \begin{array}{rcl}
    && 
  \pdr ut(t,x,v)+v\cdot\nabla_xu(t,x,v)+\sigma(x,v)u(t,x,v) \\
  &&= \dint_{\S^{n-1}}k(x,v',v)u(t,x,v')dv',\ (t,x,v)\in 
    (0,T)\times X\times\S^{n-1},\\[3mm]
  &&u_{|(0,T)\times \Gamma_-}(t,x,v)=\phi(t,x,v),\\[3mm]
&&u(0,x,v)=0, \ (x,v)\in X\times \S^{n-1},
  \end{array}
\end{equation}
where $\phi\in L^1((0,T),L^1(\Gamma_-,d\xi))$ and ${\rm
  supp}\phi\subseteq [0,\eta]$. Here, $d\xi(x,v)=|v\cdot \nu(x)|dv d\mu(x)$,
where $d\mu$ is the surface measure on $\partial X$ and $dv$ is the surface measure on $\S^{n-1}$.  In other words, we
assume that the initial condition is concentrated in the
$\eta$-vicinity of $t=0$ and measurements are performed for time $T$,
which we will choose sufficiently large so that particles have the
time to travel through $X$ and be measured.

\subsection{Semigroups and unbounded operators}
\label{semi}
We introduce the following space
\begin{eqnarray}
{\mathcal Z}&:=&\{f\in L^1(X\times \S^{n-1})\ |\ v\cdot\nabla_xf\in L^1(X\times \S^{n-1})\},\label{B2a}\\
\|f\|_{\mathcal Z}&:=&\|f\|_{L^1(X\times \S^{n-1})}+\|v\cdot\nabla_x f\|_{L^1(X\times \S^{n-1})};\label{B2b}
\end{eqnarray}
where $v\cdot \nabla_x$ is understood in the distributional sense.

It is known (see \cite{C-CRAS-1984,C-CRAS-1985}) that the trace map $\gamma_-$ from  $C^1(\bar X\times \S^{n-1})$ to $C(\Gamma_-)$ defined by 
\begin{equation}
\gamma_-(f)=f_{|\Gamma_-}
\end{equation}
extends to a continuous operator from ${\mathcal Z}$ onto $L^1(\Gamma_-, \tau_+(x,v)d\xi(x,v))$ and  admits a continuous lifting.
Note that $L^1(\Gamma_-,d\xi)$ is a subset of the spaces  $L^1(\Gamma_-, \tau_+(x,v)d\xi(x,v))$.

We introduce the following notation
\begin{eqnarray}
A_1f=-\sigma f,\ A_2f=\int_{\S^{n-1}}k(x,v',v)f(x,v')dv'.\label{B2c}
\end{eqnarray}
As $(\sigma,k)$ is admissible, the operators $A_1$ and $A_2$ are bounded operators in $L^1(X\times \S^{n-1})$.

Consider the following unbounded operators 
\begin{eqnarray} 
&&T_1f=-v\cdot \nabla_xf+A_1f,\ D(T_1)=\{f\in {\mathcal Z}\ |\ f_{|\Gamma_-}=0\},\label{B3}\\
&&Tf=T_1f+A_2 f,\ D(T)=D(T_1).
\end{eqnarray}
It is known that the unbounded operators $T_1$ and $T$ are generators of strongly continuous semigroups in $L^1(X\times \S^{n-1})$ $U_1(t)$, $U(t)$ respectively 
(see for example \cite[Proposition 2 pp 226]{dlen6}). In addition $U_1(t)$ and $U(t)$ preserve the cone of positive functions, and $U_1(t)$ is given explicitly by the 
following formula
\begin{equation}
U_1(t)f=e^{-\int_0^t\sigma(x-sv,v)ds}f(x-tv,v)\Theta(x,x-tv),\textrm{ for a.e. }(x,v)\in X\times \S^{n-1},\label{B4}
\end{equation}
for $f\in L^1(X\times \S^{n-1})$, where 
\begin{equation}
\Theta(x,y)=
\left\lbrace
\begin{matrix}
1\textrm{ if }x+p(y-x)\in X \textrm{ for all } p\in (0,1],\\
0\textrm{ otherwise},
\end{matrix}
\right.\label{B4.0}
\end{equation}
for $(x,y)\in \R^n\times \R^n$.

We recall the Dyson-Phillips formula
\begin{equation}
U(t)=\sum_{m=0}^{+\infty}H_m(t)\label{E1}
\end{equation}
for $t\ge 0$, where
\begin{eqnarray}
\!\!\!\!\!\!\!\!\!\!\!\!\!\!\!\!\!\!\!H_m(t)&:=&\int\limits_{s_1\ge 0,\ldots,s_m\ge0\atop s_1+\ldots+s_m\le t}U_1(t-s_1-\ldots-s_m)A_2U_1(s_1)\ldots A_2U_1(s_m)ds_1\ldots ds_m,\ m\ge 1,\label{E1a}\\
\!\!\!\!\!\!\!\!\!\!\!\!\!\!\!\!\!\!\!H_m(t)&=&\int_0^tH_{m-1}(t-s)A_2U_1(s)ds,\ m\ge 1, \label{E1b}\\
\!\!\!\!\!\!\!\!\!\!\!\!\!\!\!\!\!\!\!H_0(t)&:=&U_1(t).\label{E1c}
\end{eqnarray}

\subsection{Trace results}
We introduce the following space 
\begin{equation}
{\mathcal W}:=\{u\in L^1((0,T)\times X\times \S^{n-1})\ |\ \left({\pa \over \pa t}+v\cdot\nabla_x\right)u\in L^1((0,T)\times X\times \S^{n-1})\},\label{B22a}
\end{equation}
\begin{equation}
\|u\|_{\mathcal W}:=\|u\|_{L^1((0,T)\times X\times \S^{n-1})}+\left\|\left({\pa\over \pa  t}+v\cdot\nabla_x\right)u\right\|_{L^1((0,T)\times X\times \S^{n-1})};\label{B22b}
\end{equation}
where ${\pa \over \pa t}$ and $v\cdot\nabla_x$ are understood in the distributional sense.

It is known (see \cite{C-CRAS-1984,C-CRAS-1985}) that the trace map $\gamma_-$ (respectively $\gamma_+$) from  $C^1([0,T]\times \bar X\times \S^{n-1})$ to $C(X\times \S^{n-1})\times C((0,T)\times \Gamma_\pm)$ defined by 
\begin{equation}
\gamma_-(\psi)=(\psi(0,.), \psi_{|(0,T)\times \Gamma_-})\textrm{ (respectively }\gamma_+(\psi)=
(\psi(T,.),\psi_{|(0,T)\times \Gamma_+}{\rm )}
\end{equation}
extends to a continuous operator from ${\cal W}$ onto $L^1(X\times \S^{n-1}, \tau_+(x,v)dx dv)\times L^1((0,T)\times \Gamma_-, \min(T-t,\tau_+(x,v))dtd\xi(x,v))$ (respectively
$L^1(X\times \S^{n-1},\tau_-(x,v)dxdv)\times L^1((0,T)\times \Gamma_+,$ 

\noindent $\min(t,\tau_-(x,v))dtd\xi(x,v))$). In addition $\gamma_{\pm}$ admits a continuous lifting.
Note that $L^1(X\times \S^{n-1})$ is a subset of $L^1(X\times \S^{n-1}, \tau_+(x,v)dx dv)$.
Note also that $L^1((0,T)\times \Gamma_-,dt d\xi)$ (resp. $L^1((0,T)\times \Gamma_+,dt d\xi)$) is a subset of $L^1((0,T)\times \Gamma_-, \min(T-t,\tau_+(x,v))dtd\xi(x,v))$
(resp. $L^1((0,T)\times \Gamma_+, \min(t,$ $\tau_-(x,v))dtd\xi(x,v))$).

We now introduce the space 
\begin{equation}
\tilde{\mathcal W}:=\{u\in {\mathcal W}\ |\ \gamma_-(u)\in L^1(X\times \S^{n-1})\times L^1((0,T)\times \Gamma_-,dt d\xi)\}.\label{B33}
\end{equation}
We recall the following trace results (owed to \cite{C-CRAS-1984,C-CRAS-1985} in a more general setting).

\begin{lemma}
\label{lem:traceW}
The following equality is valid 
\begin{equation}
\tilde{\mathcal W}=\{u\in {\mathcal W}\ |\ \gamma_+(u)\in L^1(X\times \S^{n-1})\times L^1((0,T)\times \Gamma_+,dt d\xi)\}.\label{B4a}
\end{equation}
In addition the trace maps 
\begin{equation}
\gamma_{\pm}:\tilde{\mathcal W}\to L^1(X\times \S^{n-1})\times L^1((0,T)\times \Gamma_\pm,dt d\xi)\textrm{ are continuous, onto, and admit continuous lifting.}\label{B4b}
\end{equation}

\end{lemma}

\subsection{Solution to equation \eqref{eq:B1}}
We identify the space $L^1((0,r),L^1(\Gamma_\pm,d\xi))$ with the space $L^1((0,r)\times\Gamma_\pm,dtd\xi)$ for any $r>0$. We extend by $0$ on $\R$ outside the interval 
$(0,\eta)$ any function $\phi\in L^1((0,\eta), L^1(\Gamma_-,d\xi))$.

Let $\phi\in L^1((0,\eta),L^1(\Gamma_-,d\xi))$.
Then we consider the lifting  $G_-(t)\phi\in \tilde{\mathcal W}$ of $(0,\phi)$  defined by
\begin{equation}
G_-(t)\phi(x,v):=e^{-\int_0^{\tau_-(x,v)}\sigma(x-sv,v)ds}\phi_-(t-\tau_-(x,v),x-\tau_-(x,v)v,v),\textrm{ for a.e. }(t,x,v)\in (0,T)\times X\times \S^{n-1}.\label{B5}
\end{equation}
Note that $G_-(.)\phi$ is a solution in the distributional sense of the equation $({\pa \over \pa t}+v\cdot\nabla_x)u+\sigma u=0$ in $(0,T)\times X\times \S^{n-1}$ and 
\begin{equation}
\|G_-(.)\phi\|_{\mathcal W}\le (1+\|\sigma\|_{\infty})\|G_-(.)\phi\|_{L^1((0,T)\times X\times \S^{n-1})}\le (1+\|\sigma\|_{\infty})T\|\phi_-\|_{L^1((0,\eta)\times\Gamma_-,dt d\xi)}.\label{B6}
\end{equation}
To prove this two latter statements, one can use the following change of variables (see \cite{CS-OSAKA-99}).

\begin{lemma}
\label{lemchar}
We have
\begin{equation}
\int_{X\times \S^{n-1}}f(x,v)dxdv=\int_{\Gamma_\mp}\int_0^{\tau_\pm(x,v)}f(x\pm tv)dtd\xi(x,v),\label{eq:lemchar}
\end{equation}
for $f\in L^1(X\times V)$.
\end{lemma}

From \eqref{B6} we obtain that the map $i:L^1((0,\eta), L^1(\Gamma_-,d\xi))\to \tilde{\mathcal W}$ defined by
\begin{equation}
i(\phi)=G_-(.)\phi,\ \phi\in L^1((0,\eta), L^1(\Gamma_-,d\xi)),\label{B7}
\end{equation}
is continuous.

The following result holds (see \cite[Theorem 3 p. 229]{dlen6}).
\begin{lemma}
\label{lem:sol} 
The equation \eqref{eq:B1} admits a unique solution $u$ in $\tilde {\mathcal W}$ which is given by
\begin{equation}  
u(t)=G_-(t)\phi+\int_0^tU(t-s)A_2G_-(s)\phi ds.\label{B8}
\end{equation}
where $U(t)$ is the strongly continuous semigroup in $L^1(X\times \S^{n-1})$  introduced in subsection \ref{semi}.
\end{lemma}

Using \eqref{B8} and  the Dyson-Phillips expansion \eqref{E1} we obtain that the solution $u$ of \eqref{eq:B1}  may be decomposed as 
\begin{equation} 
u(t)=G_-(t)\phi+\sum_{m=0}^{\infty}\int_{-\infty}^tH_m(t-s)A_2G_-(s)\phi ds,\label{DP}
\end{equation}
for $t\ge 0$ and $\phi\in L^1((0,\eta)\times \pa X, dt d\mu(x))$. The first term in the above series
$G_-(t)\phi$ is the ballistic part of $u(t)$ while the term
corresponding to $m\geq1$ is $m$-linear in the scattering kernel $k$.
The term corresponding to $m=1$ is the single scattering term.

From \eqref{B7}, Lemma \ref{lem:sol} and \eqref{B4b}, we also obtain the existence of the albedo operator.
\begin{lemma}
\label{lem:albedo}
The albedo operator $A$ given by the formula
\begin{equation}
A\phi=u_{|(0,T)\times \Gamma_+}, \textrm{ for } \phi\in L^1((0,\eta),L^1(\Gamma_-,d\xi))\textrm{ where }u\textrm{ is given by }\eqref{B8}, 
\end{equation}
is well-defined and is a bounded operator from 
$L^1((0,\eta),L^1(\Gamma_-,d\xi))$ to $L^1((0,T),L^1(\Gamma_+,d\xi))$.
\end{lemma}
We refer the reader to \cite{CS-CPDE-96} for the reconstruction of the
optical parameters when the full albedo operator is known. We assume
here that only partial knowledge of the albedo operator is available
from measurements.

\section{The operator $A_{S,W}$ and its distributional kernel}
\label{sec:albedoaveraged}
\subsection{Angularly averaged measurements}
We now define more precisely the type of measurements we consider in
this paper.  The directional behavior of the source term is determined
by a fixed function $S(x,v)$, which is bounded and continuous on $\Gamma_-$.
We assume that the incoming conditions have the following structure
\begin{equation}
\phi_S(t',x',v')=S(x',v')\phi(t',x'),\ t'\in (0,\eta),\ (x',v')\in \Gamma_-,
\label{B9c}
\end{equation}
where $\phi(t,x)$ is an arbitrary function in $L^1((0,\eta)\times \pa
X)$. We model the detectors by the kernel $W(x,v)$, which we assume is
a continuous and bounded function on $\Gamma_+$. The available
measurements are therefore modeled by the availability of the averaged
albedo operator $A_{S,W}$ from $L^1((0,\eta)\times \pa X,dt d\mu(x))$
to $L^1((0,T)\times \pa X,dtd\mu(x))$ and defined by
\begin{equation}
A_{S,W}\phi(t,x)=\int_{\S^{n-1}_{x,+}}A(\phi_S)(t,x,v)W(x,v)(\nu(x)\cdot v)dv, \textrm{ for a.e. }(t,x)\in (0,T)\times \pa X.\label{B9a}
\end{equation}
The functions $S$ and $W$ are fixed throughout the paper. The case
$W\equiv1$ corresponds to measurements of the current of exiting
particles at the domain's boundary.


The decomposition of the transport solution \eqref{DP} translates
into a similar decomposition of the albedo operator of the form
\begin{equation}
A_{S,W}\phi(t,x)
=\sum_{m=0}^{+\infty}A_{m,S,W}\phi(t,x),\label{E2a}
\end{equation}
for $(t,x)\in (0,T)\times \pa X$, where  we have defined
\begin{equation}
  A_{0,S,W}\phi(t,x)=\int_{\S^{n-1}_{x,+}}\hspace{-.0cm}
  (\nu(x)\cdot v) W(x,v)\left(G_-(.)\phi_S\right)_{|(0,T)\times \Gamma_+}
  (t,x,v)dv,\label{E2b.1}
\end{equation}
\begin{equation}
A_{m,S,W}\phi(t,x)=\int_{\S^{n-1}_{x,+}}(\hspace{-.0cm}
 \nu(x)\cdot v)W(x,v)\left(\int_{-\infty}^t \hspace{-.2cm} H_{m-1}(t-s)
  A_2G_-(s)\phi_Sds\right)_{|(0,T)\times \Gamma_+}
   \hspace{-1cm}(t,x,v)dv,\label{E2b}
\end{equation}
for a.e. $(t,x)\in(0,T)\times \pa X$ where $\phi_S$ is defined by
\eqref{B9c}.  The kernels of the operators $A_{m,S,W}$ can be written
explicitly.

\subsection{Distributional kernel of the operators $A_{m,S,W}$}
Consider the nonnegative measurable $E$ from $\pa X\times\pa X\to \R$ defined by 
\begin{equation}
E(x_1,x_2)=
\left\lbrace
\begin{matrix}
e^{-\int_0^{|x_1-x_2|}\sigma(x_1-s{x_1-x_2\over |x_1-x_2|},{x_1-x_2\over|x_1-x_2|})ds}\textrm{ if }x_1+p(x_2-x_1)\in X \textrm{ for all } p\in (0,1),\\
0\textrm{ otherwise},\label{paXtimespaX_N1}
\end{matrix}\right.
\end{equation}
for a.e. $(x_1,x_2)\in \pa X\times \pa X$.
For $m\ge 3$, we also define the nonnegative measurable real function $E(x_1,\ldots,x_m)$ by the formula
\begin{equation} 
E(x_1,\ldots, x_m)=e^{-\sum_{i=1}^{m-1}\int_0^{|x_i-x_{i+1}|}\sigma(x_i-s{x_i-x_{i+1}\over |x_i-x_{i+1}|},{x_i-x_{i+1}\over|x_i-x_{i+1}|})ds}\Theta(x_m,x_{m-1})\Pi_{i=1}^{m-2}\Theta(x_i,x_{i+1}),\label{bar N2}
\end{equation}
for a.e. $(x_1,\ldots,x_m)\in \pa X\times(\R^n)^{m-2}\times\pa X$, where $\Theta$ is defined by \eqref{B4.0}.
The function $E(x_1,\ldots,x_m)$ measures the total attenuation along the broken path
$(x_1,\ldots,x_m)\in \pa X\times \R^{m-2}\times\pa X$ provided $(x_2,\ldots, x_{m-1})\in X^{m-2}$.

For $m\in \N$, $m\ge 1$ and for any subset $U$ of $\R^m$ we denote by
$\chi_U$ the characteristic function from $\R^m$ to $\R$ defined by
$\chi_U(y)=1$ when $y\in U$ and $\chi_U(y)=0$ otherwise.  Using \eqref{E2b.1}--\eqref{E2b}, \eqref{B5} and \eqref{E1b}--\eqref{E1c} we then
obtain the following result on the structure of the kernels of the
albedo operator.
\begin{proposition}
\label{prop:kernelA}
We have 
\begin{equation}
  A_{m,S,W}(\phi)(t,x)=\int_{(0,\eta)\times \pa X}
  \gamma_m(t-t',x,x')\phi(t',x')dt'd\mu(x'),\label{E8}
\end{equation}
for $m\ge 0$ and for a.e. $(t,x)\in (0,T)\times \pa X$, where
\begin{eqnarray}
\gamma_0(\tau,x,x')&:=&{E(x,x')\over |x-x'|^{n-1}}\left[W(x,v)S(x',v)(\nu(x)\cdot v)|\nu(x')\cdot v|\right]_{v={x-x'\over |x-x'|}}\delta(\tau-|x-x'|),\label{E8.0}\\
\gamma_1(\tau,x,x')&:=&\chi_{(0,+\infty)}(\tau-|x'-x|)\int_{\S^{n-1}_{x,+}}(\nu(x)\cdot v)W(x,v)
\left[E(x,x-sv,x')k(x-s v,v',v)\right.\nonumber\\
&&\!\!\!\!\!\!\!\!\!\!\!\!\!\!\!\!\!\!\!\!\!\!\!\!\!\!\!\!\!\!\!\!\!\!\!\!\!\!\!\!\!\!\times \left. \chi_{(0,\tau_-(x,v))}(s)S(x',v')|\nu(x')\cdot v'|\right]_
  {\big|v'={x-x'-sv\over |x-x'-sv|}\,;\,s={\tau^2-|x-x'|^2\over 2(\tau-v\cdot (x-x'))}}
{2^{n-2}(\tau-(x-x')\cdot v)^{n-3}\over |x-x'-\tau v|^{2n-4}} dv,\label{E8a}
\end{eqnarray}
for $(\tau,x,x')\in\R\times \partial X\times\partial X$  and where
$\gamma_m$ for $m\geq2$ admits a similar, more complex, expression
given in Section \ref{proof_thm_normemultiplescat} (see \eqref{E8b}--\eqref{E8d}).
\end{proposition}
Because the above formulas are central in our uniqueness and stability
results, we briefly present their derivation and refer the reader 
to Section \ref{proof_kernelA} for the rest of the proof of Proposition \ref{prop:kernelA}.
\begin{proof}[Derivation of \eqref{E8.0} and \eqref{E8a}]
  From \eqref{E2b.1} and the definition of $G_-$,  we obtain
  $A_{0,S,W}\phi(t,x)=\int_{\S^{n-1}_{x,+}}(\nu(x)\cdot v)
  W(x,v)E(x,x-\tau_-(x,v)v)S(x-\tau_-(x,v)v,v)\phi(t-\tau_-(x,v),x-\tau_-(x,v)v)dv$,
  $(t,x)\in (0,T)\times\pa X$ and for $\phi\in L^1((0,\eta)\times\pa
  X)$. Therefore, performing the change of variables
  ``$x'$''$=x-\tau(x,v)v$ ($dv={|\nu(x')\cdot v|\over
    |x-x'|^{n-1}}d\mu(x')$ and $\tau(x,v)=|x-x'|$), we obtain \eqref{E8.0}.
  
  From the definition of $A_2$ and $G_-$ we note that
  $A_2G_-(s)\phi_S(z,w):=\int_{\S^{n-1}}k(z,v',w)
  E(z,z-\tau_-(z,v')v')S(z-\tau_-(z,v')v',v')
  \phi(s-\tau_-(z,v'),z-\tau_-(z,v')v')dv',$ for a.e. $(z,w)\in
  X\times \S^{n-1}$ and for $\phi\in L^1((0,\eta)\times\pa X)$.
  Performing the change of variables ``$x'=z-\tau_-(z,v')v'$'', we
  obtain the equality $\left(A_2G_-(s)\phi_S\right)(z,w)=\int_{\pa
    X}\left[k(z,v',w)S(x',v')|\nu(x')\cdot v'| \right]_{v'={z-x'\over
      |z-x'|}}{E(z,x')\over |z-x'|^{n-1}}\phi(s-|z-x'|,x')d\mu(x')$,
  for a.e. $(z,w)\in X\times \S^{n-1}$ and $\phi\in
  L^1((0,\eta)\times\pa X)$.  Using also the definition of $A_{1,S,W}$
  (see \eqref{E2b} for $m=1$) we obtain the following equality for any
  $\phi\in L^1((0,\eta)\times\pa X)$ and for a.e. $(t,x)\in
  (0,T)\times\pa X$
\begin{eqnarray}
&&\!\!\!\!\!\!\!\!\!\!\!\!\!\!\!A_{1,S,W}(\phi)(t,x)=\int_{\S^{n-1}_{x,+}}\int_{-\infty}^t
\int_{\pa X}\left[k(x-(t-s)v,v',v)S(x',v')|\nu(x')\cdot v'|\right]_{v'={x-(t-s)v-x'\over |x-(t-s)v-x'|}}(\nu(x)\cdot v)\nonumber\\
&&\!\!\!\!\!\!\!\!\!\!\!\!\!\!\!\times{E(x,x-(t-s)v,x')\over |x-(t-s)v-x'|^{n-1}}\Theta(x,x-(t-s)v)\phi(s-|x-(t-s)v-x'|,x')W(x,v)d\mu(x')ds dv.\label{E11a} 
\end{eqnarray}
Then performing the changes of variables  ``$s$''$=t-s$ and  ``$t'$''$=t-s-|x-sv-x'|$ ($s={(t-t')^2-|x-x'|^2\over 2(t-t'-v\cdot (x-x'))}$, ${dt'\over ds}={2((t-t')-(x-x')\cdot v)^2\over |x-x'-(t-t')v|^2}$), we obtain \eqref{E8a}.
\end{proof}

To simplify notation, we define the multiple scattering kernels
\begin{equation}
  \label{eq:msop}
  \Gamma_k = \dsum_{m=k}^\infty \gamma_m.
\end{equation}

\subsection{Regularity of the albedo kernels}

The reconstruction of the optical parameters is based on an analysis
of the behavior in time of the kernels of the albedo operator. Our
first result in this direction is the following.
\begin{theorem}
\label{theorem:normesimplescat}
Assume that $k\in L^\infty(X\times\S^{n-1}\times\S^{n-1})$. Then the
following holds:
\begin{eqnarray}
  \label{A6b}
  \sqrt{\tau^2-|x-x'|^2}\gamma_1(\tau,x,x')\in L^\infty((0,T)\times\pa X\times\pa X) && \mbox{ when } n=2; \\
  \label{A6c}
  {\tau|x-x'|\over \ln\left({\tau+|x-x'|\over \tau-|x-x'|}\right)}\gamma_1(\tau,x,x')\in L^\infty((0,T)\times\pa X\times\pa X) && \mbox{ when }n=3; \\
  \tau|x-x'|^{n-2}\gamma_1(\tau,x,x')\in L^\infty((0,T)\times\pa X\times\pa X)
  && \mbox{ when } n\geq4 \label{A6d}.
\end{eqnarray}
In addition, assume that $k\in
L^\infty(X\times\S^{n-1}\times\S^{n-1})$ and that there exists
$\delta>0$ such that ${\rm supp}k\subseteq\{y\in X\ |\ \inf_{x\in \pa
  X}|x-y|\ge \delta\}$, i.e., the scattering coefficient vanishes in
the vicinity of $\partial X$. Then, the following holds
\begin{equation}
(\tau-|x-x'|)^{{3-n\over 2}}\gamma_1(\tau,x,x')\in L^\infty((0,T)\times 
  \pa X\times \pa X) \qquad \mbox{ when } n\ge 2.\label{A6e}
\end{equation}
\end{theorem}
Theorem
\ref{theorem:normesimplescat} is proved in Section
\ref{proof_thm_normesimplescat}. The results \eqref{A6c} and
\eqref{A6d} of Theorem \ref{theorem:normesimplescat}  correspond to singularities of
the single scattering contribution that depend on the values of $k$ on
$\partial X$. The above theorem shows that the structure of the single
scattering coefficient is quite different depending on whether $k$
vanishes on $\partial X$ or not.

The following result describes some regularity properties of the
multiple scattering. It is because multiple scattering is {\em more
  regular} than single scattering, in an appropriate sense, that we
can reconstruct the scattering coefficient in a stable manner.
\begin{theorem}
\label{theorem:normemultiplescat}
Assume that $k\in L^\infty(X\times\S^{n-1}\times\S^{n-1})$. 
Then the following holds:
\begin{eqnarray}
  \label{A6b-}
  \Gamma_2(\tau,x,x')\in L^\infty((0,T)\times\pa 
  X\times\pa X), && \mbox{ when } n=2; \\
  {\tau |x-x'|\Gamma_2(\tau,x,x')\over (\tau-|x-x'|)\left(1+\ln\left({\tau+|x-x'|\over \tau-|x-x'|}\right)\right)^2}\in L^\infty((0,T)\times\pa X\times\pa X),\label{A6c-} 
  && \mbox{ when } n= 3;  \\
  {\tau|x-x'|^{n-2}\over \tau-|x-x'|}\Gamma_2(\tau,x,x')\in L^\infty((0,T)\times\pa X\times \pa X), && \mbox{ when } n\geq4.\label{A6d-}
\end{eqnarray}

In addition, assume that $k\in
L^\infty(X\times\S^{n-1}\times\S^{n-1})$ and that there exists
$\delta>0$ such that ${\rm supp}k\subseteq\{y\in X\ |\ \inf_{x\in \pa
  X}|x-y|\ge \delta\}$. Then the following holds:
\begin{equation}
(\tau-|x-x'|)^{-1}\left(1+\ln\left({\tau+|x-x'|\over \tau-|x-x'|}\right)\right)^{-1}\Gamma_2(\tau,x,x')\in L^\infty((0,T)\times \pa X\times \pa X),\ \mbox{ when } n=3;\label{A6e-3}
\end{equation}
\begin{equation}
(\tau-|x-x'|)^{{1-n\over 2}}\Gamma_2(\tau,x,x')\in L^\infty((0,T)\times \pa X\times \pa X),\ \mbox{ when } n\geq4 .\label{A6e-}
\end{equation}
\end{theorem}
Theorem \ref{theorem:normemultiplescat} is proved in Section
\ref{proof_thm_normemultiplescat}. These results quantify how
``smoother'' multiple scattering is compared to the single scattering
contribution considered in Theorem \ref{theorem:normesimplescat}.

\subsection{Asymptotics of the single scattering term}

In this subsection we assume that $X$ is also convex. We give limits
for the single scattering term in two configurations given by:
\begin{equation}
\begin{array}{l}
\textrm{the nonnegative function }\sigma \textrm{ is bounded and  continuous on }X\times \S^{n-1},\\
\textrm{the nonnegative function }k \textrm{ is continuous on }\bar X\times\S^{n-1}\times\S^{n-1};
\end{array}
\label{H1}
\end{equation}
or 
\begin{equation}
\begin{array}{l}
\textrm{there exists a convex open subset }Y\subseteq X\textrm{ with }
C^1\textrm{ boundary such that }
\sigma(x,v)=0 \\\textrm{for }(x,v)\in (X\b Y)\times\S^{n-1}\textrm{ and }
\textrm{the nonnegative function }
\sigma\textrm{ is bounded and}\\\textrm{continuous on }
Y\times\S^{n-1}; \textrm{ and }
\textrm{there exists a convex open subset }Z\subseteq Y\subseteq X 
\\
\textrm{with }C^1\textrm{ boundary such that }
\delta:=\inf_{(x,z)\in \pa X\times Z}|x-z|>0 \textrm{ and }k(x,v',v)=0 \\
\textrm{for }(x,v',v)\in (X\b Z)\times\S^{n-1}\times\S^{n-1},
\textrm{ and the nonnegative function }k \\
\textrm{is bounded and continuous on }Z\times\S^{n-1}\times\S^{n-1},
\end{array}
\label{H2}
\end{equation}

When either \eqref{H1} or \eqref{H2} is satisfied, we want to analyze
the behavior of the function $\gamma_1(\tau,x,x')$ given by the right
hand side of \eqref{E8a} for all $(\tau,x,x')\in \R\times \pa
X\times\pa X$.  We need to introduce some notation.  Let
$\vartheta_0:\S^{n-1}\times X\to \R$ be the function defined by
\begin{equation}
\vartheta_0(v,x)=(\tau_-(x,v)\tau_+(x,v))^{-{n-1\over 2}},\ (v,x)\in \S^{n-1}\times X,\label{poids1.0}
\end{equation}
and consider the weighted X-ray transform $P_{\vartheta_0}$ defined by
\begin{equation} 
P_{\vartheta_0}f(v,x)=\int_{\tau_-(x,v)}^{\tau_+(x,v)}\vartheta_0(v,tv+x)f(tv+x)dt,\label{poids1}
\end{equation}
for a.e. $(v,x)\in \S^{n-1}\times \pa X$ and $f\in
L^2(X,\sup_{v\in \S^{n-1}}\vartheta_0(v,x)dx)$.  The first result
analyzes the behavior of $\gamma_1$ under hypothesis \eqref{H1}.
\begin{theorem}
\label{theorem:limH1}
Assume that the open subset $X$ of $\R^n$ with $C^1$ boundary is
convex.  Let $(x,x_0')\in \pa X^2$ be such that $x+s(x-x_0')\in X$ for
some $s\in (0,1)$. Set $v_0={x-x_0'\over |x-x_0'|}$ and
$t_0=|x-x_0'|$.  Then under condition \eqref{H1}, we have the
following results.
When $n=2$, then 
\begin{equation}
\begin{array}{ll}
\displaystyle\gamma_1(\tau,x,x_0')=&\displaystyle
{1\over \sqrt{\tau-t_0}}{\sqrt{2}W(x,v_0)S(x_0',v_0)(\nu(x)\cdot v_0)|\nu(x_0')\cdot v_0| E(x,x_0')\over \sqrt{t_0}}\\
&\displaystyle\times P_{\vartheta_0}k_{v_0}(v_0,x)+o\Big({1\over\sqrt{\tau-t_0}}\Big),\ \textrm{ as }\tau\to t_0^+,
\end{array}\label{A7b}
\end{equation}
where $P_{\vartheta_0}$ is defined by \eqref{poids1} and $k_{v_0}(y):=k(y,v_0,v_0)$ for $y\in X$.\\
When $n=3$, then
\begin{eqnarray}
\gamma_1(\tau,x,x_0')&=&\ln({1\over \tau-t_0}){\pi\over t_0^2}W(x,v_0)S(x_0',v_0)(\nu(x)\cdot v_0)|\nu(x_0')\cdot v_0|E(x,x_0')\nonumber\\
&&\times\left(k(x,v_0,v_0)+k(x_0',v_0,v_0)\right)+o\Big(\ln({1\over \tau-t_0})\Big),\textrm{ as }\tau\to t_0^+.\label{A7c}
\end{eqnarray}
When $n\ge 4$, then
\begin{eqnarray}
\gamma_1(\tau,x,x_0')&=&t_0^{1-n}E(x,x_0')\left[S(x_0',v_0)|\nu(x_0')\cdot v_0|\int_{\S_{x,+}^{n-1}}
\dfrac{W(x,v)(\nu(x)\cdot v)k(x,v_0,v)}{1-v\cdot v_0} dv \right.
\nonumber\\
&&\left.+W(x,v_0)(\nu(x)\cdot v_0)\int_{\S_{x_0',-}^{n-1}}
\dfrac{k(x_0',v',v_0)S(x_0',v')|\nu(x_0')\cdot v'|}{1-v'\cdot v_0}
dv'\right]\nonumber\\
&&+o(1),\ \textrm{as }\tau\to t_0^+. \label{A7d}
\end{eqnarray}
\end{theorem}
Theorem \ref{theorem:limH1} is proved in Section
\ref{proof_thm_H1H2}.   Note that $\gamma_1$ depends on the
value of $k$ on $\partial X$ in dimension $n\geq3$.
Under hypothesis \eqref{H2}, i.e., when the scattering coefficient vanishes
in the vicinity of where measurements are collected, we have the 
quite different behavior:
\begin{theorem}
\label{theorem:limH2}
Assume that the open subset $X$ of $\R^n$ with $C^1$ boundary is also
convex and assume that condition \eqref{H2} is fulfilled.  Let
$(x,x_0')\in \pa X^2$ be such that $x_0'+s(x-x_0')\in Z$ for some
$s\in (0,1)$.  Set $v_0={x-x_0'\over |x-x_0'|}$ and $t_0=|x-x_0'|$.
Then we have the following. 

When $n=2$, then \eqref{A7b} still holds.

When $n\ge3$, then
\begin{eqnarray}
\begin{array}{l}
\displaystyle\gamma_1(\tau,x,x_0')=(\tau-t_0)^{n-3\over 2}
(2t_0)^{{1-n\over 2}}{\rm Vol}_{n-2}(\S^{n-2})S(x_0',v_0)W(x,v_0)|\nu(x_0')\cdot v_0|(\nu(x)\cdot v_0)\\[3mm]
\displaystyle\times E(x,x_0')P_{\vartheta_0}k_{v_0}(v_0,x)+o((\tau-t_0)^{n-3\over 2}),
\ \textrm{as }\tau\to t_0^+,
\end{array}\label{A7e}
\end{eqnarray}
where $P_{\vartheta_0}$ is defined by \eqref{poids1} and $k_{v_0}(y):=k(y,v_0,v_0)$ for $y\in X$.
\end{theorem}
Theorem \ref{theorem:limH2} is proved in Section
\ref{proof_thm_H1H2}.  Theorem \ref{theorem:limH2} may remain valid
under different conditions from those stated in \eqref{H2}.  For
instance, when $\sigma$ is bounded and continuous on $X$ and $k$ is
continuous on $X\times\S^{n-1}\times\S^{n-1}$ and $k(x,.,.)$ decays
sufficiently rapidly as $x$ get closer and closer to the boundary $\pa
X$ for any $x\in X$, then the same asymptotics of $\gamma_1$ holds.

\section{Uniqueness and stability results}
\label{sec:stab}
We denote by $\gamma:=\Gamma_0=\sum_{m=0}^{+\infty}\gamma_m$ the
distributional kernel of $A_{S,W}$. Then $\gamma-\gamma_0=\Gamma_1$
denotes the distributional kernel of the multiple scattering of
$A_{S,W}$. For the rest of the paper, we assume that the duration of
measurement $T>\diam:=\sup_{(x,y)\in X^2}|x-y|$ so that the singularities of the ballistic and
single scattering contributions are indeed captured by the available
measurements.

Let $(\tilde \sigma,\tilde k)$ be a pair of absorption and scattering
coefficients that also satisfy \eqref{eq:hyp1p}.  We denote by a
superscript $\tilde {}$ any object (such as the albedo operator
$\tilde A$ or the distributional kernels $\tilde \gamma$ and $\tilde
\gamma_0$) associated to $(\tilde \sigma,\tilde k)$.  Moreover if
$(\sigma,k)$ satisfies \eqref{H2} for some $(Y,Z)$ and
$(\tilde\sigma, \tilde k)$ also satisfies \eqref{H2} for some $(\tilde
Y, \tilde Z)$, then we always make the additional assumption $Y=\tilde
Y$ and $Z=\tilde Z$.   Let $\|.\|_{\eta,T}:=\|.\|_{{\cal
    L}(L^1((0,\eta)\times \pa X)), L^1((0,T)\times \pa X))}$.

\subsection{Stability estimates under condition \eqref{H1} or \eqref{H2}}
\begin{theorem}
\label{theorem:stabH1}
Assume that the open subset $X$ of $\R^n$ with $C^1$ boundary is also
convex. Let $(\sigma,k)$ and $(\tilde \sigma, \tilde k)$ satisfy
condition \eqref{H1}.  Let $x_0'\in \pa X$.  Then we have:
\begin{equation}
\int_{\pa X}\left[{|E-\tilde E|(x,x_0')\over |x-x_0'|^{n-1}}
W(x,v_0)S(x',v_0)(\nu(x)\cdot v_0)|\nu(x_0')\cdot v_0|\right]_{t_0=
|x-x_0'|\atop v_0={x-x_0'\over |x-x_0'|}}\!\!\!\!\!\!\!\!\!\!\!\!\!\!\!\!
d\mu(x)\le\|A_{S,W}-\tilde A_{S,W}\|_{\eta,T}.\label{stab1.a}
\end{equation}
Let $x\in \pa X$ be such that $px_0'+(1-p)x\in X$ for some
$p\in(0,1)$. Set $v_0={x-x_0'\over |x-x_0'|}$ and $t_0=|x-x_0'|$.
When $n=2$, we have
\begin{eqnarray}
&&W(x,v_0)S(x_0',v_0)(\nu(x)\cdot v_0)|\nu(x_0')\cdot v_0|
\left|E(x,x_0')P_{\vartheta_0}k_{v_0}(v_0,x)-\tilde E(x,x_0')P_{\vartheta_0}\tilde k_{v_0}(v_0,x)\right|\nonumber\\
&&\le {1\over 2}\left\|\sqrt{\tau^2-|z-z'|^2}
  (\Gamma_1-\tilde \Gamma_1)(\tau,z,z')
  \right\|_{L^\infty},\label{stab1.b}
\end{eqnarray}
where $\|\cdot\|_{L^\infty} := \|\cdot \|_{L^\infty((0,T)\times \pa
  X\times\pa X)}$, $P_{\vartheta_0}$ is defined by \eqref{poids1} and
$k_{v_0}(y):=k(y,v_0,v_0)$ for $y\in X$ ($\tilde
k_{v_0}$ is defined similarly).

When $n=3$, then
\begin{eqnarray}
&&\left|E(x,x_0')(k(x,v_0,v_0)+k(x_0',v_0,v_0))-\tilde E(x,x_0') 
  (\tilde k(x,v_0,v_0)+ \tilde k(x_0',v_0,v_0))\right|\nonumber\\
&&\times W(x,v_0)S(x_0',v_0)(\nu(x)\cdot v_0)
  |\nu(x_0')\cdot v_0| 
  \le {1\over \pi}\left\|{\tau|z-z'|\over 
  \ln\left({\tau+|z-z'|\over \tau-|z-z'|}\right)}
(\Gamma_1-\tilde \Gamma_1)(\tau,z,z')\right\|
  _{L^\infty}.\label{stab1.c}
\end{eqnarray}

When $n\ge 4$, then
\begin{eqnarray}
&&S(x_0',v_0)|\nu(x_0')\cdot v_0|\left|\int_{\S_{x,+}^{n-1}}{W(x,v)(\nu(x)\cdot v)\over 1-v\cdot v_0}\right.\left(E(x,x_0')k(x,v_0,v)-\tilde E(x,x_0')\tilde k(x,v_0,v)\right)dv \nonumber\\
&&+W(x,v_0)(\nu(x)\cdot v_0)\left.\int_{\S_{x_0',-}^{n-1}}{S(x_0',v')|\nu(x_0')\cdot v'|\over 1-v'\cdot v_0}
\left(E(x,x_0')k(x_0',v',v_0)-\tilde E(x,x_0')\tilde k(x_0',v',v_0)\right)dv'\right|
\nonumber\\
&&\le\left\|\tau |z-z'|^{n-2}(\Gamma_1-\tilde \Gamma_1)
  (\tau,z,z')\right\|_{L^\infty}.\label{stab1.d}
\end{eqnarray}
\end{theorem}
Theorem \ref{theorem:stabH1} is proved in Section \ref{proof_thm_normesimplescat}. It
shows that the spatial structure of $k$ may be stably reconstructed at
the domain's boundary. More interesting is the following theorem,
which provides some stability of the reconstruction of the scattering
coefficient when it vanishes in the vicinity of the boundary $\partial
X$.

\begin{theorem}
\label{theorem:stabH2}
Assume that the open subset $X$ of $\R^n$ with $C^1$ boundary is also
convex. Assume also that $\inf_{(x',v')\in \Gamma_-}S(x',v')>0$ and
$\inf_{(x,v)\in \Gamma_+}W(x,v)>0$.  Let $(\sigma,k)$ and $(\tilde
\sigma,\tilde k)$ satisfy condition \eqref{H2}. Let $x_0'\in \pa X$.
Then there exist constants $C_1=C_1(S,W,X,Y)$ and 
$C_2=C_2(S,W,X,Z)$ such that
\begin{equation}
\int_{\S^{n-1}_{x_0',-}}|E-\tilde E|(x_0'+\tau_+(x_0',v_0)v_0,x_0')|\nu(x_0')\cdot v_0|dv_0\le
C_1\|A_{S,W}-\tilde A_{S,W}\|_{\eta,T},\label{stab2.a}
\end{equation}
\begin{equation}
\left|E(x,x_0')P_{\vartheta_0}k_{v_0'}(v_0',x_0')-\tilde E(x,x_0')
P_{\vartheta_0}\tilde k_{v_0'}(v_0',x_0') \right|
\le C_2\left\|(\tau-|z-z'|)^{3-n\over 2}
(\Gamma_1-\tilde \Gamma_1)(\tau,z,z')
\right\|_{L^\infty},\label{stab2.b}
\end{equation}
for $x\in \pa X$ such that $px_0'+(1-p)x\in Z$ for some $p\in (0,1)$
where $v_0'={x-x_0'\over |x-x_0'|}$, $P_{\vartheta_0}$ is defined in
\eqref{poids1}, and $k_{v_0'}(y):=k(y,v_0',v_0')$ for $y\in X$ ($\tilde k_{v_0'}$ is defined similarly).
\end{theorem}
Theorem \ref{theorem:stabH2}, which is one of the main results of this
paper, is proved in Section \ref{proof_thm_normemultiplescat}.

\subsection{The case when $X$ is a ball of $\R^n$}
When $X$ is an open Euclidean ball of $\R^n$, which is important from
the practical point of view in medical imaging as it is relatively
straightforward to place sources and detectors on a sphere, we are
able to invert the weighted X-ray transform $P_{\vartheta_0}f$,
$f(x,v):=f(x)\in L^2(X,\sup_{v\in \S^{n-1}}\vartheta_0(v,x)dx)$ using
the classical inverse X-ray transform (inverse Radon transform in
dimension $n=2$). In the next subsection we shall consider a larger
class of domains $X$, which requires one to solve more complex
weighted X-ray transforms.

Up to rescaling, we assume $X=B_n(0,1)$, the ball in $\R^n$ centered
at $0$ of radius $1$.  Consider the X-ray transform $P$ defined by
\begin{equation}
Pf(v,x)=\int_{\tau_-(x,v)}^{\tau_+(x,v)}f(sv+x)ds\textrm{ for a.e. }(v,x)\in \S^{n-1}\times \pa X,\label{x-ray}
\end{equation}
for $f\in L^2(X)$ (we extend $f$ by 0 outside $X$).
We have the following Proposition \ref{linkx-ray}.

\begin{proposition}
\label{linkx-ray}
When $X=B_n(0,1)$ we have
\begin{equation}
P_{\vartheta_0}f(v,x)=P(\varrho f)(v,x), \textrm{ for a.e. }(v,x)\in \S^{n-1}\times \pa X,\label{link}
\end{equation}
for $f\in L^2(X,\sup_{v\in \S^{n-1}}\vartheta_0(v,x)dx)$ where $\varrho(y):=(1-|y|^2)^{-{n-1\over 2}}$, $y\in X$.
\end{proposition}
\begin{proof}[Proof of Proposition \ref{linkx-ray}] 
It is easy to see that
\begin{eqnarray}
\tau_\pm(tv+qv^\bot,v)&=&\sqrt{1-q^2}\mp t,\label{ball1}\\
\vartheta_0(v,x)&=&(1-q^2-t^2)^{-{n-1\over2}}=(1-|x|^2)^{-{n-1\over 2}},
\label{ball2}
\end{eqnarray}
for $(t,q)\in \R^2$, $t^2+q^2\le 1$ and for $(v,v^\bot)\in \S^{n-1}\times\S^{n-1}$, $v\cdot v^\bot=0$, where $x=tv+qv^\bot$ (we remind that $\vartheta_0$ is defined by \eqref{poids1.0}).
Then Proposition \ref{linkx-ray} follows from the definition \eqref{poids1}.
\end{proof}

Assume that $(\sigma,k)$ satisfies condition \eqref{H1} when $n=2$ or
\eqref{H2} when $n\ge 2$. Assume also that $k(x,v,v')=k_0(x)g(v,v')$
for a.e. $(x,v,v')\in X\times\S^{n-1}\times\S^{n-1}$ where $g$ is a
given continuous function on $\S^{n-1}\times\S^{n-1}$,
$\inf_{v\in \S^{n-1}}g(v,v)>0$, and where $k_0\in
L^\infty(X)$.  Then from the decomposition of the angularly averaged
albedo operator $A_{S,W}$ (Proposition \ref{prop:kernelA}) and from
Theorems \ref{theorem:normesimplescat},
\ref{theorem:normemultiplescat}, \ref{theorem:limH1} and
\ref{theorem:limH2}, and from Proposition \ref{linkx-ray} and methods
of reconstruction of a function from its X-ray transform, it follows
that $(\sigma, k_0)$ can be reconstructed from the asymptotic expansion in time of $A_{S,W}$ provided that $\sigma=\sigma(x)$ and
$\inf_{(x',v')\in \Gamma_-}S(x',v')>0$ and $\inf_{(x,v)\in
  \Gamma_+}W(x,v)>0$.  In addition we have the following stability
estimates.
\begin{theorem} 
\label{cor_ball}
Assume $X=B_n(0,1)$ and $\inf_{(x',v')\in \Gamma_-}S(x',v')>0$ and
$\inf_{(x,v)\in \Gamma_+}W(x,v)>0$.  Let $(\sigma,k)$ and $(\tilde
\sigma,\tilde k)$ satisfy either condition \eqref{H2} or \eqref{H1}.
Assume that $\sigma$, $\tilde \sigma$ do not depend on the velocity
variable ($\sigma(x,v)=\sigma(x)$) and ${\rm supp}\sigma\cup{\rm
  supp}\tilde \sigma\subseteq Y$, where $Y\subseteq X$ is a convex open subset of $\R^n$ with
$C^1$ boundary, and let $M=\max(\|\sigma\|_{L^\infty(Y)},\|\tilde \sigma\|_{L^\infty(Y)})$.  Assume $k(x,v,v')=k_0(x)g(v,v')$ and $\tilde
k(x,v,v')=\tilde k_0(x)g(v,v')$, $g(v,v)>0$, for $(x,v,v')\in X\times
\S^{n-1}\times\S^{n-1}$ where $g$ is an a priori known continuous
function on $\S^{n-1}\times \S^{n-1}$.  

Then
there exists $C_3=C_3(S,W,X,Y,M)$ such that
\begin{equation}
\|\sigma-\tilde\sigma\|_{H^{-{1\over 2}}(Y)}\le C_3\|\sigma-\tilde\sigma\|_{L^\infty(Y)}^{1\over 2}\|A_{S,W}-\tilde A_{S,W}\|_{\eta,T}^{1\over 2}.\label{eq:cor1}
\end{equation}
When $n\ge 2$ and $(\sigma,k)$ satisfies \eqref{H2}, there
  exists $C_{4,1}=C_{4,1}(S,W,X,Y,Z,M,g)$ such that
\begin{eqnarray}
\|\varrho(k_0-\tilde k_0)\|_{H^{-{1\over 2}}(Z)}&\le&  
 C_{4,1}\|k_0-\tilde k_0\|_\infty^{1\over 2}\left(\|\tilde k_0\|_\infty\|A_{S,W}
 -\tilde A_{S,W}\|_{\eta,T}\right.\label{eq:cor2H2}\\
&&\left.+\left\|(\tau-|z-z'|)^{3-n\over 2}
  (\Gamma_1-\tilde \Gamma_1)(\tau,z,z')
\right\|_{L^\infty}\right)^{{1\over 2}}.\nonumber
\end{eqnarray}
When $n=2$ and $(\sigma,k)$ satisfies \eqref{H1}, there exists $C_{4,2}=C_{4,2}(S,W,X,M,g)$ such that
\begin{eqnarray}
\|\varrho(k_0-\tilde k_0)\|_{H^{-{1\over 2}}(X)}&\le&  C_{4,2}\|k_0-\tilde k_0\|_\infty^{3\over 4}\left(\|\tilde k_0\|_\infty\|A_{S,W}-\tilde A_{S,W}\|_{\eta,T}\right.\label{eq:cor2H1}\\
&&\left.+\left\|\sqrt{\tau^2-|z-z'|^2}(\Gamma_1-\tilde \Gamma_1)(\tau,z,z')
\right\|_{L^\infty}\right)^{{1\over 4}}.\nonumber
\end{eqnarray}
\end{theorem}
Theorem \ref{cor_ball} can be proved by mimicking the proof of Theorem
\ref{cor} given below for a larger class of domains $X$.  
However we give a proof of estimate \eqref{eq:cor2H1} in Section \ref{proof_thm_stabH1H2_cor}.

Note that
the left-hand side $\|\varrho(k_0-\tilde k_0)\|_{H^{-{1\over 2}}(Z)}$
of \eqref{eq:cor2H2} can be replaced by $\|k_0-\tilde
k_0\|_{H^{-{1\over 2}}(Z)}$ since $\varrho^{-1}\in C^\infty(\bar Z)$
and the operator $f\to \varrho^{-1}f$ is bounded in $H^{-{1\over
    2}}(Z)$ for any open convex subset $Z$ (with $C^1$ boundary) of
$X$ which satisfies $\bar Z\subseteq X$.

Under the assumptions of Theorem \ref{cor_ball} and additional
regularity assumptions on $(\sigma,k)$ one obtains stability estimates
similar to those given in Corollary \ref{cor2} given below for a
larger class of domains $X$.

\subsection{Uniqueness and stability estimates for more general domains $X$}

\begin{theorem}
\label{cor}
Assume that the open subset $X$ of $\R^n$ is convex with a real
analytic boundary and that $\inf_{(x',v')\in \Gamma_-}S(x',v')>0$ and
$\inf_{(x,v)\in \Gamma_+}W(x,v)>0$.  Let $(\sigma,k)$ and $(\tilde
\sigma,\tilde k)$ satisfy condition \eqref{H2}. Assume also that
$\sigma$, $\tilde \sigma$ do not depend on the velocity variable
($\sigma(x,v)=\sigma(x)$) and $k(x,v,v')=k_0(x)g(x,v,v')$ and $\tilde
k(x,v,v')=\tilde k_0(x)g(x,v,v')$, $g(x,v,v')>0$, for $(x,v,v')\in
X\times \S^{n-1}\times\S^{n-1}$ where $g$ is an a priori known real
analytic function on $X\times \S^{n-1}\times \S^{n-1}$ and where ${\rm
  supp}k_0\cup{\rm supp}\tilde k_0\subseteq \bar Z$, $(k_0,\tilde
k_0)\in L^\infty(Z)$.  Then estimate \eqref{eq:cor1} still holds and
there exists $C_4=C_4(S,W,X,Y,Z,M,g)$ such that
\begin{eqnarray}
\|k_0-\tilde k_0\|_{H^{-{1\over 2}}(Z)}&\le&  C_4\|k-\tilde k\|_\infty^{1\over 2}\left(\|\tilde k\|_\infty\|A_{S,W}-\tilde A_{S,W}\|_{\eta,T}\right.\label{eq:cor2}\\
&&\left.+\left\|(\tau-|z-z'|)^{3-n\over 2}(\Gamma_1-\tilde \Gamma_1)(\tau,z,z')
\right\|_{L^\infty}\right)^{{1\over 2}},\nonumber
\end{eqnarray}
where $M=\max(\|\sigma\|_{L^\infty(Y)},\|\tilde \sigma\|_{L^\infty(Y)})$.
\end{theorem}
Theorem \ref{cor} is proved in Section \ref{proof_thm_stabH1H2_cor}.  Assume
that $X$ is convex with a real analytic boundary and that
$\inf_{(x',v')\in \Gamma_-}S(x',v')>0$ and $\inf_{(x,v)\in
  \Gamma_+}W(x,v)>0$.  Let $Y$ and $Z$ be open convex subsets of
$X$, $\bar Z\subset X$, $Z\subseteq Y\subseteq X$, with a $C^1$
boundary.  Let $g$ be an a priori known real analytic function on
$X\times \S^{n-1}\times \S^{n-1}$, $g(x,v,v')>0$ for $(x,v,v')\in
X\times \S^{n-1}\times\S^{n-1}$.  Let $r_1>0$, $r_2>0$.  Consider the
class
\begin{eqnarray}
N&:=&\big\{(\sigma, k) \in H^{{n\over 2}+r_1}(Y)\times L^\infty(Z\times\S^{n-1}\times\S^{n-1}) \ |\ \|\sigma\|_{H^{{n\over 2}+r_1}(Y)}\le M_1,\nonumber\\
&&k=k_0g,\ {\rm supp}k_0\subseteq\bar Z,\ \|k_0\|_{H^{{n\over 2}+r_2}(Z)}\le M_2\big\}.\label{eq:cor4}
\end{eqnarray}
Note that there exist a function $D_1:\N\times(0,+\infty)\to (0,+\infty)$ such
that
\begin{equation}
\begin{array}{l}
\|\sigma\|_{L^\infty(Y)}\le D_1(n,r_1)\|\sigma\|_{H^{{n\over 2}+r_1}(Y)}\le D_1(n,r_1)M_1,\\
\|k_0\|_{L^\infty(Z)}\le D_1(n,r_2)\|k_0\|_{H^{{n\over 2}+r_2}(Z)}\le D_1(n,r_2)M_2,\\
\|k\|_{L^\infty(Z)}\le \|g\|_{L^\infty(Z)}\|k_0\|_{L^\infty(Z)}\le D_1(n,r_2)M_2\|g\|_{L^\infty(Z)},
\end{array}
\label{eq:cor4b}
\end{equation}
for $(\sigma,k)\in N$.
We also use the interpolation formula
\begin{equation}
\|f\|_{H^s(O)}\le \|f\|_{H^{s_1}(O)}^{s_2-s\over s_2-s_1}\|f\|_{H^{s_2}(O)}^{s-s_1\over s_2-s_1},\label{eq:cor4c}
\end{equation}
for $s_1<s<s_2$ and for $(O,s_1,s_2)\in \{(Y,-{1\over 2},{n\over 2}+r_1),\ (Z,-{1\over 2},{n\over 2}+r_2)\}$.
Using Theorem \ref{cor} and \eqref{eq:cor4b}, and applying \eqref{eq:cor4c} on $f=\sigma-\tilde \sigma$ and $f=k_0-\tilde k_0$ we obtain the following result.
\begin{corollary}
\label{cor2}
Let $(\sigma, k)$, $(\tilde \sigma,\tilde k)\in N$.
Then, for $-{1\over 2}\le s\le {n\over 2}+r_1$ and for $0<r<r_1$, there exists $C_5=C_5(S,W,X,Y,M_1,r_1,s)$ such that
\begin{eqnarray}
\|\sigma-\tilde\sigma\|_{H^s(Y)}&\le& C_5\|\sigma-\tilde\sigma \|_{L^\infty(Y)}^{\kappa\over 2}\|A_{S,W}-\tilde A_{S,W}\|_{\eta,T}^{\kappa\over 2},\label{eq:cor5}\\
\|\sigma-\tilde\sigma\|_{H^{{n\over 2}+r}(Y)}&\le& C_6\|A_{S,W}-\tilde A_{S,W}\|_{\eta,T}^{\kappa'\over 2-\kappa'},\label{eq:cor5b}
\end{eqnarray}
where $(\kappa,\kappa')=\left({n+2(r_1-s)\over n+1+2r_1},{2(r_1-r)\over n+1+2r_1}\right)$ and $C_6=C_5^{2\over 2-\kappa'}D_1(n,r)^{\kappa'\over 2-\kappa'}$ ($D_1(n,r)$ is defined by 
\eqref{eq:cor4b}).  
In addition, there exists
$C_7=C_7(S,W,X,Y,Z,g,M_1,r_1,M_2,r_2,s)$ such that
\begin{eqnarray}
\|k_0-\tilde k_0\|_{H^s(Z)}&\le& C_7\|k_0-\tilde k_0\|_{L^\infty(Z)}^{\kappa\over 2}\left(\|A_{S,W}-\tilde A_{S,W}\|_{\eta,T}
\right.
\label{eq:cor6}
\\
&&\left.
+\left\|(\tau-|z-z'|)^{3-n\over 2}(\Gamma_1-\tilde \Gamma_1)(\tau,z,z')
\right\|_{L^\infty}\right)^{\kappa\over 2},\nonumber
\end{eqnarray}
\begin{equation}
\|k_0-\tilde k_0\|_{H^{{n\over 2}+r}(Z)}\le C_8\left(\|A_{S,W}-\tilde A_{S,W}\|_{\eta,T}
\label{eq:cor6b}
+\left\|(\tau-|z-z'|)^{3-n\over 2}(\Gamma_1-\tilde \Gamma_1)(\tau,z,z')
\right\|_{L^\infty}\right)^{\kappa'\over 2-\kappa'},
\end{equation}
for $-{1\over 2}\le s\le {n\over 2}+r_2$ and $0<r<r_2$, where $(\kappa,\kappa')=\left({n+2(r_2-s)\over n+1+2r_2},{2(r_2-r)\over n+1+2r_2}\right)$
and $C_8=C_7^{2\over 2-\kappa'}D_1(n,r)^{\kappa'\over 2-\kappa'}$ ($D_1(n,r)$ is defined by 
\eqref{eq:cor4b}). 

\end{corollary}

\begin{remark}
\label{rem:cor2}
\rm
(i.) Theorem \ref{cor} and Corollary \ref{cor2} remain valid when: $X$ is
only assumed to be convex with $C^2$ boundary; the weight
$\vartheta_o$ defined by \eqref{poids1.0} (resp. the function $g$
which appears in the assumptions of Theorem \ref{cor} and Corollary
\ref{cor2}) is sufficiently close (in the $C^2$ norm) to an analytic
weight $\theta_{0,a}$ on the vicinity of $\bar Z\times \S^{n-1}$
(resp. an analytic function $g_a$ on the vicinity of $\bar
Z\times\S^{n-1}\times\S^{n-1}$); see proof of Theorem \ref{cor} and
\cite[Theorem 2.3]{FSU-JGA-08}.

(ii.) When $n=3$ then under hypothesis \eqref{H2}, we have
\begin{eqnarray*}
\left\|(\tau-|z-z'|)^{3-n\over 2}(\Gamma_1-\tilde \Gamma_1)(\tau,z,z')
\right\|_{L^\infty}
=\left\|\sum_{m=1}^\infty (A_{m,S,W}-\tilde A_{m,S,W})\right\|_{{\cal L}(L^1((0,\eta)\times\pa X),L^{\infty}((0,T)\times\pa X))}.
\end{eqnarray*}
where the distributional kernel of the bounded operator $\sum_{m=1}^{+\infty}(A_{m,S,W}-\tilde A_{m,S,W})$ from $L^1((0,\eta)\times\pa X)$ to $L^1((0,T)\times \pa X)$ is given by $\Gamma_1-\tilde\Gamma_1$. Therefore when $n=3$ and under condition \eqref{H2}, the
right-hand side of the stability estimates \eqref{eq:cor2} and
\eqref{eq:cor6} can be expressed with operator norms only (instead of
using a norm on the distributional kernel of the multiple scattering).
\end{remark}

\section{Proof of Theorems \ref{theorem:limH1}, \ref{theorem:limH2}}
\label{proof_thm_H1H2}
\begin{proof}[Proof of Theorem \ref{theorem:limH2}.]
For the sake of simplicity and without loss of generality we assume
  $v_0=(1,0,\ldots,0)$.  Assume that condition \eqref{H2} is
  satisfied.  For $n\ge 2$ consider the following open subset of
  $(0,+\infty)\times \S^{n-1}\times\S^{n-1}$
\begin{equation}
{\cal D}:=\{(s,v,v')\in (0,+\infty)\times \S^{n-1}_{x,+}\times \S^{n-1}_{x_0',-}\ |\ s\in (0,\tau_-(x,v))\}.\label{1.9a}
\end{equation}
Then we introduce the bounded function $\Psi_n$ on ${\cal D}$ defined by
\begin{equation} 
\Psi_n(s,v,v')=2^{n-2}W(x,v)(\nu(x)\cdot v) E(x,x-s v,x_0')k(x-s v,v',v)S(x_0',v')|\nu(x_0')\cdot v'|,\label{1.9b}
\end{equation}
for $(s,v,v')\in {\cal D}$. Note that from convexity of $X$ it follows that $\tau_{\pm}$ is continuous on 
$\Gamma_{\mp}$ and 
$E(x,x-sv,x_0')=e^{-\int_0^s\sigma(x-pv,v)dp-\int_0^{|x-x_0'-sv|}\sigma(x-sv-p{x-x_0'-sv\over |x-x_0'-sv|},{x-x_0'-sv\over |x-x_0'-sv|})dp}$ for $v\in \S^{n-1}_{x,+}$ and $0<s<\tau_-(x,v)$. 
Under \eqref{H2} we obtain that
\begin{equation}
\begin{array}{l}
\Psi_n(s,v,v')=0\textrm{ for }(s,v,v')\in (0,+\infty)\times\S_{x,+}^{n-1}\times\S_{x_0',-}^{n-1}\textrm{ such that }x-sv\not\in \bar Z,\\
\textrm{and the function }\Psi_n \textrm{ is continuous at any point }(s,v,v')\in {\cal D} \textrm{ such that }x-sv\in Z.
\end{array}
\label{limH2.1}
\end{equation}

We first prove \eqref{A7b} for $n=2$. Let $\tau>t_0$.  From
\eqref{1.9b}, \eqref{E8a}, it follows that
\begin{eqnarray}
\gamma_1(\tau,x,x_0')&=&\int_{-\alpha_0}^{\pi-\alpha_0}{1\over \tau-t_0\cos(\Omega)}\left[\chi_{(0,\tau_-(x,v))}(s)\Psi_2(s,v,v')\right]_{{v=(\cos \Omega,\sin\Omega)\atop 
v'={t_0(1,0)-sv \over \tau-s}}
\atop s={\tau^2-t_0^2\over 2(\tau-t_0\cos(\Omega))}}
d\Omega\nonumber\\
&=&\gamma_{1,1}(\tau,x,x_0')+\gamma_{1,2}(\tau,x,x_0'),\label{1.9c}
\end{eqnarray}
where $\S^{n-1}_{x,+}=\{(\cos\Omega,\sin \Omega)\ | \ -\alpha_0<\Omega<\pi -\alpha_0\}$ ($0<\alpha_0<\pi$) and 
\begin{eqnarray}
\gamma_{1,1}(\tau,x,x_0')&=&\int_0^{\pi}{\chi_{(0,\pi-\alpha_0)}(\Omega)\over \tau-t_0\cos(\Omega)}\left[\chi_{(0,\tau_-(x,v))}(s)\Psi_2(s,v,v')\right]_{{v=(\cos\Omega,\sin \Omega)
\atop v'={t_0(1,0)-sv \over \tau-s}}
\atop s={\tau^2-t_0^2\over 2(\tau-t_0\cos(\Omega))}}d\Omega,\label{1.9d}\\
\gamma_{1,2}(\tau,x,x_0')&=&\int_{-\pi}^0{\chi_{(-\alpha_0,0)}(\Omega)\over \tau-t_0\cos(\Omega)}\left[\chi_{(0,\tau_-(x,v))}(s)\Psi_2(s,v,v')\right]_{{v=(\cos\Omega,\sin \Omega)
\atop v'={t_0(1,0)-s v\over \tau-s}}
\atop s={\tau^2-t_0^2\over 2(\tau-t_0\cos(\Omega))}}d\Omega.\label{1.9e}
\end{eqnarray}

We shall prove that
\begin{eqnarray}
&&\sqrt{\tau-t_0}\gamma_{1,i}(\tau,x,x_0')\to {W(x,v_0)S(x_0',v_0)(\nu(x)\cdot v_0)|\nu(x_0')\cdot v_0|E(x_0',x)\over \sqrt{2t_0}}
\int_0^{t_0}{k(x-s v_0,v_0,v_0)\over \sqrt{s(t_0-s)}}ds,\nonumber\\
&&{\rm as}\ \tau\to t_0^+,\label{1.9f}
\end{eqnarray}
for $i=1,2$. Then adding \eqref{1.9f} for $i=1$ and $i=2$, we obtain
\eqref{A7b}.  We only prove \eqref{1.9f} for $i=1$ since the proof for
$i=2$ is similar.  Let $\tau>t_0$. Using the change of variables
$s={\tau^2-t_0^2\over 2(\tau-t_0\cos(\Omega))}-{\tau-t_0\over 2}$, we
obtain
\begin{equation}
\gamma_{1,1}(\tau,x,x_0')={1\over \sqrt{\tau^2-t_0^2}}\int_0^{t_0}\chi_{(0,\pi-\alpha_0)}(\Omega(s,\tau)){\chi_{(0,\tau_-(x,v(s,\tau)))}(s+{\tau-t_0\over 2})
\Psi_2(s,v(s,\tau),v'(s,\tau))\over\sqrt{s(t_0-s)}}d\tau,\label{1.11}
\end{equation}
where
\begin{eqnarray}
\begin{array}{l}
\displaystyle v(s,\tau)=(\cos\Omega(s,\tau),\sin\Omega(s,\tau)),\ \Omega(s,\tau)={\rm arccos}\left({\tau-{\tau^2-t_0^2\over 2s+\tau-t_0}\over t_0}\right),\\
\displaystyle v'(s,\tau)={t_0(1,0)-\left(s+{\tau-t_0\over 2}\right) v(s,\tau)\over {\tau+t_0\over 2}-s}.
\end{array}
\label{1.10}
\end{eqnarray}

Let $s\in (0,t_0)$. From \eqref{1.10}, it follows that
\begin{equation}
v(s,\tau)\to (1,0)\ {\rm as}\ \tau\to t_0^+,\ 
v'(s,\tau)\to (1,0)\ {\rm as}\ \tau\to t_0^+.\label{1.11b}
\end{equation}
Note that using the definition of $v_0$ and using the assumption
$x_0'+\ep (x-x_0')\in X$ for some $\ep\in (0,1)$ we obtain
$t_0=\tau_-(x,v_0)$.  Note also that the function $s\mapsto {1\over
  \sqrt{s(t_0-s)}},$ $s\in (0,t_0)$, is integrable in $(0,t_0)$.
Therefore, using \eqref{limH2.1}, the boundedness of $\Psi_2$ on
${\cal D}$ and the Lebesgue dominated convergence theorem, we obtain
\eqref{1.9f}.  This proves \eqref{A7b} when $n=2$.


Let $n\ge 3$ and prove \eqref{A7e}. From \eqref{1.9b} and
\eqref{E8a}, it follows that
\begin{equation}
\gamma_1(\tau,x,x_0')
=\int_{\S^{n-1}}{(\tau -t_0v_0\cdot v)^{n-3}\over |t_0v_0-\tau v|^{2n-4}}\chi_{(0,+\infty)}(\nu(x)\cdot v)\Psi_n(s,v,v')_{v'={t_0v_0-s v\over \tau-s}\atop s={\tau^2-t_0^2\over 2(\tau-t_0v\cdot v_0)}}dv,
\label{11.2}
\end{equation}
for $\tau>|x-x_0'|$.

Let $\Phi(\Omega, \omega)=(\sin \Omega, \cos(\Omega)\omega_1,\ldots,\cos(\Omega)\omega_{n-1})$ for $\Omega\in (-{\pi\over 2}, {\pi \over 2})$ and $\omega=(\omega_1,\ldots,
\omega_{n-1})\in \S^{n-2}$. Using spherical coordinates we obtain
\begin{eqnarray}
\gamma_1(\tau,x,x_0')
&=&\int_{-\pi/2}^{\pi/2}\cos(\Omega)^{n-2}{(\tau-t_0\sin(\Omega))^{n-3}\over (t_0^2+\tau^2-2t_0\tau\sin(\Omega))^{n-2}}\label{11.4}\\
&&\int_{\S^{n-2}}\chi_{(0,+\infty)}(\nu(x)\cdot\Phi(\Omega,\omega))\Psi_n(s,\Phi(\Omega,\omega),v')_{v'={t_0v_0-s\Phi(\Omega,\omega)\over \tau-s}\atop s={\tau^2-t_0^2\over 2(\tau-t_0\sin(\Omega))}}d\omega d\Omega,
\nonumber
\end{eqnarray}
for $\tau>t_0$.
Performing the change of variables ``$r={\tau^2-t_0^2\over 2(\tau-t_0\sin(\Omega))}-{\tau-t_0\over 2}$'' on the first integral on the right-hand side of \eqref{11.4}, we obtain
\begin{eqnarray}
\gamma_1(\tau,x,x_0')
&=&2^{2-n}t_0^{2-n}(\tau^2-t_0^2)^{n-3\over 2}\int_0^{t_0}{\sqrt{r(t_0-r)}^{n-3}\over ({\tau-t_0\over 2}+r)^{n-2}({\tau+t_0\over 2}-r)^{n-2}}\label{6.5}\\
&&\int_{\S^{n-2}}\left[\chi(\Phi(\Omega,\omega))\Psi_n(r+{\tau-t_0\over 2},\Phi(\Omega,\omega),v')\right]_{{\Omega={\rm arcsin}
(t_0^{-1}(\tau-{(\tau^2-t_0^2)\over 2(r+{\tau-t_0\over 2})}))
\atop s=r+{\tau-t_0\over 2}}\atop v'={t_0v_0-s\Phi(\Omega,\omega)\over \tau-s}}d\omega dr.
\nonumber
\end{eqnarray}
Therefore using \eqref{6.5}, \eqref{limH2.1} and \eqref{1.9b} and
using Lebesgue dominated convergence theorem, we obtain \eqref{A7e}.
This concludes the proof of Theorem \ref{theorem:limH2}.
\end{proof}

\begin{proof}[Proof of Theorem \ref{theorem:limH1}]
For the sake of simplicity and without loss of generality we assume $v_0=(1,0,\ldots,0)$.
Assume that condition \eqref{H1} is satisfied.
We consider the measurable function $\Psi_n$ defined by \eqref{1.9b} for all $(s,v,v')\in {\cal D}$ where ${\cal D}$ is defined by \eqref{1.9a}.
Under \eqref{H1} we obtain that
\begin{equation}
\begin{array}{l}
\displaystyle\textrm{the function }\Psi_n \textrm{ is continuous at any point }(s,v,v')\in {\cal D}\\
\displaystyle (\textrm{i.e. for any }(s,v,v')\in (0,+\infty)\times\S_x^{n-1}\times\S_{x_0'}^{n-1},\ x-sv\in X),
\end{array}
\label{limH1.1}
\end{equation}
The proof of \eqref{A7b} under condition \eqref{H1} is actually similar to the proof of \eqref{A7b} under condition \eqref{H2}.
Note that \eqref{1.9c}--\eqref{1.9e} still hold so that we have to prove that \eqref{1.9f} still holds for $i=1,2$. Again we only sketch the proof of \eqref{1.9f} for $i=1$. 
Note also that \eqref{1.11}--\eqref{1.11b} still hold.
Then using \eqref{1.11}--\eqref{1.11b}, \eqref{limH1.1} and \eqref{1.9b} and using Lebesgue dominated convergence theorem, we obtain \eqref{1.9f} for $i=1$.
This proves \eqref{A7b}.

Let $n\ge 3$. Formula \eqref{11.4} still holds.
Now assume that $n=3$. We shall prove \eqref{A7c}. Let $\tau>t_0$.
Using the change of variables ``$\ep={\ln(t_0^2+\tau^2-2t_0\tau\sin(\Omega))-\ln ((\tau-t_0)^2)\over\ln((\tau+t_0)^2)-\ln((\tau-t_0)^2)}$'', the equality \eqref{11.4} gives
\begin{equation}
\gamma_1(\tau,x,x_0')={\ln\left({\tau+t_0\over \tau-t_0}\right)\over 2t_0\tau}(\gamma_{1,1}(\tau,x,x_0')+\gamma_{1,2}(\tau,x,x_0'))\label{11.3.1}
\end{equation}
where
\begin{equation}
\gamma_{1,1}(\tau,x,x_0')=\int_0^{1\over 2}\int_{\S^1}\chi_{(0,+\infty)}(\nu(x)\cdot\Phi(\Omega(\tau,\ep),\omega))\Psi_3(s(\tau,\ep),\Phi(\Omega(\tau,\ep),\omega),v'(\tau,\omega,\ep))
d\omega d\ep
\label{11.3.2a}
\end{equation}
\begin{equation}
\gamma_{1,2}(\tau,x,x_0')=\int_{1\over 2}^1\int_{\S^1}\chi_{(0,+\infty)}(\nu(x)\cdot\Phi(\Omega(\tau,\ep),\omega))\Psi_3(s(\tau,\ep),\Phi(\Omega(\tau,\ep),\omega),v'(\tau,\omega,\ep))
d\omega d\ep
\label{11.3.2b}
\end{equation} 
and
\begin{eqnarray}
\Omega(\tau,\ep)&:=&{\rm arcsin}\left({\tau^2+t_0^2-(\tau-t_0)^{2(1-\ep)}(\tau+t_0)^{2\ep}\over 2t_0\tau}\right),\label{11.3.3a}\\
s(\tau,\ep)&:=&{\tau^2-t_0^2\over 2(\tau-t_0\sin(\Omega(\tau,\ep)))},\label{11.3.3b}\\
v'(\tau,\omega,\ep)&:=&{t_0v_0-s(\tau,\ep)\Phi(\Omega(\tau,\ep),\omega)\over \tau-s(\tau,\ep)},\label{11.3.3c}
\end{eqnarray}
for $0<\ep<1$, $\omega\in \S^1$.

We shall give some properties of $\Omega(\tau,\ep)$, $s(\tau,\ep)$ and $v'(\tau,\omega,\ep)$ for $0<\ep<1$ and $\omega\in \S^{n-1}$.
From \eqref{11.3.3a}, it follows that
\begin{equation}
\Omega(\tau,\ep)\to {\pi\over 2},\textrm{ as }\tau\to t_0^+, \textrm{ for all }\ep\in (0,1).\label{11.3.4}
\end{equation}

From \eqref{11.3.3a} and \eqref{11.3.3b}, it follows that
\begin{equation}
s(\tau,\ep)={\tau(t_0+\tau)(\tau-t_0)^{2\ep-1}\over (t_0+\tau)(\tau-t_0)^{2\ep-1}+(\tau+t_0)^{2\ep}},\ s(\tau,\ep)={\tau(t_0+\tau)\over t_0+\tau+(\tau-t_0)^{1-2\ep}(\tau+t_0)^{2\ep}}
\label{11.3.5b}
\end{equation}
for $\ep\in(0,1)$.
From \eqref{11.3.5b} it follows that 
\begin{equation}
s(\tau,\ep)\to 0^+ \textrm{ as }\tau\to t_0^+, \textrm{ when }\ep\in(1/2,1),\ 
s(\tau,\ep)\to t_0^- \textrm{ as }\tau\to t_0^+, \textrm{ when }\ep\in(0,1/2).\label{11.3.6b}
\end{equation}

In addition, from \eqref{11.3.3a}--\eqref{11.3.3c}, it follows that
\begin{eqnarray}
v'(\tau,\omega,\ep)&=&\left({-\tau(\tau+t_0)^2(\tau-t_0)^{2\ep}+2t_0^2\tau(\tau+t_0)^{2\ep}+\tau(\tau+t_0)^{1+2\ep}(\tau-t_0)\over 2t_0\tau^2(\tau+t_0)^{2\ep}},\right.\nonumber\\
&&\sqrt{-(\tau-t_0)^{4\ep}(\tau+t_0)^2-(\tau-t_0)^2(\tau+t_0)^{4\ep}+2(\tau^2+t_0^2)(\tau-t_0)^{2\ep}(\tau+t_0)^{2\ep}}\nonumber\\
&&\times\left.{(\tau+t_0+(\tau-t_0)^{1-2\ep}(\tau+t_0)^{2\ep})\over2t_0\tau(\tau+t_0)^{2\ep}}
\omega\right)\label{11.3.7a}
\end{eqnarray}
for $0<\ep<1$.
Therefore
\begin{equation}
v'(\tau,\omega,\ep)\to v_0= (1,0,0)\textrm{ as }\tau\to t_0^+,\label{11.3.8}
\end{equation}
for $0<\ep<1$.

Using \eqref{11.3.2a}, \eqref{11.3.4}, \eqref{11.3.6b}, \eqref{11.3.8} and Lebesgue dominated convergence theorem, we obtain
\begin{eqnarray}
\gamma_{1,1}(\tau,x,x_0')&\underset{\tau\to t_0^+}{\to}&\int_0^{1\over 2}\int_{\S^1}\lim_{s\to t_0^-}\Psi_3(s,v_0,v_0)d\omega d\ep\nonumber\\
&=&2\pi E(x_0',x)W(x,v_0)S(x_0',v_0)(\nu(x_0')v_0)(\nu(x)v_0)k(x_0',v_0,v_0)\label{11.3.9a}
\end{eqnarray}
(we also used \eqref{1.9b}).

Similarly, using \eqref{11.3.2a}, \eqref{11.3.4}, \eqref{11.3.6b}, \eqref{11.3.8} and Lebesgue dominated convergence theorem, we obtain
\begin{equation}
\gamma_{1,2}(\tau,x,x_0')\to 2\pi E(x_0',x)W(x,v_0)S(x_0',v_0)(\nu(x_0')v_0)(\nu(x)v_0)k(x,v_0,v_0),\textrm{ as }\tau\to t_0^+.\label{11.3.9b}
\end{equation}

Statement \eqref{A7c} follows from \eqref{11.3.1} and \eqref{11.3.9a}--\eqref{11.3.9b}.\\

Let $n\ge 4$. We shall prove \eqref{A7d}.
From \eqref{11.4} it follows that
\begin{equation}
\gamma_1(\tau,x,x_0')=\gamma_{1,1}(\tau,x,x_0')+\gamma_{1,2}(\tau,x,x_0'),\label{11.56}
\end{equation}
where
\begin{eqnarray}
\gamma_{1,1}(\tau,x,x_0')&=&\int_{-{\pi\over2}}^{\pi\over 2}\chi_{\left(-{\pi\over 2},\arcsin\left(t_0\over \tau\right)\right)}(\Omega)\cos(\Omega)^{n-2}
{(\tau-t_0\sin(\Omega))^{n-3}\over (t_0^2+\tau^2-2t_0\tau\sin(\Omega))^{n-2}}\label{11.57a}\\
&&\!\!\!\!\!\!\!\!\!\!\!\!\!\!\!\!\!\!\!\!\!\!\!\!\times\int_{\S^{n-2}}\chi_{(0,+\infty)}(\Phi(\Omega,\omega)\cdot \nu(x))\chi_{(0,\tau_-(x,\Phi(\Omega,\omega))))}(s)
\Psi_n(s,\Phi(\Omega,\omega),v')_{v'={t_0v_0-s\Phi(\Omega,\omega)\over \tau-s}\atop s={\tau^2-t_0^2\over 2(\tau-t_0\sin(\Omega))}}d\omega d\Omega,\nonumber\\
\gamma_{1,2}(\tau,x,x_0')&=&\int_{\arcsin\left(t_0\over \tau\right)}^{{\pi\over 2}}\cos(\Omega)^{n-2}{(\tau-t_0\sin(\Omega))^{n-3}\over (t_0^2+\tau^2-2t_0\tau\sin(\Omega))^{n-2}}
\label{11.57b}\\
&&\!\!\!\!\!\!\!\!\!\!\!\!\!\!\!\!\!\!\!\!\!\!\!\!\times\int_{\S^{n-2}}\chi_{(0,+\infty)}(\nu(x)\cdot\Phi(\Omega,\omega))\chi_{(0,\tau_-(x,\Phi(\Omega,\omega))))}(s)
\Psi_n(s,\Phi(\Omega,\omega),v')_{v'={t_0v_0-s\Phi(\Omega,\omega)\over \tau-s}\atop s={\tau^2-t_0^2\over 2(\tau-t_0\sin(\Omega))}}d\omega d\Omega.\nonumber
\end{eqnarray}

First we study $\gamma_{1,1}$. Note that
\begin{equation}
{\tau-t_0\sin(\Omega_1)\over t_0^2+\tau^2-2t_0\tau\sin(\Omega_1)}<{\tau-t_0\sin(\Omega_2)\over t_0^2+\tau^2-2t_0\tau\sin(\Omega_2)},\label{11.58}
\end{equation}
for $-{\pi\over 2}\le \Omega_1<\Omega_2\le {\pi\over 2}$ and for $\tau>t_0$.
Therefore using also the estimate $\cos(\Omega)\le 1$ we obtain
\begin{eqnarray}
&&\cos(\Omega)^{n-2}{(\tau-t_0\sin(\Omega))^{n-3}\over (t_0^2+\tau^2-2t_0\tau\sin(\Omega))^{n-2}}\nonumber\\
&&\le\cos(\Omega)^{n-4}{\cos(\Omega)^2\over t_0^2+\tau^2-2t_0\tau\sin(\Omega)}
\left({\tau-t_0\sin(\Omega)\over t_0^2+\tau^2-2t_0\tau\sin(\Omega)}\right)^{n-3} 
\le{C_0^2\over 2t_0\tau^{n-2}},\label{11.59}
\end{eqnarray}
for $\Omega\in (-{\pi\over 2}, {\pi\over 2})$ and  $\sin(\Omega)\le {t_0\over \tau}$ (we used \eqref{11.58} with ``$\Omega_1$''$=\Omega$ and ``$\Omega_2$''$={t_0\over \tau}$,
and we used the estimate $t_0^2+\tau^2-2t_0\tau\sin(\Omega)\ge 2t_0\tau(1-\sin(\Omega))$), 
where
\begin{equation}
C_0:=\sup_{\varphi\in (0,2\pi)}{\sin^2(\varphi)\over 1-\cos(\varphi)}=\sup_{\Omega\in (-{3\pi\over 2},{\pi\over 2})}{\cos^2(\Omega)\over 1-\sin(\Omega)}.\label{P3}
\end{equation}
Using \eqref{11.57a}, \eqref{11.59}, \eqref{1.9b} and Lebesgue dominated convergence theorem, we obtain
\begin{eqnarray}
&&\!\!\!\!\!\!\!\!\!\!\!\!\!\gamma_{1,1}(\tau,x,x_0')\underset{\tau\to t_0^+}{\longrightarrow}2^{2-n}t_0^{1-n}\int_{-{\pi\over2}}^{\pi\over 2}{\cos(\Omega)^{n-2}\over 1-\sin(\Omega)}
\int_{\S^{n-2}}\chi(\nu(x)\cdot\Phi(\Omega,\omega))\lim_{s\to 0^+}\Psi_n(s,\Phi(\Omega,\omega),v_0)d\omega d\Omega\nonumber\\
&&=2^{2-n}t_0^{1-n}\int_{\S^{n-1}_{x,+}}{1\over 1-v\cdot v_0}\lim_{s\to 0^+}\Psi_n(s,v,v_0)dv\nonumber\\
&&=t_0^{1-n}E(x_0',x)S(x_0',v_0)(\nu(x_0')\cdot v_0)\int_{\S^{n-1}_{x,+}}{1\over 1-v\cdot v_0}W(x,v)(\nu(x)\cdot v)k(x,v_0,v)dv.\label{11.60}
\end{eqnarray}

Now we shall study $\gamma_{1,2}$ defined by \eqref{11.57b}. Note that using the convexity of $X$ we obtain $v'={x-sv-x'_0\over |x-sv-x_0'|}\in \S_{x_0',-}^{n-1}$ whenever 
$v\in \S_{x,+}^{n-1}$ and $s\in (0,\tau_-(x,v))$. Therefore using the change of variables ``$\sin(\Omega')={2\tau t_0-(\tau^2+t_0^2)\sin(\Omega)\over \tau^2+t_0^2-2t_0\tau\sin(\Omega)}$'', $\Omega\in (-{\pi\over 2},{\pi\over 2})$
(``$\cos(\Omega){d\Omega\over d\Omega'}={(\tau^2-t_0^2)^2\cos(\Omega')\over (\tau^2+t_0^2-2t_0\tau\sin(\Omega'))^2}$''), we obtain
\begin{eqnarray}
&&\!\!\!\!\!\!\!\!\!\gamma_{1,2}(\tau,x,x_0'):=\int_{-{\pi\over 2}}^{\arcsin\left(t_0\over \tau\right)}\!\!\!\!\!\cos(\Omega')^{n-2}{(\tau-t_0\sin(\Omega'))^{n-3}\over (t_0^2+\tau^2-2t_0\tau\sin(\Omega'))^{n-2}}\label{11.9}\\
&&\!\!\!\!\!\!\int_{{\omega\in\S^{n-2}\atop \nu(x)\cdot\Phi(\Omega(\tau,\Omega'),\omega)>0}\atop -\nu(x_0')\cdot\Phi(\Omega',-\omega)>0}\!\!\!\!\!\chi_{(0,\tau_-(x,\Phi(\Omega(\tau,\Omega'),\omega)))}(s(\tau,\Omega'))\Psi_n(s(\tau,\Omega'),\Phi(\Omega(\tau,\Omega'),\omega),v')_{v'=\Phi(\Omega',-\omega)}d\omega d\Omega',\nonumber
\end{eqnarray}
where
\begin{equation}
\Omega(\tau,\Omega')={\arcsin}\left({2\tau t_0-(\tau^2+t_0^2)\sin(\Omega')\over 
\tau^2+t_0^2-2t_0\tau\sin(\Omega')}\right),\ 
s(\tau,\Omega')={\tau^2+t_0^2-2t_0\tau\sin(\Omega')\over 2(\tau-t_0\sin(\Omega'))},\label{11.9b}
\end{equation}
for $\Omega'\in (-{\pi\over 2},{\pi\over 2})$.
Note that
\begin{eqnarray}
\Omega(\tau,\Omega')&\underset{\tau\to t_0^+}{\to}&{\pi\over 2},\ \Phi(\Omega(\tau,\Omega'),\omega)\underset{\tau\to t_0^+}{\to}v_0,\label{12b}\\
s(\tau,\Omega')&\underset{\tau\to t_0^+}{\to}&t_0,\label{12c}
\end{eqnarray}
for $(\Omega',\omega)\in (-{\pi\over 2},{\pi\over 2})\times\S^{n-2}$. Note also that from \eqref{12b}, it follows that at fixed $v'=\Phi(\Omega',-\omega)\in \S_{x_0',-}^{n-1}$ the condition $\chi_{(0,\tau_-(x,\Phi(\Omega(\tau,\Omega'),\omega)))}
(s(\tau,\Omega'))=1$  
(which is equivalent to $x-s(\tau,\Omega')\Phi(\Omega(\tau,\Omega'),\omega)\in X$  due to convexity of $X$) for $\Phi(\Omega(\tau,\Omega'),\omega)\in \S_{x,+}^{n-1}$
is satisfied when $\tau-t_0>0$ is sufficiently small. Therefore from \eqref{11.59} 
(with ``$\Omega$'' replaced by ``$\Omega'$'') and from \eqref{11.9} and Lebesgue dominated convergence theorem we obtain
\begin{eqnarray}
&&\!\!\!\!\!\!\!\!\!\!\!\!\!\!\!\!\!\!\!\gamma_{1,2}(\tau,x,x_0')\underset{\tau\to t_0^+}{\longrightarrow}2^{2-n}t_0^{1-n}\int_{-{\pi\over2}}^{\pi\over 2}{\cos(\Omega')^{n-2}\over 1-\sin(\Omega')}
\int_{\S^{n-2}}\chi_{(0,+\infty)}(-\nu(x_0')\cdot\Phi(\Omega',-\omega))\nonumber\\
&&\!\!\!\!\!\!\!\!\!\!\times\lim_{s\to t_0^-}\Psi_n(s,v_0,\Phi(\Omega',-\omega))d\omega d\Omega'=2^{2-n}t_0^{1-n}\int_{\S^{n-1}_{x_0',-}}{1\over 1-v'\cdot v_0}\lim_{s\to t_0^-}\Psi_n(s,v_0,v')dv'\nonumber\\
&&\!\!\!\!\!\!\!\!\!\!=t_0^{1-n}E(x_0',x)W(x,v_0)(\nu(x)\cdot v_0)\int_{\S^{n-1}_{x_0',-}}{1\over 1-v'\cdot v_0}S(x_0',v')|\nu(x_0')\cdot v'|k(x,v',v_0)dv'.\label{11.60b}
\end{eqnarray}
Statement \eqref{A7d} follows from \eqref{11.56}, \eqref{11.60} and \eqref{11.60b}.
Theorem \ref{theorem:limH1} is proved.
\end{proof}

\section{Proof of Theorems \ref{theorem:stabH1}, \ref{theorem:stabH2}, \ref{cor} and Theorem \ref{cor_ball} \eqref{eq:cor2H1}}
\label{proof_thm_stabH1H2_cor}
\begin{proof}[Proof of Theorem \ref{theorem:stabH2}]
 We now prove \eqref{stab2.a}. Let $x_0'\in \pa X$. For
  $\ep=(\ep_1,\ep_2)\in (0,+\infty)^2$ and $\ep_3\in (0,+\infty)$ let
  $(f_{\ep_1},g_{\ep_2})\in C^1(\pa X)\times C^1(\R)$ satisfy
\begin{eqnarray}
&&g_{\ep_2}\ge 0, f_{\ep_1}\ge 0, \ {\rm supp}g_{\ep_2}\subseteq (0,\min(\ep_2,\eta)),\label{PH6a}\\
&&{\rm supp}f_{\ep_1}\subseteq\{x'\in \pa X\ |\ |x'-x_0'|<\ep_1\},\label{PH6b}\\
&&\int_0^\eta g_{\ep_2}(t')dt'=1,\ \int_{\pa X}f_{\ep_1}(x')d\mu(x')=1,\label{PH6c}
\end{eqnarray}
for $\ep=(\ep_1,\ep_2)\in (0,+\infty)^2$. Therefore $\phi_\ep :=g_{\ep_2} f_{\ep_1}$ is an approximation of the delta function at $(0,x_0')\in \R\times\pa X$ for $\ep:=(\ep_1,\ep_2)\in (0,+\infty)^2$.
Let $\psi_{\ep_3}\in L^\infty((0,T)\times\pa X)$ be defined by
\begin{equation}
\psi_{\ep_3}(t,x)=\chi_{(-\ep_3,\ep_3)}(t-|x-x_0'|)(2\chi_{(0,+\infty)}((E-\tilde E)(x,x_0'))-1),\ (t,x)\in (0,T)\times\pa X,\label{PH11}
\end{equation}
for $\ep_3>0$. From \eqref{E8} and \eqref{eq:msop} it follows that
\begin{eqnarray}
&&\int_{(0,T)\times\pa X}\psi_{\ep_3}(t,x)(A_{S,W}-\tilde A_{S,W})\phi_\ep(t,x)dt d\mu(x)=I_0(\psi_{\ep_3},\phi_\ep)\nonumber\\
&&+\int_{(0,T)\times \pa X\times (0,\eta)\times\pa X}\psi_{\ep_3}(t,x)\phi_{\ep}(t',x')(\Gamma_1-\tilde\Gamma_1)(t-t',x,x')dt 
d\mu(x) dt'd\mu(x'),\label{PH3}
\end{eqnarray}
for $\ep=(\ep_1,\ep_2)\in(0,+\infty)$ and $\ep_3\in (0,+\infty)$, where
\begin{eqnarray}
I_0(\psi_{\ep_3},\phi_\ep)&=&\int_{(0,T)_t\times\pa X_x\times \pa X_{x'}\atop |x-x'|<t}\psi_{\ep_3}(t,x)\phi_\ep(t-|x-x'|,x')
{E(x,x')-\tilde E(x,x')\over |x-x'|^{n-1}}\nonumber\\
&&\times\left[W(x,v)S(x',v)(\nu(x)\cdot v)|\nu(x')\cdot v|\right]_{v={x-x'\over |x-x'|}}dtd\mu(x)d\mu(x').\label{PH2}
\end{eqnarray}
From \eqref{A6e}, \eqref{A6b-}, \eqref{A6e-3} and \eqref{A6e-} it follows that
\begin{equation}
(\tau-|x-x'|)^{3-n\over 2}(\Gamma_1-\tilde \Gamma_1)(\tau,x,x')\in L^\infty((0,T)\times\pa X\times\pa X).\label{PH2.4}
\end{equation}
Combining \eqref{PH3} and the equality $\|\phi_\ep\|_{L^1((0,\eta)\times \pa X)}=1$ and the estimate $\|\psi_{\ep_3}\|_{L^\infty((0,T)\times\pa X)}\le 1$  and \eqref{PH2.4} 
we obtain
\begin{equation}
I_0(\psi_{\ep_3},\phi_\ep)\le \|A_{S,W}-\tilde A_{S,W}\|_{\eta,T}+C\Delta_1(\psi_{\ep_3},\phi_\ep),
\label{PH2.5}
\end{equation}
for $\ep=(\ep_1,\ep_2)\in (0,+\infty)$ and $\ep_3\in (0,+\infty)$,
where $C=\|(\tau-|x-x'|)^{3-n\over
  2}(\Gamma_1-\tilde\Gamma_1)(\tau,x,x')$ $\|_{L^\infty((0,T)\times\pa
  X_x\times\pa X_{x'})}$ and
\begin{equation}
\Delta_1(\psi_{\ep_3},\phi_\ep)=\int_{(0,T)_t\times\pa X_x\times (0,\eta)_{t'}\times \pa X_{x'}\atop |x-x'|<t-t'}\psi_{\ep_3}(t,x)\phi_\ep(t',x')(t-t'-|x-x'|)^{n-3\over 2}dt d\mu(x)dt'd\mu(x').
\label{PH2.5b}
\end{equation}
Note that the function $\Phi_{1,\ep_3}:[0,\eta)\times\pa X \to \R$ defined by 
\begin{equation}
\Phi_{1,\ep_3}(t',x'):=\int_{(0,T)_t\times\pa X_x\atop |x-x'|<t-t'}\psi_{\ep_3}(t,x)(t-t'-|x-x'|)^{n-3\over 2}dt d\mu(x),\ (t',x')\in [0,\eta)\times\pa X,\label{PH2.7}
\end{equation} 
is continuous on $[0,\eta)\times\pa X$ for $\ep_3\in (0,+\infty)$.
Therefore from \eqref{PH6a}--\eqref{PH6c} and the equality $\Delta_1(\psi_{\ep_3},\phi_{\ep})=\int_{(0,\eta)\times\pa X}\phi_{\ep}(t',x')\Phi_{1,\ep_3}(t',x')dt'd\mu(x')$ it follows that
\begin{equation}
\lim_{\ep_3\to0^+}\lim_{\ep_2\to 0^+}\lim_{\ep_1\to 0^+}\Delta_1(\psi_{\ep_3},\phi_\ep)=\lim_{\ep_3\to0^+}\Phi_{1,\ep_3}(0,x_0')=0\label{PH2.13}
\end{equation}
(we also used \eqref{PH2.7}, \eqref{PH11} and the Lebesgue dominated
convergence theorem to prove that $\lim_{\ep_3\to
  0^+}\Phi_{1,\ep_3}(0,x_0')=0$).  Note that under condition
\eqref{H2} the function $\Phi_{0,\ep_2,\ep_3}:\pa X\to \R$ defined by
\begin{eqnarray}
\Phi_{0,\ep_2,\ep_3}(x')&=&\int_{(0,T)_t\times\pa X_x\atop |x-x'|<t}\psi_{\ep_3}(t,x)g_{\ep_2}(t-|x-x'|)
{E(x,x')-\tilde E(x,x')\over |x-x'|^{n-1}}\nonumber\\
&&\times\left[W(x,v)S(x',v)(\nu(x)\cdot v)|\nu(x')\cdot v|\right]_{v={x-x'\over |x-x'|}}dtd\mu(x),\label{PH2.9}
\end{eqnarray}
is continuous on $\pa X$, for $(\ep_2,\ep_3)\in (0,+\infty)^2$. Therefore from the equality $I_0(\psi_{\ep_3},\phi_\ep)=
\int_{\pa X}\Phi_{0,\ep_2,\ep_3}(x')$ $\times f_{\ep_1}(x')d\mu(x')$ (see \eqref{PH2}) it follows that
\begin{equation}
\lim_{\ep_1\to 0^+}I_0(\psi_{\ep_3},\phi_\ep)=\Phi_{0,\ep_2,\ep_3}(x_0'), \textrm{ for }(\ep_2,\ep_3)\in (0,+\infty)^2.\label{PH2.10}
\end{equation}
Therefore using the Lebesgue dominated convergence theorem and
\eqref{PH2.9} we obtain
\begin{equation}
\lim_{\ep_3\to0^+}\lim_{\ep_2\to 0^+}\lim_{\ep_1\to 0^+}I_0(\psi_{\ep_3},\phi_\ep)=\int_{\pa X_x}\!\!\!\!\!\!
{\left|E(x,x_0')-\tilde E(x,x_0')\right|\over |x-x_0'|^{n-1}}\left[W(x,v)S(x_0',v)(\nu(x)\cdot v)|\nu(x_0')\cdot v|\right]_{v={x-x_0'\over |x-x_0'|}}
d\mu(x).\label{PH2.12}
\end{equation}
Combining \eqref{PH2.12}, \eqref{PH2.13} and \eqref{PH2.5} we obtain
the formula \eqref{stab1.a}. Using \eqref{stab1.a} and the estimates
$\inf_{(x',v')\in \Gamma_-}S(x',v')>0$ and
$\inf_{(x,v)\in\Gamma_+}W(x,v)>0$ and the change of variables
$x=x_0'+\tau_+(x_0',v_0)v_0$ (${\nu(x)\cdot v_0\over
  |x-x_0'|^{n-1}}d\mu(x)=dv_0$) we obtain \eqref{stab2.a} where the
constant $C_1$ which appears on the right-hand side of \eqref{stab2.a}
is given by $C_1=\left(\inf_{(x',v')\in
    \Gamma_-}S(x',v')\inf_{(x,v)\in\Gamma_+}W(x,v)\right)^{-1}$.

We now prove \eqref{stab2.b}. Let $x\in \pa X$ be such that
$px_0'+(1-p)x\in Z$ for some $p\in (0,1)$. We set $t_0=|x-x_0'|$ and
$v_0={x-x_0'\over |x-x_0'|}$.  From \eqref{A6e}, \eqref{A6b-}, \eqref{A6e-3} and \eqref{A6e-} it
follows that
\begin{eqnarray}
&&(\tau-|x-x_0'|)^{3-n\over 2}|\gamma_1-\tilde \gamma_1|(\tau,x,x_0')\le (\tau-|x-x_0'|)^{3-n\over 2}|\Gamma_2-\tilde\Gamma_2|(\tau,z,z')
\label{pstabh2.2}\\
&&+\|(s-|z-z'|)^{3-n\over 2}(\Gamma_1-\tilde\Gamma_1)(s,z,z')
\|_{L^\infty((0,T)_s\times\pa X_z\times\pa X_{z'})}
,\nonumber
\end{eqnarray}
for $\tau>|x-x_0'|$. From \eqref{A6b-}, \eqref{A6e-3}--\eqref{A6e-} it turns out that $\lim_{\tau\to|x-x_0'|^+}(\tau-|x-x_0'|)^{3-n\over 2}|\Gamma_2-\tilde\Gamma_2|(\tau,z,z')=0$.
Therefore applying \eqref{A7b} and \eqref{A7e} on the left-hand side of \eqref{pstabh2.2} we obtain
\begin{eqnarray}
&&2^{1-n\over 2}|x-x_0'|^{-{n-1\over 2}}C_nS(x_0',v_0)W(x,v_0)|\nu(x_0')\cdot v_0|(\nu(x)\cdot v_0)\nonumber\\
&&\times
\left|\int_0^{t_0}{e^{-\int_0^{t_0}\sigma(x_0'+sv_0,v_0)ds}k(x-pv_0,v_0,v_0)-e^{-\int_0^{t_0}\tilde \sigma(x_0'+sv_0,v_0)ds}\tilde k(x-pv_0,v_0,v_0)\over p^{n-1\over 2}(t_0-p)^{n-1\over 2}} dp
\right|\nonumber\\
&&\le \|(s-|z-z'|)^{3-n\over 2}(\Gamma_1-\tilde\Gamma_1)(s,z,z'))\|_{L^\infty((0,T)_s\times\pa X_z\times\pa X_{z'})},\label{pstabh2.3}
\end{eqnarray}
where $C_n=2$ if $n=2$ and $C_n={\rm Vol}_{n-2}(\S^{n-2})$ if $n\ge
3$.  Then note that $C_X:=\inf_{x_1\in \pa X,\ z\in \bar
  Z}\nu(x_1)\cdot {x_1-z\over |x_1-z|}>0$ since $X$ is a bounded
convex subset of $\R^n$ with $C^1$ boundary and $\bar Z\subset X$.
Therefore \eqref{stab2.b} follows from \eqref{pstabh2.3} where the
constant $C_2$ which appears on the right-hand side of \eqref{stab2.b}
is given by $C_2={2^{n-1\over 2}\diam^{n-1\over 2}\over
  C_nC_X^2\inf_{(x',v')\in \Gamma_-}S(x',v')\inf_{(x_1,v_1)\in
    \Gamma_+}W(x_1,v_1)}$.  Theorem \ref{theorem:stabH2} is proved.
\end{proof}

\begin{proof}[Proof of Theorem \ref{theorem:stabH1}]
We prove \eqref{stab1.a}.  Let $x_0'\in \pa X$. For $\ep=(\ep_1,\ep_2)\in (0,+\infty)^2$ and $\ep_3\in (0,+\infty)$ let $(f_{\ep_1},g_{\ep_2})\in C^1(\pa X)\times C^1(\R)$ satisfy \eqref{PH6a}--\eqref{PH6c} and $\psi_{\ep_3}$ be defined by \eqref{PH11}. First note that \eqref{PH3}--\eqref{PH2} still hold. From Theorems \ref{theorem:normesimplescat} and \ref{theorem:normemultiplescat} it follows that
\begin{equation}
|x-x'|^{n-{7\over 4}}(\tau-|x-x'|)^{3\over 4}(\Gamma_1-\tilde\Gamma_1)(\tau,x,x')\in L^\infty((0,T)\times\pa X\times\pa X).\label{PH4}
\end{equation}
Combining \eqref{PH3}, \eqref{PH4} and the equality $\|\phi_\ep\|_{L^1((0,\eta)\times \pa X)}$ and the estimate $\|\psi_{\ep_3}\|_{L^\infty((0,T)\times\pa X)}\le 1$  we obtain
\begin{equation}
I_0(\psi_{\ep_3},\phi_\ep)\le \|A_{S,W}-\tilde A_{S,W}\|_{\eta,T}+C\Delta_2(\psi_{\ep_3},\phi_\ep),
\label{PH5}
\end{equation}
for $\ep=(\ep_1,\ep_2)\in (0,+\infty)$, $\ep_3\in (0,+\infty)$, where
$C=\||x-x'|^{n-{7\over 4}}(\tau-|x-x'|)^{3\over 4}(\Gamma_1-\tilde\Gamma_1)(\tau,x,x')$ $\|_{L^\infty((0,T)\times\pa X_x\times\pa X_{x'}}$
and 
\begin{equation}
\Delta_2(\psi_{\ep_3},\phi_\ep)=\int_{(0,T)_t\times\pa X_x\times (0,\eta)_{t'}\times \pa X_{x'}\atop |x-x'|<t-t'}
{\psi_{\ep_3}(t,x)\phi_\ep(t',x')\over |x-x'|^{n-{7\over 4}}(t-t'-|x-x'|)^{3\over 4}}dt d\mu(x)dt'd\mu(x').
\label{PH5b}
\end{equation}
Note that the function $\Phi_{1,\ep_3}:[0,T)\times\pa X \to \R$ defined by 
\begin{equation}
\Phi_{3,\ep_3}(t',x'):=\int_{(0,T)_t\times\pa X_x\atop |x-x'|<t-t'}{\psi_{\ep_3}(t,x)
\over |x-x'|^{n-{7\over 4}}(t-t'-|x-x'|)^{3\over 4}}dt d\mu(x),\ (t',x')\in [0,\eta)\times\pa X,\label{PH7}
\end{equation} 
is continuous on $[0,\eta)\times\pa X$.
Therefore using \eqref{PH5b}, \eqref{PH6a}--\eqref{PH6c} and using the equality 
$\Delta_2(\psi_{\ep_3},\phi_\ep)=\int_{(0,\eta)_{t'}\times \pa X_{x'}}\Phi_{3,\ep_3}(t',x')\phi_{\ep}(t',x')dt'd\mu(x')$
we obtain
\begin{equation}
\lim_{\ep_3\to0^+}\lim_{\ep_2\to 0^+}\lim_{\ep_1\to 0^+}\Delta_2(\psi_{\ep_3},\phi_\ep)=\lim_{\ep_3\to 0^+}\Phi_{3,\ep_3}(0,x_0')=0.\label{PH13}
\end{equation}
(we also used \eqref{PH7}, \eqref{PH11} and Lebesgue dominated convergence theorem to prove $\lim_{\ep_3\to 0^+}\Phi_{3,\ep_3}(0,x_0')=0$). Note that under condition \eqref{H1} the function $\Phi_{4,\ep_3,\ep_2}:\pa X\to \R$ defined by
\begin{eqnarray}
\Phi_{4,\ep_3,\ep_2}(x')&=&\int_{(0,T)_t\times\pa X_x\atop |x-x'|<t}\psi_{\ep_3}(t,x)g_{\ep_2}(t-|x-x'|)
{E_{\pa X\times\pa X}(x,x')-\tilde E_{\pa X\times\pa X}(x,x')\over |x-x'|^{n-1}}\nonumber\\
&&\times\left[W(x,v)S(x',v)(\nu(x)\cdot v)|\nu(x')\cdot v|\right]_{v={x-x'\over |x-x'|}}dtd\mu(x),\label{PH9}
\end{eqnarray}
is continuous on $\pa X$. Therefore from \eqref{PH6a}--\eqref{PH6c} and the equality $I_0(\psi_{\ep_3},\phi_\ep)=\int_{\pa X}f_{\ep_1}(x')\Phi_{4,\ep_3,\ep_2}(x')d\mu(x')$ (see \eqref{PH2}) it follows that
\begin{equation}
\lim_{\ep_1\to 0^+}I_0(\psi_{\ep_3},\phi_\ep)=\Phi_{4,\ep_3,\ep_2}(x_0'),\ \textrm{for }(\ep_2,\ep_3)\in (0,+\infty).\label{PH10}
\end{equation}
Then using Lebesgue dominated convergence theorem and \eqref{PH9} we obtain
\begin{eqnarray}
\lim_{\ep_3\to 0^+}\lim_{\ep_2\to 0^+}\lim_{\ep_1\to 0^+}I_0(\psi_{\ep_3},\phi_\ep)&=&\int_{\pa X_x}
{\left|E_{\pa X\times\pa X}(x,x_0')-\tilde E_{\pa X\times\pa X}(x,x_0')\right|\over |x-x_0'|^{n-1}}\label{PH12}\\
&&\times\left[W(x,v)S(x_0',v)(\nu(x)\cdot v)|\nu(x_0')\cdot v|\right]_{v={x-x_0'\over |x-x_0'|}}
d\mu(x).\nonumber
\end{eqnarray}
Combining \eqref{PH12}, \eqref{PH13} and \eqref{PH5} we obtain \eqref{stab1.a}.

We prove \eqref{stab1.b}--\eqref{stab1.d}. Let $x\in \pa X$ be such that $px_0'+(1-p)x\in X$ for some $p\in (0,1)$.
Let $\beta_n:\{(\tau,z,z')\in (0,T)\times\pa X\times\pa X\ |\ z\not=z',\ \tau>|z-z'|\}\to \R$ be defined by
\begin{equation}
\beta_n(\tau,z,z')=
\left\lbrace
\begin{array}{l}
\sqrt{\tau^2-|z-z'|^2}, \textrm{ if }n=2,\\
{\tau|z-z'|\over \ln\left({\tau+|z-z'|\over \tau-|z-z'|}\right)},\textrm{ if }n=3,\\
\tau|z-z'|^{n-2}, \textrm{ if }n\ge 4,
\end{array}
\right.\label{pstab1}
\end{equation}
for $(\tau,z,z') \in (0,T)\times\pa X\times\pa X$, $z\not=z'$, $\tau>|z-z'|$.
From \eqref{A6b}--\eqref{A6d} and \eqref{A6b-}--\eqref{A6d-} it follows that
\begin{eqnarray}
\beta_n(\tau,x,x_0')|(\gamma_1-\tilde \gamma_1)(\tau,x,x_0')|&\le& \beta_n(\tau,x,x_0')|(\Gamma_2-\tilde\Gamma_2)(\tau,z,z')|
\label{pstab2}\\
&&+\|\beta_n(s,z,z')(\Gamma_1-\tilde\Gamma_1)(s,z,z')\|_{L^\infty((0,T)_s\times\pa X_z\times\pa X_{z'})}.\nonumber
\end{eqnarray}
From \eqref{pstab1} and \eqref{A6b-}--\eqref{A6d-} it turns out that $\lim_{\tau\to|x-x_0'|^+}\beta_n(\tau,x,x_0')|(\Gamma_2-\tilde\Gamma_2)(\tau,z,z')|=0$.
Therefore applying \eqref{A7b} (resp. \eqref{A7c}, \eqref{A7d}) on the left-hand side of \eqref{pstab2} we obtain \eqref{stab1.b} (resp. \eqref{stab1.c}, \eqref{stab1.d}).
Theorem \ref{theorem:stabH1} is proved.
\end{proof}

\begin{proof}[Proof of Theorem \ref{cor}]

We first prove \eqref{eq:cor1}. We extend $\sigma$ and $\tilde\sigma$ by $0$ outside $Y$.
For a bounded and continuous function $f$ on $Y$  consider the X-ray transform 
$Pf:\S^{n-1}\times \R^n\to \R$ defined by \eqref{x-ray}
(we extend $f$ by $0$ outside $Y$).
We recall the following estimate
\begin{equation}
\|f\|_{H^{-{1\over 2}}(Y)}\le \left(\int_{\S^{n-1}}\int_{\Pi_v}|Pf(v,x)|^2dx dv\right)^{1\over 2},\label{x-raystab}
\end{equation}
where $\Pi_v:=\{x\in \R^n\ |\ v\cdot x=0\}$ for $v\in \S^{n-1}$.
Note that using the estimate $\|\sigma\|_\infty\le M$, we obtain
\begin{equation}
\int_0^{\tau_+(x_0',v)}\sigma(x_0'+sv,v)ds\le M\tau_+(x_0',v)\le M\diam,\textrm{ for }(x_0',v)\in \Gamma_-.\label{pcor1}
\end{equation} 
Replacing $\sigma$ by $\tilde \sigma$ on the left-hand side of \eqref{pcor1} we obtain an estimate similar to \eqref{pcor1} for $\tilde\sigma$. 
Therefore using the estimate $|e^{t_1}-e^{t_2}|\ge e^{-M\diam}|t_1-t_2|$ for $(t_1,t_2)\in [0,+\infty)^2$, $\max(t_1,t_2)\le M\diam$, we obtain
\begin{equation}
\left|e^{-\int_0^{\tau_+(x_0',v)}\sigma(x_0'+sv,v)ds}-e^{-\int_0^{\tau_+(x_0',v)}\tilde \sigma(x_0'+sv,v)ds}\right|\ge e^{-M\diam}\left|P(\sigma-\tilde \sigma)(v,x_0')\right|,
\label{pcor2}
\end{equation}
for $(x_0',v)\in \Gamma_-$.
Integrating the left-hand side of \eqref{stab2.a} over $\pa X$ and using \eqref{pcor2}, we obtain
\begin{equation}
\int_{\Gamma_-}\left|P(\sigma-\tilde \sigma)(v,x_0')\right|d\xi(v,x_0')\le e^{M\diam}{\rm Vol}(\pa X)C_1\|A_{S,W}-\tilde A_{S,W}\|_{\eta,T},\label{pcor3}
\end{equation}
where $C_1$ is the constant that appears on the right-hand side of \eqref{stab2.a}.
Note that using that $X$ is a convex open subset of $\R^n$ with $C^1$ boundary we obtain 
$\int_{\Gamma_-}\left|P(\sigma-\tilde \sigma)(v,x_0')\right|d\xi(v,x_0')=\int_{\S^{n-1}}\int_{\Pi_v}|P(\sigma-\tilde\sigma)(v,x)|dx dv$. 
Therefore using \eqref{pcor3} and the estimate
$|P(\sigma-\tilde\sigma)(v,x)|^2\le \|\sigma-\tilde\sigma\|_{L^\infty(Y)}\diam|P(\sigma-\tilde\sigma)(v,x)|$ 
for $(v,x)\in T\S^{n-1}$ (see \eqref{pcor1} and the estimates $\sigma\ge 0$, $\tilde\sigma\ge 0$) we obtain
\begin{equation}
\left(\int_{\S^{n-1}}\int_{\Pi_v}|P(\sigma-\tilde\sigma)(v,x)|^2dx dv\right)^{1\over 2}\le C_3 \|\sigma-\tilde\sigma\|_{\infty}^{1\over 2}\|A_{S,W}-\tilde A_{S,W}\|_{\eta,T}^{1\over 2}.\label{pcor4}
\end{equation}
where $C_3=\left(\diam e^{M\diam}{\rm Vol}(\pa X)C_1\right)^{1\over 2}$. Combining \eqref{pcor4} and \eqref{x-raystab} we obtain \eqref{eq:cor1}.

We now prove \eqref{eq:cor2}. Let $f\in L^2(X)$, ${\rm supp}f\subseteq
\bar Z$. We consider the weighted X-ray transform of $f$, $P_{\vartheta}f$,
defined by
\begin{equation}
P_{\vartheta}f(x,v)=\int_0^{\tau_+(v,x)}f(pv+x)\vartheta(pv+x,v)dp,\textrm{ for a.e. }(x,v)\in \Gamma_-,\label{pcor5}
\end{equation}
where $\vartheta:X\times\S^{n-1}\to (0,+\infty)$ is the analytic function given by
\begin{equation}
\vartheta(x,v)=(\tau_-(x,v)\tau_+(x,v))^{-{n-1\over 2}}g(x,v,v), \textrm{ for }(x,v)\in X\times \S^{n-1}.\label{pcor6}
\end{equation}
From \cite[theorem 2.2]{FSU-JGA-08} and from 
\cite[theorem 4]{Rullgard-IP04} we obtain
\begin{equation}
\|f\|_{H^{-{1\over 2}}(Z)}\le C\|P_\vartheta f\|_{L^2(\Gamma_-,d\xi)},\label{pcor7}
\end{equation}
where $C=C(X,Z,g)$ is a constant that does not depend on $f$.  Let
$x_0'\in \pa X$ and let $x\in \pa X$ such that $px_0'+(1-p)x\in Z$ for
some $p\in (0,1)$ where $v_0={x-x_0'\over |x-x_0'|}$ and
$t_0=|x-x_0'|$.  Note that using \eqref{H2} (since $\tilde k \in
L^\infty(Z)$ and ${\rm supp}\tilde k\subseteq \bar Z\subseteq\{x\in X\ |\ 
\inf_{x'\in \pa X}|x-x'|\ge \delta\}$), we obtain
\begin{eqnarray}
\int_0^{\tau_+(x_0,v_0')}{\tilde k(x_0'+p v_0',v_0',v_0')\over p^{n-1\over 2}(\tau_+(x_0',v_0')-p)^{n-1\over 2}}dp 
&\le& \|\tilde k\|_{L^\infty(Z)}\int_{\delta}^{\tau_+(x_0,v_0')-\delta}{1\over p^{n-1\over 2}(\tau_+(x_0',v_0')-p)^{n-1\over 2}}dp\nonumber\\
\le\,\,\,\|\tilde k\|_{L^\infty(Z)}\delta^{-(n-1)}\tau_+(x_0',v_0')
  &\le& 
 \|\tilde k\|_{L^\infty(Z)}\delta^{-(n-1)}\diam.\label{pcor8}
\end{eqnarray}
We use the estimate 
\begin{eqnarray}
|P_\vartheta(k_0-\tilde k_0)(x_0',v_0')|&\le&e^{P\sigma(v_0',x_0')}|P_\vartheta\tilde k_0(x_0',v_0')|\left|e^{-P\sigma(v_0',x_0')}-e^{-P\tilde\sigma(v_0',x_0')}\right| 
\label{pcor9}\\
&&+e^{P\sigma(v_0',x_0')}\left|e^{-P\sigma(v_0',x_0')}P_\vartheta k_0(x_0',v_0')-e^{-P\tilde\sigma(v_0',x_0')}P_\vartheta\tilde k_0(x_0',v_0')\right|.\nonumber
\end{eqnarray}
Integrating both sides of inequality \eqref{pcor9} over $v_0'\in\S_{x_0',-}^{n-1}$ and using the estimate $e^{P\sigma(v_0',x_0')}\le e^{M\diam}$, and using \eqref{pcor8}, 
\eqref{stab2.a}--\eqref{stab2.b},
we obtain
\begin{eqnarray}
&&\int_{\S^{n-1}_{x_0',-}}|P_\vartheta(k_0-\tilde k_0)|(x_0',v_0')|\nu(x_0')\cdot v|dv\le\delta^{-(n-1)}\diam e^{M\diam}C_1\|\tilde k\|_\infty\|A_{S,W}-\tilde A_{S,W}\|_{\eta,T}
\nonumber\\
&&+{{\rm Vol}(\S^{n-1})e^{M\diam}C_2\over 2}
\left\|(\tau-|z-z'|)^{n-3\over 2}(\Gamma_1-\tilde \Gamma_1)(\tau,z,z')
\right\|_{L^\infty((0,T)\times \pa X\times\pa X)},\label{pcor10}
\end{eqnarray}
where $C_1$ and $C_2$ are the constants that appear on the right-hand side of \eqref{stab2.a} and \eqref{stab2.b}.

From the estimate $|P_\vartheta (k_0-\tilde k_0)(v_0',x_0')|\le \|k-\tilde k\|_{L^\infty(Z)}
\delta^{-(n-1)}\diam$ for a.e. $(x_0',v_0')\in \Gamma_-$ (see \eqref{pcor8}), it follows that
\begin{equation}
\|P_\vartheta(k_0-\tilde k_0)\|_{L^2(\Gamma_-,d\xi)}^2\le \|k-\tilde k\|_{L^\infty(Z)}
\delta^{-(n-1)}\diam\int_{\pa X}\int_{\S^{n-1}_{x_0',-}}|P_\vartheta(k_0-\tilde k_0)(x_0',v_0')||\nu(x_0')\cdot v|dvd\mu(x_0').\label{pcor11}
\end{equation}
Combining \eqref{pcor10}--\eqref{pcor11} and \eqref{pcor7} we obtain \eqref{eq:cor2}.
\end{proof}

\begin{proof}[Proof of Theorem \ref{cor_ball} \eqref{eq:cor2H1}]
We first prove \eqref{1/4.4} given below.
Note that from \eqref{poids1.0}--\eqref{poids1}, it follows that
\begin{equation}
|P_{\vartheta_0}f(v,x)|\le \|f\|_\infty\int_0^{\tau_+(x,v)}{1\over \sqrt{t(\tau_+(x,v)-t)}}dt=C\|f\|_\infty,\label{1/4.1}
\end{equation}
for $(v,x)\in \Gamma_-$ and for $f\in C(X)\cap L^\infty(X)$, where $C=\int_0^1{1\over \sqrt{t(1-t)}}dt$.

Note that $\nu(x)=x$  and $\nu(x)\cdot (x-x_0')=|\nu(x_0')\cdot (x-x_0')|$ for $(x,x_0')\in \pa X=\S^1$. Therefore from \eqref{stab1.b}, it follows that
\begin{equation}
|\nu(x_0')\cdot v_0'|^2
\left|E(x,x_0')P_{\vartheta_0}k_0(v_0',x)-\tilde E(x,x_0')P_{\vartheta_0}\tilde k_0'(v_0',x)\right|
\le C'\left\|\sqrt{\tau^2-|z-z'|^2}
  (\Gamma_1-\tilde \Gamma_1)(\tau,z,z')
  \right\|_{L^\infty},\label{1/4.2}
\end{equation}
for $(x,x_0')\in \pa X^2$, $x\not=x_0'$ and $v_0'={x-x_0'\over |x-x_0'|}$ 
(we also used $P_{\vartheta_0}k_{v_0}(v_0,x)=g(v_0,v_0)P_{\vartheta_0}k_0(v_0,x)$ and the similar identity for $\tilde k$), where $C'={1\over 2\inf_{\Gamma_-}S\inf_{\Gamma_+}W\inf_{v\in \S^{n-1}}g(v,v)}$.
In addition, from \eqref{1/4.2}, \eqref{pcor9} (with $P_{\vartheta_0}$ in place of ``$P_\vartheta$''), and from \eqref{pcor1} and \eqref{1/4.1} (with $\tilde k_0$ in place of $f$), it follows that
\begin{eqnarray}
|\nu(x_0')\cdot v_0'|^2|P_{\vartheta_0}(k_0-\tilde k_0)(x_0',v_0')|&\le&e^{M\diam}C\|\tilde k_0\|_\infty\left|e^{-P\sigma(v_0',x_0')}-e^{-P\tilde\sigma(v_0',x_0')}\right| |\nu(x_0')\cdot v_0|
\nonumber\\
&&\!\!\!\!\!\!\!\!\!\!\!\!\!\!\!\!\!\!\!\!\!\!\!\!\!\!\!\!\!\!\!\!\!\!\!\!\!\!\!\!\!\!\!\!\!\!\!\!\!\!\!\!\!\!\!\!\!\!\!\!\!\!\!\!\!\!\!\!\!\!\!\!\!\!\!\!\!\!\!\!\!\!
+e^{M\diam}\left|e^{-P\sigma(v_0',x_0')}P_\vartheta k_0(x_0',v_0')-e^{-P\tilde\sigma(v_0',x_0')}P_\vartheta\tilde k_0(x_0',v_0')\right||\nu(x_0')\cdot v_0'|^2.\label{1/4.3}
\end{eqnarray}
for $(x_0',v_0')\in \Gamma_-$ (we also used the estimate $|\nu(x_0')\cdot v_0'|\le 1$).
Performing the change of variables ``$x=x_0'+\tau_+(x_0',v_0')v_0'$'' (${\nu(x)\cdot v_0'\over |x-x_0'|^{n-1}}d\mu(x)=dv_0'$) on the left-hand side of \eqref{stab1.a}, we obtain
that the estimate \eqref{stab2.a} still holds. Using \eqref{stab2.a}, \eqref{1/4.3} and \eqref{1/4.2} (and \eqref{link}), we obtain that there exists a constant $C''$ such 
that
\begin{eqnarray}
\int_{\S_{x_0'}^{n-1}}|\nu(x_0')\cdot v_0'|^2|P(\rho (k_0-\tilde k_0))(x_0',v_0')|dv_0'
&\le& C''\left(\|\tilde k_0\|_\infty\|A_{S,W}-\tilde A_{S,W}\|_{\eta,T}\right.\nonumber\\
&&\!\!\!\!\!\!\!\!\!\!\!\!\!\!\!\!\!\!\!\!\!\!\!\!\!\!\!\!+ \left.\left\|\sqrt{\tau^2-|z-z'|^2}
  (\Gamma_1-\tilde \Gamma_1)(\tau,z,z')
  \right\|_{L^\infty}\right)\label{1/4.4}
\end{eqnarray}
for $x_0'\in \pa X$. 
Moreover, using \eqref{1/4.1} (with $k_0-\tilde k_0$ in place of ``$f$'') and Cauchy-Bunyakovski-Schwarz estimate, we obtain
\begin{eqnarray}
&&\!\!\!\!\!\!\!\!\!\!\!\!\!\!\left(\int_{\Gamma_-}|P(\rho(k_0-\tilde k_0))|^2(v_0,x_0')d\xi(x_0',v_0)\right)^{1\over 2}\nonumber\\
&&\!\!\!\!\!\!\!\!\!\!\!\!\!\!\le C\|k_0-\tilde k_0\|_{\infty}^{3\over 4}\left(\int_{\pa X}\int_{\S_{x_0',-}^1}\left(|P(\rho(k_0-\tilde k_0))|(v_0,x_0')|v_0\cdot \nu(x_0')|^2\right)^{1\over 2}dv_0'd\mu(x_0')
\right)^{1\over 2}\nonumber\\
&&\!\!\!\!\!\!\!\!\!\!\!\!\!\!\le C\|k_0-\tilde k_0\|_{\infty}^{3\over 4}\sqrt{2}\pi\left(\int_{\pa X}\int_{\S_{x_0',-}^1}|P(\rho(k_0-\tilde k_0))|(v_0,x_0')|v_0\cdot \nu(x_0')|^2dv_0'd\mu(x_0')
\right)^{1\over 4}.\label{1/4.5}
\end{eqnarray}
Finally combining  \eqref{1/4.4}--\eqref{1/4.5}, \eqref{x-raystab} (and the identity $\int_{\Gamma_-}\left|P(\rho(k_0-\tilde k_0)(v,x_0')\right|^2d\xi(v,x_0')=\int_{\S^{n-1}}\int_{\Pi_v}|P(k_0-\tilde k_0)(v,x)|^2dx dv$), we obtain \eqref{eq:cor2H1}. 
\end{proof}

\section{Proof of Theorem \ref{theorem:normesimplescat}}
\label{proof_thm_normesimplescat}
For $0<b<a$ we remind that
\begin{equation}
\int_0^{2\pi}{1\over a-b\sin(\Omega)}d\Omega={2\pi\over \sqrt{a^2-b^2}}.\label{P1}
\end{equation}
We will use the following Lemma \ref{lem:normeN} to prove Theorem \ref{theorem:normesimplescat} \eqref{A6b}, \eqref{A6c}  and \eqref{A6d}.

\begin{lemma}
\label{lem:normeN}
Let $n\ge 2$. Let $N$ denote the nonnegative measurable function from 
$(0,T)\times \pa X \times\R^n$ to $[0,+\infty[$ defined by
\begin{equation}
N(\tau,x,x')=\chi_{(0,+\infty)}(\tau-|x-x'|)\int_{\S^{n-1}}{(\tau-(x-x')\cdot v)^{n-3}\over |x-x'-\tau v|^{2n-4}}dv,\label{l9.3.1}
\end{equation}
for $(\tau,x,x')\in (0,T)\times \pa X\times \R^n$.
When $n=2$, then
\begin{equation}
N(\tau,x,x')=\chi_{(0,+\infty)}(\tau-|x-x'|){2\pi\over \sqrt{\tau^2-|x-x'|^2}},\label{l9.3.2a}
\end{equation}
for $(\tau,x,x')\in (0,T)\times\pa X\times \R^n$.
When $n=3$, then
\begin{equation}
N(\tau,x,x')=2\pi{\chi_{(0,+\infty)}(\tau-|x-x'|)\over \tau|x-x'|}\ln\left({\tau+|x-x'|\over \tau-|x-x'|}\right),\label{l9.3.2b}
\end{equation}
for $(\tau,x,x')\in (0,T)\times\pa X\times\R^n$.
When $n\ge 4$, then
\begin{equation}
\sup_{(\tau,x,x')\in (0,T)\times\pa X\times\R^n}\tau|x-x'|^{n-2}N(\tau,x,x')<\infty.\label{l9.3.2}
\end{equation}
\end{lemma}

\begin{proof}[Proof of Lemma \ref{lem:normeN}.]
Let $(\tau,x,x')\in (0,T)\times\pa X\times \R^n$. 
We first prove \eqref{l9.3.2a}. Let $n=2$.
Note that
$$
N(\tau,x,x')=\chi_{(0,+\infty)}(\tau-|x-x'|)
\int_0^{2\pi}{1\over \tau-|x-x'|\sin(\Omega)}d\Omega.
$$
Therefore using \eqref{P1} we obtain \eqref{l9.3.2a}.
We prove \eqref{l9.3.2b}. Let $n=3$.
Note that
$$
N(\tau,x,x')=2\pi{\chi_{(0,+\infty)}(\tau-|x-x'|)\over 2\tau|x-x'|}
\int_{-{\pi\over 2}}^{\pi\over 2}{d\over d\Omega}\ln\left(\tau^2+|x-x'|^2-2\tau|x-x'|\sin(\Omega)\right)d\Omega,
$$
which gives \eqref{l9.3.2b}.

We prove \eqref{l9.3.2}. Let $n\ge 4$ and let $(\tau,x,x')\in (0,T)\times \pa X\times\R^n$ be such that $\tau>|x-x'|$ (we remind that $N(\tau,x,x')=0$ if $\tau\le |x-x'|$).
Using spherical coordinates, we obtain
\begin{equation}
N(\tau,x,x')={\rm Vol}_{n-2}(\S^{n-2})
\int_{-{\pi\over 2}}^{\pi\over 2}{(\tau-|x-x'|\sin (\Omega))^{n-3}\over \left(|x-x'|^2+\tau^2-2\tau|x-x'| \sin(\Omega)\right)^{n-2}}\cos(\Omega)^{n-2}d\Omega.
\label{l9.3.3}
\end{equation}
Performing the change of variables ``$r={\tau^2-|x-x'|^2\over 2(\tau-|x-x'|\sin(\Omega))}-{\tau-|x-x'|\over 2}$'', we obtain
\begin{eqnarray}
N(\tau,x,x')&=&{{\rm Vol}_{n-2}(\S^{n-2})(\tau^2-|x-x'|^2)^{n-3\over 2}\over |x-x'|^{n-2}}
\int_0^{|x-x'|}{\sqrt{r(|x-x'|-r)}^{n-3}\over({\tau-|x-x'|\over 2}+r)^{n-2}({\tau+|x-x'|\over 2}-r)^{n-2}}dr\nonumber\\
&=&{2{\rm Vol}_{n-2}(\S^{n-2})(\tau^2-|x-x'|^2)^{n-3\over 2}\over |x-x'|^{n-2}}
\int_0^{|x-x'|\over 2}{\sqrt{r(|x-x'|-r)}^{n-3}\over({\tau-|x-x'|\over 2}+r)^{n-2}({\tau+|x-x'|\over 2}-r)^{n-2}}dr\nonumber\\
&\le&{2{\rm Vol}_{n-2}(\S^{n-2})|(\tau^2-|x-x'|^2)^{n-3\over 2}\over |x-x'|^{n-2}}
\int_0^{|x-x'|\over 2}{1\over({\tau-|x-x'|\over 2}+r)^{n-1\over 2}({\tau+|x-x'|\over 2}-r)^{n-1\over 2}}dr\nonumber
\end{eqnarray}
\begin{eqnarray}
&\le&{2{\rm Vol}_{n-2}(\S^{n-2})(\tau^2-|x-x'|^2)^{n-3\over 2}\over |x-x'|^{n-2}}{\tau\over 2}^{1-n\over 2}
\int_0^{|x-x'|\over 2}{1\over({\tau-|x-x'|\over 2}+r)^{n-1\over 2}}dr\nonumber\\
&\le&{2^{n-1}\over n-3}{\rm Vol}_{n-2}(\S^{n-2})|x-x'|^{2-n}\left({\tau+|x-x'|\over 2\tau}\right)^{n-3\over 2}\tau^{-1}\nonumber\\
&\le&{2^{n-1}\over n-3}{\rm Vol}_{n-2}(\S^{n-2})|x-x'|^{2-n}\tau^{-1},\label{l9.3.4}
\end{eqnarray}
which proves \eqref{l9.3.2}.
\end{proof}

We are ready to prove Theorem \ref{theorem:normesimplescat}.
First we give an estimate on the simple scattering term.
From \eqref{E8a} it follows that
\begin{equation}
|\gamma_1(\tau,x,x')|\le2^{n-2}\|W\|_{\infty}\|S\|_{\infty}\|k\|_{\infty}I_1(\tau,x,x')\label{S_0.1}
\end{equation}
for a.e. $(\tau,x,x')\in \R\times\pa X\times \pa X$, where
\begin{eqnarray}
I_1(\tau,x,x')&=&\chi_{(0,+\infty)}(\tau-|x-x'|)\int_{\S^{n-1}}\chi_{{\rm supp}k}(x-sv)_{|s={\tau^2-|x-x'|^2\over 2(\tau-v\cdot (x-x'))}}\nonumber\\
&&\times {(\tau-(x-x')\cdot v)^{n-3}\over |x-x'-\tau v|^{2n-4}} dv.\label{S_0.2}
\end{eqnarray}
Let $(\tau,x,x')\in (0,T)\times\pa X\times \pa X$ be such that $x\not=x'$ and $\tau>|x-x'|$. 
Assume without loss of generality $x'-x=|x'-x|(1,0\ldots 0)$.

First we prove \eqref{A6b}--\eqref{A6d}.
From \eqref{S_0.2} and \eqref{l9.3.1}, it follows that
\begin{equation}
I_1(\tau,x,x')\le N(\tau,x,x').\label{S_0.4b*}
\end{equation}
Combining \eqref{l9.3.2a} (respectively \eqref{l9.3.2b}, \eqref{l9.3.2}) with \eqref{S_0.1} and \eqref{S_0.4b*}, we obtain \eqref{A6b} (respectively \eqref{A6c}, \eqref{A6d}).

Now assume that $k\in L^\infty(X\times\S^{n-1}\times\S^{n-1})$ and
\begin{equation}
{\rm supp}k\subseteq \{x\in X\ |\ \inf_{y\in \pa X}|y-x|\ge \delta\}\textrm{ for some }0<\delta<\infty.\label{l9.1.0}
\end{equation}
Let $v\in \S^{n-1}$ and $s:={\tau^2-|x-x'|^2\over 2(\tau-(x-x')\cdot v)}$. Straightforward computations give
$s+|x-x'-sv|=\tau$.
Using \eqref{l9.1.0} we obtain that
\begin{equation}
\textrm{if }\tau<\delta \textrm{ or }s>\tau-\delta, \textrm{ then }x-sv\not\in {\rm supp}k.\label{l9.1.1}
\end{equation}

Using \eqref{S_0.2} and \eqref{l9.1.1}, we obtain
\begin{equation}
\textrm{if }\tau<\delta \textrm{ then }I_1(\tau,x,x')=0.\label{S_0.4a}
\end{equation} 
We prove \eqref{A6e} for $n=2$. 
Using \eqref{S_0.4b*},  we obtain that
$I_1(\tau,x,x')\le{2\pi\over \sqrt{\delta}\sqrt{\tau-|x-x'|}}$ for $\tau\ge \delta$.
Combining \eqref{S_0.1} with this latter estimate and \eqref{S_0.4a}, we obtain \eqref{A6e} for $n=2$.

We now prove \eqref{A6e} for $n\ge 3$.
Let $n\ge 3$ and $\tau\ge \delta$ (the case $\tau <\delta$ is already considered in \eqref{S_0.4a}).
Performing the change of variables ``$r={\tau^2-|x-x'|^2\over 2(\tau-|x-x'|\sin(\Omega))}-{\tau-|x-x'|\over 2}$'' with 
``$v=\Phi(\Omega,\omega):=(\sin (\Omega),\cos(\Omega) \omega)$, $\Omega\in (-{\pi\over 2}, {\pi\over 2})$, 
$\omega\in \S^{n-2}$'' on the right-hand side of \eqref{S_0.2}, we obtain
\begin{eqnarray}
I_1(\tau,x,x')
&=&2^{2-n}{(\tau^2-|x-x'|^2)^{n-3\over 2}\over |x-x'|^{n-2}}\int_0^{|x-x'|}{\sqrt{r(|x-x'|-r)}^{n-3}\over ({\tau-|x-x'|\over 2}+r)^{n-2}({\tau+|x-x'|\over 2}-r)^{n-2}}\nonumber\\
&&\!\!\!\!\!\!\!\!\!\!\!\!\!\int_{\S^{n-2}}\left[\chi_{{\rm supp}k}(x-sv)\right]_{{\Omega={\rm arcsin}
(|x-x'|^{-1}(\tau-{(\tau^2-|x-x'|^2)\over 2(r+{\tau-|x-x'|\over 2})}))
\atop s=r+{\tau-|x-x'|\over 2}}\atop v=\Phi(\Omega,\omega)} d\omega dr.
\label{S_0.6}
\end{eqnarray}

Now assume $\tau>{\delta\over 2}+|x-x'|$. 
Then 
\begin{eqnarray*}
&&|x-x'|^{2-n}\int_0^{|x-x'|}{\sqrt{r(|x-x'|-r)}^{n-3}\over ({\tau-|x-x'|\over 2}+r)^{n-2}({\tau+|x-x'|\over 2}-r)^{n-2}}dr\\
&&\le \left({\delta\over 4}\right)^{4-2n}|x-x'|^{2-n}\int_0^{|x-x'|}\sqrt{r(|x-x'|-r)}^{n-3}dr=
\left({\delta\over 4}\right)^{4-2n}\int_0^1\sqrt{r(1-r)}dr\le\left({\delta\over 4}\right)^{4-2n}.
\end{eqnarray*}
Therefore using \eqref{S_0.6} we obtain
\begin{equation}
(\tau-|x-x'|)^{-{n-3\over 2}}I_1(\tau,x,x')
\le 2^{n-2}{\rm Vol}_{n-2}(\S^{n-2})(T+\diam)^{n-3\over 2}\left({\delta\over 2}\right)^{4-2n}.\label{S_0.8b}
\end{equation}
Finally assume $\delta\le\tau\le{\delta\over 2}+|x-x'|$ and $|x-x'|<\tau\le T$. From \eqref{l9.1.1}, it follows that
\begin{equation}
(\tau-|x-x'|)^{-{n-3\over 2}}I_1(\tau,x,x')\le{{\rm Vol}_{n-2}(\S^{n-2})(T+\diam)^{n-3\over 2}\over 2^{n-2}|x-x'|^{n-2}}\int\limits_{r_-(\tau,x,x')}^{r_+(\tau,x,x')}\!\!\!\!\!\!\!\!\!\!
{\sqrt{r(|x-x'|-r)}^{n-3}\over ({\tau-|x-x'|\over 2}
+r)^{n-2}({\tau+|x-x'|\over 2}-r)^{n-2}}dr,\label{S_0.9}
\end{equation}
where
\begin{equation}
r_-(\tau,x,x'):={|x-x'|+\delta-\tau\over 2},\ 
r_+(\tau,x,x'):={\tau-\delta+|x-x'|\over 2}.\label{S_0.10b}
\end{equation}
Note that 
\begin{equation*}
\int_{r_-(\tau,x,x')}^{r_+(\tau,x,x')}{\sqrt{r(|x-x'|-r)}^{n-3}\over ({\tau-|x-x'|\over 2}
+r)^{n-2}({\tau+|x-x'|\over 2}-r)^{n-2}}dr
=2\int_{r_-(\tau,x,x')}^{|x-x'|\over 2}{\sqrt{r(|x-x'|-r)}^{n-3}\over ({\tau-|x-x'|\over 2}
+r)^{n-2}({\tau+|x-x'|\over 2}-r)^{n-2}}dr
\end{equation*}
\begin{equation}
\le  2\left({\tau\over 2}\right)^{2-n}|x-x'|^{n-3}\int_{r_-(\tau,x,x')}^{|x-x'|\over 2}{1\over ({\tau-|x-x'|\over 2}
+r)^{n-2}}dr=2^{n-1}\tau^{2-n}|x-x'|^{n-3}\int_{r_-(\tau,x,x')}^{|x-x'|\over 2}{1\over ({\tau-|x-x'|\over 2}
+r)^{n-2}}dr.\label{S_0.11}
\end{equation}
Using \eqref{S_0.10b} we obtain 
\begin{equation}
\int_{r_-(\tau,x,x')}^{|x-x'|\over 2}{1\over ({\tau-|x-x'|\over 2}
+r)^{n-2}}dr=C(n,\tau):=\left\lbrace
\begin{matrix}
\ln\left({\tau\over \delta}\right),\textrm{ if }n=3,\\
{1\over n-3}\left(\left({\delta\over 2}\right)^{3-n}-\left({\tau\over 2}\right)^{3-n}\right) \textrm{ otherwise.}
\end{matrix}
\right.
\label{S_0.12}
\end{equation}
From \eqref{S_0.9}, \eqref{S_0.11}, \eqref{S_0.12} and  the estimates $\delta\le \tau<{\delta\over 2}+|x-x'|$, it follows that
\begin{equation}
(\tau-|x-x'|)^{-{n-3\over 2}}I_1(\tau,x,x')\le2^n{\rm Vol}_{n-2}(\S^{n-2})\delta^{-(n-1)}(T+\diam)^{n-3\over 2}C(n,T),\label{S_0.13}
\end{equation}
where the constant $C(n,T)$ is defined in \eqref{S_0.12}.
Combining \eqref{S_0.1} with \eqref{S_0.4a}, \eqref{S_0.8b} and \eqref{S_0.13}, we obtain \eqref{A6e} for $n\ge 3$.
\hfill$\Box$


\section{Proof of Theorem \ref{theorem:normemultiplescat} }
\label{proof_thm_normemultiplescat}
We shall use the following Lemmas \ref{lem:prep}, \ref{lem:tildeJ}, \ref{lem:vol} and \ref{lem:n=3nulaubord}.  Lemmas \ref{lem:prep}, \ref{lem:tildeJ}, \ref{lem:vol} and \ref{lem:n=3nulaubord} are proved in Section \ref{proof_N}.

We introduce some notation first. Let $m\ge 1$ and $z',z\in \R^n$
such that $z\not= z'$.  Let $\mu \ge 0$. We denote by ${\cal
  E}_{m,n}(\mu,z,z')$ the subset of $(\R^n)^m$ defined by
\begin{equation}
{\cal E}_{m,n}(\mu,z,z')=\{(y_1,\ldots,y_m)\in (\R^n)^m\ 
|\ |y_1|+\ldots +|y_m|+|z-z'-y_1-\ldots -y_m|< \mu\}.\label{ellipse}
\end{equation}
When $\mu \le |z-z'|$, then ${\cal E}_{m,n}(\mu,z,z')=\emptyset$.
\begin{lemma}
\label{lem:prep}
Let $J_2$  be the function from $(0,T)\times\pa X\times\R^n$ defined by
\begin{equation}
J_2(\mu,z,z')=\int_{{\cal E}_{1,n}(\mu,z,z')}{1\over |y|^{n-1}}N(\mu-|y|,z,z'+y)dy,\label{l7.1.1}
\end{equation}
where $N$ is defined by \eqref{l9.3.1}.
Then the following statements are valid:
\begin{equation}
\sup_{(\mu,z,z')\in (0,T)\times\pa X\times\R^n\atop \mu>|z-z'|}J_2(\mu,z,z')<\infty,\ \mbox{when}\ n=2;\label{l7.1.2}
\end{equation}
\begin{equation}
\sup_{(\mu,z,z')\in (0,T)\times\pa X\times\R^n\atop \mu>|z-z'|}(\mu-|z-z'|)^{-1}\mu|z-z'|\left(1+\ln\left({\mu+|z-z'|\over \mu-|z-z'|}\right)\right)^{-2}J_2(\mu,z,z')<\infty,
\ \mbox{when}\ n=3;\label{l7.1.3}
\end{equation}
\begin{equation}
\sup_{(\mu,z,z')\in (0,T)\times\pa X\times\R^n\atop \mu>|z-z'|}(\mu-|z-z'|)^{-1}\mu|z-z'|^{n-2}J_2(\mu,z,z')<\infty, \ \mbox{when}\ n\ge 4.\label{l7.1.4}
\end{equation}
\end{lemma}

\begin{lemma}
\label{lem:tildeJ}
Let $m\ge 3$ and let $\tilde J_m$ be the function from $\{(\tau,x,x')\in (0,T)\times\R^n\times\R^n\ |\ 0<|x-x'|<\tau\}$ to $\R$ defined by
\begin{equation}
\tilde J_m(\tau,x,x')=\int_{{\cal E}_{m-2,n}(\tau,x,x')}{dy_2\ldots dy_{m-1}\over |y_2|^{n-1}\ldots |y_{m-1}|^{n-1}|x-x'-y_2-\ldots-y_{m-1}|^{n-2}},\label{th3.2pr8b}
\end{equation}
for $(\tau,x,x')\in (0,T)\times \R^n\times\R^n$, $0<|x-x'|<\tau$. 
When $n=3$, then there exists a constant $\tilde C$ which does not depend on $m$ such that
\begin{equation}
\tilde J_m(\tau,x,x')\le \tilde C{\tau-|x-x'|\over|x-x'|}\left(1+\ln\left({\tau+|x-x'|\over \tau-|x-x'|}\right)\right){m^{n-1}\left({\rm Vol}_{n-1}(\S^{n-1})\tau\right)^{m-3}\over (m-3)!},\label{th3.2pr9a}
\end{equation}
for $(\tau,x,x')\in (0,T)\times\R^n\times\R^n$, $0<|x-x'|<\tau$.
When $n\ge 4$, then there exists a constant $\tilde C$ which does not depend on $m$ such that
\begin{equation}
\tilde J_m(\tau,x,x')\le \tilde C(\tau-|x-x'|)|x-x'|^{2-n}{m^{n-1}\left({\rm Vol}_{n-1}(\S^{n-1})\tau\right)^{m-3}\over (m-3)!},\label{th3.2pr9b}
\end{equation}
for $(\tau,x,x')\in (0,T)\times\R^n\times\R^n$, $0<|x-x'|<\tau$.
\end{lemma}

\begin{lemma}
\label{lem:vol}
Let $n\ge 2$. Let $(\tau,x,x')\in \R\times \R^n\times\R^n$ be such that $\tau>|x-x'|>0$, the following estimate is valid:
\begin{equation}
{\rm Vol}_{n}({\cal E}_{1,n}(\tau,x,x'))\le{{\rm Vol}_{n-2}(\S^{n-2})\pi(\tau+|x-x'|)\over 4}
\left({\sqrt{\tau^2-|x-x'|^2}\over 2}\right)^{n-1},\label{lA1b}
\end{equation}
where ${\cal E}_{1,n}$ is defined by \eqref{ellipse}.
\end{lemma}

\begin{lemma} 
\label{lem:n=3nulaubord}Let $B$ be the function from $\{(\mu,z,z')\in (0,T)\times \pa X\times \R^3\ |\ \mu>|z-z'|>0\}$ to $\R$ defined by
\begin{equation}
B(\mu,z,z'):= \int_{{\cal E}_{1,3}(\mu,z,z')}\ln\left({\mu-|y|+|z-z'-y|\over \mu-|y|-|z-z'-y|}\right)dy
=
\int_{{\cal E}_{1,3}(\mu,t_0(1,0,0),0)}\ln\left({\mu-|y|+|(t_0,0,0)-y|\over \mu-|y|-|(t_0,0,0)-y|}\right)dy,\label{lemp1}
\end{equation}
for $(\mu,z,z')\in (0,T)\times\pa X\times \R^n$ where $t_0=|z-z'|$, $z\not= z'$, $\mu>|z-z'|$.
Then we have:
\begin{equation}
\sup_{(\mu,z,z')\in(0,T)\times\pa X\times\R^3 \atop \mu>|z-z'|>0}(\mu-|z-z'|)^{-1}\left(1+\ln\left({\mu+|z-z'|\over \mu-|z-z'|}\right)\right)^{-1}B(\mu,z,z')<\infty.\label{lemp0}
\end{equation}
\end{lemma}

We also need the explicit expression of $\gamma_m$, $m\ge 2$, to prove Theorem \ref{theorem:normemultiplescat}
\begin{eqnarray}
&&\gamma_2(\tau,x,x'):=\int_{y\in {\cal E}_{1,n}(\tau, x,x')\atop x'+y\in X}\int_{\S^{n-1}_{x,+}}(\nu(x)\cdot v)W(x,v)\left[E(x,x-(\tau-|y|-s_1)v,x'+y,x')\right.\nonumber\\
&&\chi_{(0,\tau_-(x,v))}(\tau-|y|-s_1)k(x-(\tau-s_1-|y|)v,v_1,v)k(x'+y,v',v_1)S(x',v')\nonumber\\
&&\left.|\nu(x')\cdot v'|\right]_{{s_1={|x-x'-y-(\tau-|y|)v|^2\over 2(\tau-|y|-(x-x'-y)\cdot v)}\atop v_1={x-x'-y-(\tau-s_1-|y|)v\over s_1}}\atop v'={y\over |y|}}
{2^{n-2}(\tau-|y|-(x-x'-y)\cdot v)^{n-3}\over |x-x'-y-(\tau-|y|)v|^{2n-4}|y|^{n-1}} dy dv,\label{E8b}
\end{eqnarray}
and
\begin{eqnarray}
&&\gamma_m(\tau,x,x'):=
\int_{(y_2,\ldots,y_m)\in {\cal E}_{m-1,n}(\tau,x,x')\atop (x'+y_m,\ldots,x'+y_m+\ldots+y_2)\in X^{m-1}}\int_{\S^{n-1}_{x,+}}(\nu(x)\cdot v)W(x,v)\nonumber\\
&&\times {2^{n-2}\left(\tau-|y_2|-\ldots-|y_m|-(x-x'-y_2-\ldots-y_m)\cdot v\right)^{n-3}\over |y_2|^{n-1}\ldots|y_m|^{n-1}|x-x'-y_2-\ldots-y_m-
(\tau-|y_2|-\ldots-|y_m|)v|^{2n-4}}\nonumber\\
&&\times \left[\chi_{(0,\tau_-(x,v))}(\tau-s_1-|y_2|-\ldots-|y_m|)E(x,x-(\tau-s_1-|y_2|-\ldots-|y_m|)v,\right.\nonumber\\
&&x'+y_m\ldots+y_2,\ldots, x'+y_m,x') k(x-(\tau-s_1-|y_2|-\ldots-|y_m|)v,v_1,v)\nonumber\\
&&\times k(x'+y_m+\ldots+y_2,v_2,v_1)\ldots k(x'+y_m+\ldots+y_{i+1},v_{i+1},v_i)\ldots \nonumber\\
&&k(x'+y_m+y_{m-1},v_{m-1},v_{m-2})k(x'+y_m,v',v_{m-1})S(x',v')\nonumber\\
&&\left.|\nu(x')\cdot v'|
\right]_{{{v_1={x-x'-y_2-\ldots-y_m-(\tau-s_1-|y_2|-\ldots-|y_m|)v\over s_1}\atop s_1={|x-x'-y_2-\ldots-y_m-(\tau-|y_2|-\ldots-|y_m|)v|^2\over 2(t-|y_2|-
\ldots-|y_{m-1}|-(x-x'-y_2-\ldots y_{m-1})\cdot v)}}\atop v'={y_m\over |y_m|}}\atop v_i={y_i\over |y_i|},\ i=2\ldots m-1}
dy_2\ldots dy_m dv,\label{E8d}
\end{eqnarray}
for $\tau \in \R$ and a.e. $(x,x')\in \pa X\times\pa X$ and for $m\ge 3$.

We are ready to prove Theorem \ref{theorem:normemultiplescat}.
We prove \eqref{A6b-}, \eqref{A6c-} and \eqref{A6d-}.
Let $\tau\in (0,T)$ and let $x\in \pa X$, $x'\in \pa X$ and $x\not=x'$. Set $t_0=|x-x'|$.
We first look for an upper bound on $|\gamma_2(\tau,x,x')|$.
Using \eqref{E8b} and the fact that $\sigma$ is a nonnegative function, we obtain 
\begin{equation}
|\gamma_2(\tau,x,x')|\le2^{n-2}\|W\|_{\infty}\|S\|_{\infty}\|k\|_{\infty}^2J_2(\tau,x,x'),\label{th3.2pr1}
\end{equation}
where $J_2$ and ${\cal E}_{1,n}(\tau,x,x')$ are defined by \eqref{l7.1.1} and \eqref{ellipse}.
From \eqref{th3.2pr1} and \eqref{l7.1.2}--\eqref{l7.1.4} it follows that there exists a real constant $C$ such that
\begin{equation}
|\gamma_2(\tau,x,x')|\le C\|W\|_{\infty}\|S\|_{\infty}\|k\|_{\infty}^2\sup_{(s,z,z')\in (0,T)\times\R^n\times\R^n\atop s>|z-z'|}J_2(s,z,z'),\textrm{ when }
n=2,\label{th3.2pr2a}
\end{equation}
and
\begin{equation}
{\tau|x-x'|\over(\tau-|x-x'|)\left(1+\ln\left({\tau+|x-x'|\over \tau-|x-x'|}\right)\right)^2}|\gamma_2(\tau,x,x')|
\le C\|W\|_{\infty}\|S\|_{\infty}\|k\|_{\infty}^2, \textrm{ when }n=3,
\label{th3.2pr2b}
\end{equation}
and 
\begin{equation}
{\tau|x-x'|^{n-2}\over \tau-|x-x'|}|\gamma_2(\tau,x,x')|\le C\|W\|_{\infty}\|S\|_{\infty}\|k\|_{\infty}^2,
\textrm{ when }n\ge 4.\label{th3.2pr2c} 
\end{equation}

Let $m\ge 3$. 
Using \eqref{E8d} we obtain
\begin{equation}
|\gamma_m(\tau,x,x')|\le 2^{n-2}\|W\|_{\infty}\|S\|_{\infty}\|k\|_{\infty}^{m}J_m(\tau,x,x'),\label{th3.2pr3}
\end{equation}
where 
\begin{equation}
J_m(\tau,x,x')=\int_{{\cal E}_{m-2,n}(\tau,x,x')}{J_2(\tau(\bar y), w(\bar y),0)\over |y_2|^{n-1}\ldots |y_{m-1}|^{n-1}}dy_2\ldots dy_{m-1}\label{th3.2pr4},
\end{equation}
and $J_2$ (resp. ${\cal E}_{m-2,n}(\tau,x,x')$) is defined by \eqref{l7.1.1} (resp. \eqref{ellipse}) and where $\bar y=(y_2,\ldots, y_{m-1})$,
$\tau(\bar y)=\tau-|y_2|-\ldots-|y_{m-1}|$, $t_0(\bar y)=|x-x'-y_2-\ldots-y_{m-1}|$ and $w(\bar y)=x-x'-y_2-\ldots-y_{m-1}$ for $\bar y\in (\R^n)^{m-2}$.

Assume $n=2$. 
Then using \eqref{th3.2pr4}, \eqref{l7.1.2} and spherical coordinates (and \eqref{ellipse}), we obtain 
\begin{eqnarray}
J_m(\tau,x,x')&\le&\sup_{(s,z,z')\in (0,T)\times\R^n\times\R^n\atop s>|z-z'|}J_2(s,z,z')\int_{{\cal E}_{m-2,n}(\tau,x,x')}{1\over |y_2|\ldots |y_{m-1}|}d\bar y\nonumber\\
&\le&(2\pi)^{m-2}\sup_{(s,z,z')\in (0,T)\times\R^n\times\R^n\atop s>|z-z'|}J_2(s,z,z')\int_{s_2+\ldots+s_{m-1}\le \tau\atop s_i\ge 0,\ i=2\ldots m-1}ds_2\ldots ds_{m-1}\nonumber\\
&=&(2\pi)^{m-2}{\tau^{m-2}\over (m-2)!}\sup_{(s,z,z')\in (0,T)\times\R^n\times\R^n\atop s>|z-z'|}J_2(s,z,z').\label{th3.2pr5}
\end{eqnarray}

Finally combining \eqref{th3.2pr5} and \eqref{th3.2pr3}, we obtain
\begin{equation}
|\gamma_m(\tau,x,x')|\le(2\pi)^{m-2}\|W\|_{\infty}\|S\|_{\infty}\|k\|_{\infty}^{m}{\tau^{m-2}\over (m-2)!}\sup_{(s,z,z')\in (0,T)\times\R^n\times\R^n\atop s>|z-z'|}J_2(s,z,z'),\label{th3.2pr6}
\end{equation}
Statement \eqref{A6b-} follows from  \eqref{th3.2pr2a}, \eqref{th3.2pr6}. 

Assume $n\ge 3$.
Note that
\begin{equation}
{\mu-|z-z'|\over \mu}=1-{|z-z'|\over \mu}\le 1, \label{th3.2pr7a}
\end{equation}
and 
\begin{equation}
{\mu-|z-z'|\over \mu}\left(1+\ln\left({\mu+|z-z'|\over \mu-|z-z'|}\right)\right)^2\le 
\sup_{s\in (0,1)}(1-s)\left(1+\ln\left({1+s\over 1-s}\right)\right)^2,\label{th3.2pr7b}
\end{equation}
for $(\mu,z,z')\in (0,T)\times\R^n\times\R^n$ such that $|z-z'|<\mu$.

From \eqref{th3.2pr4}, \eqref{th3.2pr7a} and \eqref{l7.1.3}, \eqref{th3.2pr7b} and \eqref{l7.1.4}, it follows that there exists a real constant $C$ such that
\begin{equation}
J_m(\tau,x,x')\le C\tilde J_m(\tau,x,x'),\label{th3.2pr8a}
\end{equation}
where $\tilde J_m$ is defined by \eqref{th3.2pr8b}.

Assume $n=3$. Combining \eqref{th3.2pr3}, \eqref{th3.2pr8a}, \eqref{th3.2pr9a}, we obtain that there exists a real constant $C'$ (which does not depend on $\tau$, $x$, $x'$ and $m$) such that
\begin{equation}
|\gamma_m(\tau,x,x')|\le C'\|W\|_{\infty}\|S\|_{\infty}\|k\|_{\infty}^{m}{\tau-|x-x'|\over |x-x'|}\left(1+\ln\left({\tau+|x-x'|\over \tau-|x-x'|}\right)\right)
{m^{n-1}\left({\rm Vol}_{n-1}(\S^{n-1})\tau\right)^{m-3}\over (m-3)!}.\label{th3.2pr10a}
\end{equation}
Statement \eqref{A6c-} follows from \eqref{th3.2pr2b} and \eqref{th3.2pr10a}.

Now assume $n\ge 4$. Combining \eqref{th3.2pr3}, \eqref{th3.2pr8a}, \eqref{th3.2pr9b}, we obtain that there exists a real constant $C'$ (which does not depend on $\tau$, $x$,
$x'$ and $m$) such that
\begin{equation}
|\gamma_m(\tau,x,x')|\le C'\|W\|_{\infty}\|S\|_{\infty}\|k\|_{\infty}^{m}(\tau-|x-x'|)|x-x'|^{2-n}{m^{n-1}\left({\rm Vol}_{n-1}(\S^{n-1})\tau\right)^{m-3}\over (m-3)!}.
\label{th3.2pr10ab}
\end{equation}
Statement \eqref{A6d-} follows from \eqref{th3.2pr2c} and \eqref{th3.2pr10ab}.

We now prove \eqref{A6e-3}--\eqref{A6e-}.  Let $n\ge 3$ and $m\ge 2$.  From the
expression of $\gamma_m$ (see \eqref{E8b}--\eqref{E8d}), it follows that
\begin{equation}
|\gamma_m(\tau,x,x')|\le 2^{n-2}\|W\|_{\infty}\|S\|_{\infty}\|k\|_{\infty}^mI_m(\tau,x,x')\label{9.1}
\end{equation}
where
\begin{equation}
I_m(\tau,x,x'):=\int\limits_{(y_2,\ldots,y_m)\in{\cal E}_{m-1,n}(\tau,x,x_0')\atop (x'+y_m,\ldots,x'+ \sum_{i=2}^m y_i)\in  ({\rm supp}k)^{m-1}}{N(\tau-\sum_{i=2}^m|y_i|,x,x'+\sum_{i=2}^my_i)dy_m\ldots dy_2\over  |y_2|^{n-1}\ldots |y_m|^{n-1}},\label{9.1b}
\end{equation}
where $N$ and ${\cal E}_{m-1,n}(\tau,x,x')$ are defined by \eqref{l9.3.1} and \eqref{ellipse}.
Note that
\begin{eqnarray}
&&|y_m|\ge \delta
\quad \mbox{ and } \quad 
\tau-\sum_{i=2}^m|y_i|\ge|x-x'-y_2-\ldots-y_m|\ge \delta,\label{9.2b}
\end{eqnarray}
for $(y_2,..,y_m)\in {\cal E}_{m-1,n}(\tau,x,x')$ such that $x'+y_m\in {\rm supp}k$ and $x'+y_2+\ldots +y_m\in {\rm supp}k$ since ${\rm supp}k\subseteq \{z\in X\ |\ 
\inf_{y\in \pa X}|y-z|\ge \delta\}$.

We prove \eqref{A6e-3}. Assume $n=3$. Using \eqref{9.1b}--\eqref{9.2b}, \eqref{l9.3.2b} and \eqref{lemp1} we obtain
\begin{equation}
I_2(\tau,x,x')\le 2\pi\delta^{-4}B(\tau,x,x').\label{9.8}
\end{equation}
Therefore using \eqref{lemp0} and \eqref{9.8} we obtain 
\begin{equation}
\sup_{(s,z,z')\in (0,T)\times\pa X\times\pa X\atop s>|z-z'|>0}(s-|z-z'|)^{-1}\left(1+\ln\left({s+|z-z'|\over s-|z-z'|}\right)\right)^{-1}I_2(s,z,z')<\infty.\label{9.9}
\end{equation}

Now assume $m\ge 3$. Using \eqref{9.1b}, we obtain
\begin{equation}
I_m(\tau,x,x')\le 
\int\limits_{(y_3,\ldots,y_m)\in{\cal E}_{m-2,n}(\tau,x,x_0')\atop (x'+\sum_{i=3}^m y_i,x'+y_m)\in ({\rm supp} k)^2}
{J_2(\tau-\sum_{i=3}^m|y_i|,x,x'+\sum_{i=3}^my_i)dy_m\ldots dy_3\over  |y_3|^{n-1}\ldots |y_m|^{n-1}},
\label{9.10}
\end{equation}
where $J_2$ is defined by \eqref{l7.1.1}.
Using \eqref{9.10} and \eqref{l7.1.3} and the estimate $\sup_{r\in (0,1)}r(1-\ln(r))^2<\infty$ we obtain
\begin{equation} 
I_m(\tau,x,x')\le D
\int\limits_{(y_3,\ldots,y_m)\in{\cal E}_{m-2,n}(\tau,x,x_0')\atop (x'+\sum_{i=3}^m y_i,x'+y_m)\in ({\rm supp} k)^2}
{(|x-x'-y_3-\ldots-y_m|+\tau-\sum_{i=3}^m|y_i|)dy_m\ldots dy_3\over  |y_3|^{n-1}\ldots |y_m|^{n-1}(\tau-\sum_{i=3}^m|y_i|)|x-x'-y_3-\ldots-y_m|},
\label{9.10bis}
\end{equation}
where $D:=\sup_{r\in (0,1)}r(1-\ln(r))^2\sup_{(s,z,z')\in (0,T)\times\pa X\times\pa X\atop s>|z-z'|>0}(s-|z-z'|)^{-1}s|z-z'|\left(1+\ln\left({s+|z-z'|\over s-|z-z'|}\right)\right)^{-2}$ $J_2(s,z,z')$.
If $m= 3$, then using \eqref{9.2b} with ``$(y_2,\ldots,y_m)$'' replaced by ``$(y_3,\ldots,y_m)$'', we obtain 
\begin{equation}
I_3(\tau,x,x')\le 2\tau\delta^{-4} D{\rm Vol}({\cal  E}_{1,3}(\tau,x,x'))\label{9.11a}
\end{equation}
(we also used the estimate $|x-x'-y_3|+\tau-|y_3|\le 2\tau$ for $y_3\in {\cal  E}_{1,3}(\tau,x,x')$).
If $m\ge 4$, then using \eqref{9.2b} with ``$(y_2,\ldots,y_m)$'' replaced by ``$(y_3,\ldots,y_m)$'', we obtain 
\begin{eqnarray}
I_m(\tau,x,x')&\le& 2\tau\delta^{-4} D{\rm Vol}({\cal  E}_{1,3}(\tau,x,x'))\int\limits_{(y_3,\ldots,y_{m-1})\in {\cal  E}_{m-3,3}(\tau,x,x')}
{dy_3\ldots dy_{m-1}\over  |y_3|^{n-1}\ldots |y_{m-1}|^{n-1}}\nonumber\\
&\le&
2\tau\delta^{-4} D{\rm Vol}({\cal  E}_{1,3}(\tau,x,x')){\rm Vol}(\S^{n-1})^{m-3}\int_{(s_3,\ldots,s_{m-1})\in (0,+\infty)^{m-3}\atop s_3+\ldots +s_{m-1}<\tau}\!\!\!\!\!\!\!\!\!\!\!\!\!\!\!\!\!\!\!\!\!\!\!\!\!\!\!\!\!\!\!\!\!\!\!\!\!\!\!
\!\!ds_3\ldots ds_{m-1}\nonumber\\
&=&2\tau\delta^{-4} D{\rm Vol}(\S^{n-1})^{m-3}{\tau^{m-3}\over (m-3)!}{\rm Vol}({\cal  E}_{1,3}(\tau,x,x')
\label{9.11b}
\end{eqnarray}
(we also used the estimate $|x-x'-y_3-\ldots-y_m|+\tau-|y_3|-\ldots-|y_m|\le 2\tau$ for $(y_3,\ldots,y_m)\in {\cal  E}_{m-2,3}(\tau,x,x')$ and we performed the changes of variables $y_i=s_i\omega_i$, $(s_i,\omega_i)
\in (0,+\infty)\times\S^{n-1}$).
Statement \eqref{A6e-3}  follows from \eqref{9.1}, \eqref{9.9} and \eqref{9.11a}--\eqref{9.11b} (and \eqref{lA1b}).

We prove \eqref{A6e-}. Let $n\ge 4$.
Using \eqref{9.1b} and \eqref{l9.3.2}, we obtain
\begin{eqnarray}
&&I_m(\tau,x,x')\le\|s|z-z'|^{n-2}N(s,z,z')\|_{L^\infty(\R_s\times\pa X_z\times\R^n_{z'})}\nonumber\\
&&\times\int\limits_{{(y_2,\ldots,y_m)\in{\cal E}_{m-1,n}(\tau,x,x_0')\atop x'+y_m\in {\rm supp}k}
\atop x'+\sum_{i=2}^my_i\in 
{\rm supp}k}{dy_m\ldots dy_2\over  |y_2|^{n-1}\ldots |y_m|^{n-1}|x-x'-\sum_{i=2}^my_i|^{n-2}(\tau-\sum_{i=2}^m|y_i|)}.\label{9.3b}
\end{eqnarray}

Assume $m=2$. Using \eqref{9.2b} and \eqref{9.3b}, we obtain
\begin{equation}
I_m(\tau,x,x')\le\delta^{-2n+2}\|s|z-z'|^{n-2}N(s,z,z')\|_{L^\infty(\R_s\times\pa X_z\times\R^n_{z'})}{\rm Vol}({\cal E}_{1,n}(\tau,x,x')).\label{9.3}
\end{equation}
Therefore using \eqref{lA1b}, we obtain
\begin{equation}
\footnotesize
I_m(\tau,x,x')\le\delta^{-2n+2}\|s|z-z'|^{n-2}N(s,z,z')\|_{L^\infty(\R_s\times\pa X_z\times\R^n_{z'})}{\rm Vol}_{n-2}(\S^{n-2})\pi (\tau+|x-x'|)
\left({\sqrt{\tau^2-|x-x'|^2}\over 2}\right)^{n-1}\!\!\!.\label{9.4}
\end{equation}

Assume $m\ge 3$. Using \eqref{9.2b} and \eqref{9.3b}, we obtain
\begin{equation}
I_m(\tau,x,x')\le\delta^{-2n+2}\|s|z-z'|^{n-2}N(s,z,z')\|_{L^\infty(\R_s\times\pa X_z\times\R^n_{z'})}
\hspace{-1cm}\int\limits_{{(y_2,\ldots,y_m)\in{\cal E}_{m-1,n}(\tau,x,x')}}\hspace{-1cm}{dy_m\ldots dy_2\over  |y_2|^{n-1}\ldots |y_{m-1}|^{n-1}}.\label{9.5}
\end{equation}
Note that $|y_m|+|x-x'-y_m|\le |y_2|+\ldots+|y_m|+|x-x'-y_2-\ldots-y_m|$ for $(y_2,\ldots, y_m)\in (\R^n)^{m-1}$. Hence
\begin{eqnarray}
&&|y_2|+\ldots +|y_{m-1}|< \tau-|y_m|\qquad \mbox{ and } \qquad
|y_m|+|x-x'-y_m|<\tau,\label{9.6b}
\end{eqnarray}
for $(y_2,\ldots,y_m)\in{\cal E}_{m-1,n}(\tau,x,x')$ (see
\eqref{ellipse}).  Therefore
\begin{eqnarray}
&&I_m(\tau,x,x')\le\delta^{-2n+2}\|s|z-z'|^{n-2}N(s,z,z')\|_{L^\infty(\R_s\times\pa X_z\times\R^n_{z'})}
\nonumber\\
&&\times\int_{y_m\in {\cal  E}_{1,n}(\tau,x,x')}\int_{\sum_{i=2}^{m-1}|y_i|<\tau-|y_m|}{dy_m\ldots dy_2\over |y_2|^{n-1}\ldots |y_{m-1}|^{n-1}}\nonumber\\
&&=\delta^{-2n+2}{\rm Vol}(\S^{n-1})^{m-2}\|s|z-z'|^{n-2}N(s,z,z')\|_{L^\infty(\R_s\times\pa X_z\times\R^n_{z'})}
\int_{y_m\in {\cal  E}_{1,n}(\tau,x,x')}{(\tau-|y_m|)^{m-2}\over (m-2)!}dy_m\nonumber\\
&&\le\delta^{-2n+2}{\rm Vol}(\S^{n-1})^{m-2}\|s|z-z'|^{n-2}N(s,z,z')\|_{L^\infty(\R_s\times\pa X_z\times\R^n_{z'})}
{\rm Vol}({\cal  E}_{1,n}(\tau,x,x')){\tau^{m-2}\over (m-2)!}\nonumber\\
&&=\delta^{-2n+2}{\rm Vol}(\S^{n-1})^{m-2}\|s|z-z'|^{n-2}N(s,z,z')\|_{L^\infty(\R_s\times\pa X_z\times\R^n_{z'})}
{\rm Vol}_{n-2}(\S^{n-2})\pi (\tau+|x-x'|)\nonumber\\
&&\times
\left({\sqrt{\tau^2-|x-x'|^2}\over 2}\right)^{n-1}{\tau^{m-2}\over (m-2)!}.\label{9.7}
\end{eqnarray}
Statement \eqref{A6e-} follows from \eqref{9.1}, \eqref{9.4} and \eqref{9.7}.{}\hfill$\Box$

\section{Proof of Lemmas \ref{lem:prep}, \ref{lem:tildeJ}, \ref{lem:vol} and \ref{lem:n=3nulaubord}}
\label{proof_N}
We remind the following change of variables for the proof of Lemmas \ref{lem:prep}, \ref{lem:tildeJ}, \ref{lem:vol} and \ref{lem:n=3nulaubord}.
\begin{eqnarray}
&&\int_{{\cal E}_{1,n}(\tau,t_0v,0))}f(y)dy\nonumber\\
&&=\left\lbrace\begin{array}{l}
\displaystyle\int_{(0,2\pi)\times(t_0,\tau)}f\left({t_0+s\cos(\varphi)\over 2},{\sqrt{s^2-t_0^2}\over 2}\sin\varphi\right)\\
\displaystyle\times{(s^2-t_0^2\cos^2(\varphi))\over 4\sqrt{s^2-t_0^2}}ds d\varphi,\textrm{ if }n= 2,\\
\displaystyle\int_{\S^{n-2}\times(0,\pi)\times(t_0,\tau)}f\left({t_0+s\cos(\varphi)\over 2},{\sqrt{s^2-t_0^2}\over 2}\sin\varphi\omega\right)\\
\displaystyle\times\left({\sin(\varphi)\sqrt{s^2-t_0^2}\over 2}\right)^{n-2}
{s^2-t_0^2\cos^2(\varphi)\over 4\sqrt{s^2-t_0^2}}d\omega ds d\varphi,\textrm{ if }n\ge 3,
\end{array}
\right.
\label{P2}
\end{eqnarray} 
for $f\in L^1(\R^n)$ and $(\tau,t_0,v)\in (0,+\infty)\times(0,+\infty)\times\S^{n-1}$ such that $\tau>t_0$.

\begin{proof}[Proof of Lemma \ref{lem:prep}]
We prove \eqref{l7.1.2}.
Let $(\mu,z,z')\in (0,T)\times\pa X\times\R^2$ be such that $\mu>|z-z'|$. From \eqref{l7.1.1} and \eqref{l9.3.2a} it follows that
\begin{eqnarray}
J_2(\mu,z,z')&=&\int_{{\cal E}_{1,2}(\mu,z,z')}{2\pi\over |y|\sqrt{(\mu-|y|)^2-|z-z'-y|^2}}dy\nonumber\\
&=&\int_{{\cal E}_{1,2}(\mu,t_0(1,0),0)}{2\pi\over |y|\sqrt{(\mu-|y|)^2-|t_0(1,0)-y|^2}}dy,\label{th3.2.50}
\end{eqnarray}
where $t_0=|z-z'|$.

Using the change of variables $y={t_0\over 2}(1,0)+(s\cos(\varphi),{\sqrt{s^2-t_0^2}\over 2}\sin(\varphi))$ (see \eqref{P2}), 
$\varphi\in  (0,2\pi)$, $s\in (t_0,\mu)$, we obtain 
\begin{equation}
J_2(\mu,z,z')=4\pi\int_{t_0}^\mu\int_0^{2\pi}J_{2,1}(\mu,s,\varphi)d\varphi ds,\label{th3.2.51}
\end{equation}
where
\begin{equation}
J_{2,1}(\mu,s,\varphi)={s-t_0\cos(\varphi)\over \sqrt{s^2-t_0^2}\sqrt{\mu-s}\sqrt{\mu-t_0\cos(\varphi)}},
\label{th3.2.52a}
\end{equation}
for $\varphi\in (0,2\pi)$ and $s\in (t_0,\mu)$.

We give an estimate on $J_{2,1}$.
From \eqref{th3.2.52a} and the estimates  $\mu-t_0\cos(\varphi)\ge s-t_0\cos(\varphi)$, $s+t_0\ge s-t_0\cos(\varphi)$, it follows that
\begin{equation} 
J_{2,1}(\mu,s,\varphi)\le{1\over \sqrt{s-t_0}\sqrt{\mu-s}},\label{th3.2.53a}
\end{equation}
for $\varphi\in(0,{\pi\over 2})$ and $s\in (t_0,\mu)$. 
Performing the change of variables $s=t_0+\ep(\mu-t_0)$ we have
\begin{equation}
\int_{t_0}^\mu{1\over\sqrt{s-t_0}\sqrt{\mu-s}}ds=\int_0^1{1\over \sqrt{\ep(1-\ep)}}d\ep<+\infty,\label{P4}
\end{equation}
for $s\in (t_0,\mu)$.
Combining  \eqref{th3.2.51}, \eqref{th3.2.53a}, \eqref{P4}, we obtain
\begin{equation}
\sup_{(\mu,z,z')\in (0,T)\times\pa X\times\R^2\atop \mu>|z-z'|}J_2(\mu,z,z')\le 8\pi^2\int_0^1{1\over \sqrt{\ep(1-\ep)}}d\ep<\infty.\label{th3.2.54}
\end{equation}
Statement \eqref{l7.1.2} follows from \eqref{th3.2.54}.\\

We prove \eqref{l7.1.3}.
Let $(\mu,z,z')\in (0,T)\times\pa X\times\R^3$ be such that $\mu>|z-z'|$. Set $t_0=|z-z'|$. From \eqref{l7.1.1}, \eqref{l9.3.2b} and \eqref{P2}, it follows that
\begin{eqnarray}
J_2(\mu,z,z')&\le& \int_{{\cal E}_{1,3}(\mu,z,z')}{2\pi\ln\left({\mu-|y|+|z-z'-y|\over \mu-|y|-|z-z'-y|}\right)\over |y|^2(\mu-|y|)|z-z'-y|}dy\nonumber\\
&=&\int_{{\cal E}_{1,3}(\mu,t_0(1,0,0),0)}{2\pi\ln\left({\mu-|y|+|t_0(1,0,0)-y|\over \mu-|y|-|t_0(1,0,0)-y|}\right)\over |y|^2(\mu-|y|)|t_0(1,0,0)-y|}dy\nonumber\\
&=&8\pi^2\int_{t_0}^\mu\int_0^{\pi}J_{2,1}(\mu,s,\varphi)d\varphi ds,\label{l7ii2}
\end{eqnarray}
where
\begin{eqnarray}
J_{2,1}(\mu,s,\varphi)&=&{\sin(\varphi)\ln\left({\mu-t_0\cos(\varphi)\over \mu-s}\right)\over 
(s+t_0\cos(\varphi))(2\mu-s-t_0\cos(\varphi))}\nonumber\\
&=&{\ln\left({\mu-t_0\cos(\varphi)\over \mu-s}\right)} 
\left({\sin(\varphi)\over 2\mu (s+t_0\cos(\varphi))}+{\sin(\varphi)\over 2\mu(2\mu-s-t_0\cos(\varphi))}\right),\label{l7ii3b}
\end{eqnarray}
for $\varphi\in (0,\pi)$ and $s\in (t_0,\mu)$.
From \eqref{l7ii3b} and the estimates $2\mu-s-t_0\cos(\varphi)\ge \mu-t_0\cos(\varphi)$,
$0\le \ln\left({\mu-t_0\cos(\varphi)\over \mu-s}\right)\le\ln\left({\mu+t_0\over \mu-s}\right)$,
it follows that
\begin{equation*} 
J_{2,1}(\mu,s,\varphi)\le{\ln\left({\mu+t_0\over \mu-s}\right)} 
\left({\sin(\varphi)\over 2\mu (s-t_0\cos(\varphi))}+{\sin(\varphi)\over 2\mu (\mu-t_0\cos(\varphi))}\right),
\end{equation*}
for $\varphi\in (0,\pi)$ and $s\in (t_0,\mu)$. Therefore 
\begin{eqnarray}
\int_0^{\pi}J_{2,1}(\mu,s,\varphi)d\varphi&\le&
{\ln\left({\mu+t_0\over \mu-s}\right)\over 2\mu t_0} 
\left(\ln \left({s+t_0\over s-t_0}\right)+\ln\left({\mu+t_0\over \mu-t_0}\right)\right)\nonumber\\
&\le&{\ln\left({\mu+t_0\over \mu-s}\right)\over 2\mu t_0} 
\left(\ln \left({\mu+t_0\over s-t_0}\right)+\ln\left({\mu+t_0\over \mu-t_0}\right)\right).
\label{l7ii4b}
\end{eqnarray}
We remind the following integral value
\begin{equation}
\int_{t_0}^\mu\ln\left({\mu+t_0\over \mu-s}\right)ds=(\mu-t_0)\ln\left({\mu+t_0\over \mu-t_0}\right)+\mu-t_0.\label{P5}
\end{equation}
Using the estimate $\ln \left({\mu+t_0\over \mu-s}\right)\le
\ln\left({2(\mu+t_0)\over \mu-t_0}\right)$ for $s\in (t_0,{t_0+\mu\over 2})$, we obtain 
\begin{eqnarray}
&&\int_{t_0}^\mu \ln\left(\mu+t_0\over s-t_0\right)\ln\left({\mu+t_0\over \mu-s}\right)ds
=2\int_{t_0}^{t_0+\mu\over 2} \ln\left(\mu+t_0\over s-t_0\right)\ln\left({\mu+t_0\over \mu-s}\right)ds\nonumber\\
&&\le 2 \ln\left({2(\mu+t_0)\over \mu-t_0}\right)\int_{t_0}^{t_0+\mu\over 2} \ln\left(\mu+t_0\over s-t_0\right)ds\nonumber\\
&&\le 2 (\mu-t_0)\left(\ln\left({(\mu+t_0)\over \mu-t_0}\right)+\ln(2)\right)\left(\ln\left({\mu+t_0\over \mu-t_0}\right)+1\right).\label{l7ii7}
\end{eqnarray}
Combining  \eqref{l7ii2}--\eqref{l7ii7} and \eqref{P5},  we obtain
\begin{equation}
J_2(\mu,z,z')\le 4\pi^2 {\mu-t_0\over \mu t_0}\left(3\ln\left({(\mu+t_0)\over \mu-t_0}\right)+2\ln(2)\right)\left(\ln\left({\mu+t_0\over \mu-t_0}\right)+1\right).\label{l7ii8}
\end{equation}  
Statement \eqref{l7.1.3} follows from \eqref{l7ii8}.\\

We prove \eqref{l7.1.4}.
Let $n\ge 4$. Let $(\mu,z,z')\in (0,T)\times\pa X\times\R^n$ be such that $\mu>|z-z'|$. From \eqref{l7.1.1} and \eqref{l9.3.2} it follows that
\begin{equation}
J_2(\mu,z,z')\le C\int_{{\cal E}_{1,n}(\mu,z,z')}\!\!\!\!\!\!\!\!\!\!\!\!\!\!\!\!\!\!\!|y|^{1-n}(\mu-|y|)^{-1}|z-z'-y|^{2-n}dy
=C\int_{{\cal E}_{1,n}(\mu,t_0(1,0\ldots 0),0)}\!\!\!\!\!\!\!\!\!\!\!\!\!\!\!\!\!\!\!\!\!\!|y|^{1-n}(\mu-|y|)^{-1}|t_0(1,0,0)-y|^{2-n}dy,\label{l7iii1}
\end{equation}
where $t_0=|z-z'|$ and $C=\sup_{(\tilde\mu,\tilde z,\tilde z')\in (0,T)\times \pa X\times\R^n}\tilde \mu|\tilde z-\tilde z'|^{n-2}N(\tilde \mu, \tilde z,\tilde z')$.

Using the change of variables $y={t_0\over 2}(1,0)+(s\cos(\varphi),{\sqrt{s^2-t_0^2}\over 2}\sin(\varphi)\omega)$ (see \eqref{P2}), 
$\varphi\in  (0,\pi)$, $s\in (t_0,\mu)$, $\omega\in \S^{n-2}$, we obtain 
\begin{equation}
J_2(\mu,z,z')\le C'\int_{t_0}^\mu\int_0^{\pi}J_{2,1}(\mu,s,\varphi)d\varphi ds,\label{l7iii2}
\end{equation}
where $C'=2^{n-2}{\rm Vol}_{n-2}(\S^{n-2})C$ and
\begin{equation}
J_{2,1}(\mu,s,\varphi)={(s^2-t_0^2)^{n-3\over 2}\sin^{n-2}(\varphi)\over 
(s+t_0\cos(\varphi))^{n-2}(2\mu-s-t_0\cos(\varphi))(s-t_0\cos(\varphi))^{n-3}},
\label{l7iii3b}
\end{equation}
for $\varphi\in (0,2\pi)$ and $s\in (t_0,\mu)$.

We give estimates on $J_{2,1}$.
Let $\varphi\in (0,{\pi\over 2})$ and $s\in (t_0,\mu)$.
From \eqref{l7iii3b} and the estimates $s+t_0\cos(\varphi)\ge s$, $\sqrt{s^2-t_0^2}\sin(\varphi)\le s-t_0\cos(\varphi)$ and  the estimate $2\mu-s-t_0\cos(\varphi)\ge s-t_0\cos(\varphi)$,
it follows that
\begin{eqnarray}
J_{2,1}(\mu,s,\varphi)&\le& {\sqrt{s^2-t_0^2}\sin^2(\varphi)\over 
s^{n-2}(s-t_0\cos(\varphi))^2}\nonumber\\
&\le&C_0{\sqrt{s^2-t_0^2}\over 
s^{n-1}(s-t_0\cos(\varphi))},\label{a}
\end{eqnarray}
where $C_0$ is defined by \eqref{P3} (we also used the estimate $s-t_0\cos(\varphi)\ge s(1-\cos(\varphi))$).
Let $\varphi\in ({\pi\over 2},\pi)$ and $s\in (t_0,\mu)$.
From \eqref{l7iii3b} and the estimates $s-t_0\cos(\varphi)\ge s$, $\sqrt{s^2-t_0^2}\sin(\varphi)\le s+t_0\cos(\varphi)$ and  the estimate $2\mu-s-t_0\cos(\varphi)\ge s$,
it follows that
\begin{equation}
J_{2,1}(\mu,s,\varphi)\le {\sqrt{s^2-t_0^2}\sin^2(\varphi)\over 
s^{n-2}(s+t_0\cos(\varphi))^2}
\le C_0{\sqrt{s^2-t_0^2}\over 
s^{n-1}(s+t_0\cos(\varphi))},\label{b}
\end{equation}
where $C_0$ is defined by \eqref{P3} (we also used the estimate $s+t_0\cos(\varphi)\ge s(1+\cos(\varphi))$).

Combining \eqref{a} and \eqref{b} and \eqref{P1}, we obtain
\begin{equation}
\int_0^{\pi}J_{2,1}(\mu,s,\varphi)d\varphi \le {2\pi C_0\over s^{n-1}},\textrm{ for }s\in (t_0,\mu).\label{c}
\end{equation}
Note that 
\begin{eqnarray}
\int_{t_0}^\mu{1\over s^{n-1}}ds&=&{1\over n-2}\left({\mu^{n-2}-t_0^{n-2}\over t_0^{n-2}\mu^{n-2}}\right)={\mu-t_0\over n-2}\sum_{i=0}^{n-3}{\mu^{-1-i}t_0^{i+2-n}}
\nonumber\\
&\le&{\mu-t_0\over \mu t_0^{n-2}}.\label{l7iii5}
\end{eqnarray}
(we used the estimate $t_0<\mu$ which gives $\mu^{-1-i}t_0^{i+2-n}\le \mu^{-1} t_0^{n-2}$ for $i=0\ldots n-3$).

Combining \eqref{l7iii2}, \eqref{c}--\eqref{l7iii5}, we obtain
\begin{equation}
J_2(\mu,z,z')\le C'{\mu-t_0\over \mu t_0^{n-2}}.\label{l7iii6}
\end{equation}
where $C'$ does not depend on $\mu$ and $z$, $z'$. 
Statement \eqref{l7.1.4} follows from \eqref{l7iii6}.
\end{proof}

\begin{proof}[Proof of Lemma \ref{lem:tildeJ}]
Let $(\tau,x,x')\in (0,T)\times\R^n\times\R^n$ such that $\tau>|x-x'|>0$.  Set $t_0=|x-x'|$.
Let $(y_2,\ldots, y_{m-1})\in {\cal E}_{m-2,n}(\tau,x,x')$. Then
$t_0\le |y_2|+\ldots +|y_{m-1}|+|x-x'-y_2-\ldots y_{m-1}|$. Therefore either $|x-x'-y_2-\ldots-y_{m-1}|\ge{t_0\over m-1}$ or $|x-x'-y_2-\ldots-y_{m-1}|<{t_0\over m-1}$ and there exists $j\in \N$, $j=1\ldots n$ such that
$|y_j|\ge {t_0\over m-1}$.
Therefore using \eqref{th3.2pr8b} we obtain
\begin{eqnarray}
\tilde J_m(\tau,x,x')&\le& \sum_{j=2}^{m-1}\int_{{(y_2,\ldots,y_{m-1})\in{\cal E}_{m-2,n}(\tau,x,x')\atop |x-x'-\sum_{i=2}^{m-1}y_i|<{t_0\over m-1}}\atop |y_j|\ge {t_0\over m-1}}
{dy_2\ldots dy_{m-1}\over |y_2|^{n-1}\ldots |y_{m-1}|^{n-1}|x-x'-\sum_{i=2}^{m-1}y_i|^{n-2}}\nonumber\\
&&+\tilde J_{m,0}(\tau,x,x')\nonumber\\
&=&(m-2)\tilde J_{m,1}(\tau,x,x')+\tilde J_{m,0}(\tau,x,x'),\label{th3.2pr12}
\end{eqnarray}
where
\begin{eqnarray}
\!\!\!\!\!\!\!\!\!\!\!\!\tilde J_{m,0}(\tau,x,x')&=&\int_{(y_2,\ldots,y_{m-1})\in{\cal E}_{m-2,n}(\tau,x,x')\atop |x-x'-\sum_{i=2}^{m-1}y_i|>{t_0\over m-1}}
{dy_2\ldots dy_{m-1}\over |y_2|^{n-1}\ldots |y_{m-1}|^{n-1}|x-x'-\sum_{i=2}^{m-1}y_i|^{n-2}},\label{th3.2pr13a}\\
\!\!\!\!\!\!\!\!\!\!\!\!\tilde J_{m,1}(\tau,x,x')&=&\int_{{(y_2,\ldots,y_{m-1})\in{\cal E}_{m-2,n}(\tau,x,x')\atop |x-x'-\sum_{i=2}^{m-1}y_i|<{t_0\over m-1}}\atop |y_2|\ge {t_0\over m-1}}
{dy_2\ldots dy_{m-1}\over |y_2|^{n-1}\ldots |y_{m-1}|^{n-1}|x-x'-\sum_{i=2}^{m-1}y_i|^{n-2}}.\label{th3.2pr13b}
\end{eqnarray}
We first reduce the estimate of $\tilde J_{m,0}$ and $\tilde J_{m,1}$ to an estimate on 
\begin{equation}
P_m(\tau,x,x'):=\int_{(y_2,\ldots,y_{m-1})\in{\cal E}_{m-2,n}(\tau,x,x')}
{dy_2\ldots dy_{m-1}\over |y_2|^{n-1}\ldots |y_{m-1}|^{n-1}}.\label{th3.2pr14}
\end{equation}
From \eqref{th3.2pr13a} and the estimate $|x-x'-y_2-\ldots-y_{m-1}|>{t_0\over m-1}$ it follows that
\begin{equation}
\tilde J_{m,0}(\tau,x,x')\le \left({m-1\over t_0}\right)^{n-2}P_m(\tau,x,x').\label{th3.2pr15}
\end{equation}
From \eqref{th3.2pr13b} and the estimates $|y_2|\ge {t_0\over m-1}\ge |x-x'-y_2-\ldots-y_{m-1}|$ it follows that
\begin{equation}
\tilde J_{m,1}(\tau,x,x')\le \int_{{(y_2,\ldots,y_{m-1})\in{\cal E}_{m-2,n}(\tau,x,x')\atop |x-x'-\sum_{i=2}^{m-1}y_i|<{t_0\over m-1}}\atop |y_2|\ge {t_0\over m-1}}
{dy_2\ldots dy_{m-1}\over |y_2|^{n-2}\ldots |y_{m-1}|^{n-1}|x-x'-\sum_{i=2}^{m-1}y_i|^{n-1}}.\label{th3.2pr16}
\end{equation}
Therefore performing the change of variables ``$y_2$''$=x-x'-y_2-\ldots-y_{m-1} $ we obtain
\begin{equation}
\tilde J_{m,1}(\tau,x,x')\le \tilde J_{m,0}\le \left({m-1\over t_0}\right)^{n-2}P_m(\tau,x,x').\label{th3.2pr17}
\end{equation}

Now we estimate $P_3(\tau,x,x')$.
From \eqref{P2} and \eqref{th3.2pr14} it follows that
\begin{equation}
P_3(\tau,x,x')={\rm Vol}_{n-2}(\S^{n-2})\int_0^\pi\int_{t_0}^\tau{\sin^{n-2}(\varphi)\left(s^2-t_0^2\right)^{n-3\over 2}(s-t_0\cos(\varphi))\over
2(s+t_0\cos(\varphi))^{n-2}}ds d\varphi.\label{th3.2pr18}
\end{equation}
Let $n=3$. Then using the estimate $\cos(\varphi)\ge -1$ (and the fact that ${s-t_0\cos(\varphi)\over s+t_0\cos(\varphi)}={s-t_0\over s+t_0\cos(\varphi)}+{t_0(1-\cos(\varphi))\over s+t_0\cos(\varphi))}\le 1+{2t_0\over s+t_0\cos(\varphi)}$) we obtain
\begin{eqnarray}
P_3(\tau,x,x')&\le&\pi\int_{t_0}^\tau\int_0^\pi\left(\sin(\varphi)-2{d\over d\varphi}\ln (s+t_0\cos(\varphi)\right) 
d\varphi ds\nonumber\\
&=&2\pi\left(\tau-t_0+\int_{t_0}^\tau\ln (s+t_0)ds-\int_{t_0}^\tau\ln(s-t_0)ds\right)\nonumber\\
&=&2\pi(\tau-t_0)\left(2+\ln\left({\tau+t_0\over\tau-t_0}\right)\right)\label{th3.2pr19}
\end{eqnarray}
(we used the estimate $\ln(s+t_0)\le \ln(\tau+t_0)$ for $s\in (t_0,\tau)$ and we used the integral value \eqref{P5}).
Let $n\ge 4$. Using \eqref{th3.2pr18} and using the estimates $\sqrt{s^2-t_0^2}\sin(\varphi)\le s+t_0\cos(\varphi)$, $s+t_0\cos(\varphi)\ge s(1+\cos(\varphi))$
and $s-t_0\cos(\varphi)\le s+t_0\le 2s$ for $(s,\varphi)\in (t_0,\tau)\times(0,\pi)$,  we obtain
\begin{eqnarray}
\!\!\!\!\!\!\!\!\!\!\!\!\!\!\!\!\!P_3(\tau,x,x')&\le&{\rm Vol}_{n-2}(\S^{n-2})\int_0^\pi\int_{t_0}^\tau{\sin^{2}(\varphi)\sqrt{s^2-t_0^2}(s-t_0\cos(\varphi))\over
2(s+t_0\cos(\varphi))^2}ds d\varphi\nonumber\\
\!\!\!\!\!\!\!\!\!\!\!\!\!\!\!\!\!&\le&C_0{\rm Vol}_{n-2}(\S^{n-2})\int_{t_0}^{\tau}\int_{0}^\pi{\sqrt{s^2-t_0^2}\over
s+t_0\cos(\varphi)} d\varphi ds\le\pi C_0{\rm Vol}_{n-2}(\S^{n-2})(\tau-t_0),
\label{th3.2pr20}
\end{eqnarray}
where $C_0$ is defined by \eqref{P3} (we also used \eqref{P1}).

Finally let $m\ge 4$ then
\begin{equation}
(y_2,\ldots,y_{m-1})\in{\cal E}_{m-2,n}(\tau,x,x')\Rightarrow \left(y_2\in {\cal E}_{1,n}(\tau,x,x') \textrm{ and }\sum_{i=3}^{m-1}|y_i|<\tau\right).\label{th3.2pr21}
\end{equation}
Therefore using \eqref{th3.2pr14}, spherical coordinates (``$y_i$''$=s_i\omega_i$, $s_i\in (0,+\infty)$, $\omega_i\in \S^{n-1}$ for $i=3,\ldots, m-1$) we obtain
\begin{eqnarray}
&&\!\!\!\!\!\!\!\!\!\!\!\!\!\!\!\!\!\!\!\!\!\!\!\!P_m(\tau,x,x')\le P_3(\tau,x,x')\int_{|y_3|+\ldots+|y_{m-1}|<\tau}{dy_3\ldots dy_{m-1}\over |y_3|^{n-1}\ldots |y_{m-1}|^{n-1}}\nonumber\\
&&\!\!\!\!\!\!\!\!\!\!\!\!\!\!\!\!\!\!\!\!\!\!\!\!\le P_3(\tau,x,x'){\rm Vol}_{n-1}(\S^{n-1})^{m-3}\int_{s_3+\ldots+s_{m-1}<\tau\atop 0<s_i,\ i=3,\ldots, m-1}\!\!\!\!\!\!\!\!\!\!\!\!\!\!\!\!\!\!\!\!\!\!\!\!\!\Pi_{i=3}^{m-1}ds_i
=P_3(\tau,x,x'){\rm Vol}_{n-1}(\S^{n-1})^{m-3}{\tau^{m-3}\over (m-3)!}.\label{th3.2pr22}
\end{eqnarray}

Finally statement \eqref{th3.2pr9a} follows from \eqref{th3.2pr12}, \eqref{th3.2pr15}, \eqref{th3.2pr17}, \eqref{th3.2pr19} and \eqref{th3.2pr22}, and statement 
\eqref{th3.2pr9b} follows from \eqref{th3.2pr12}, \eqref{th3.2pr15}, \eqref{th3.2pr17}, \eqref{th3.2pr20} and \eqref{th3.2pr22}.
\end{proof}

\begin{proof}[Proof of Lemma \ref{lem:vol}.] Let $n\ge 2$. Using a rotation and \eqref{ellipse}, we have 
\begin{equation}
{\rm Vol}_n({\cal E}_{1,n}(\tau,x,x'))={\rm Vol}_{n}({\cal E}_{1,n}(\tau,t_0e_1,0)),\label{L1.0}
\end{equation}
where $t_0=|x-x'|$ and $e_1=(0,\ldots,0)\in \R^n$.

From \eqref{P2}, it follows that
\begin{equation}
{\rm Vol}_n({\cal E}_{1,n}(\tau,t_0e_1,0))={\rm Vol}_{n-2}(\S^{n-2})\int_{t_0}^{\tau}\int_0^\pi\left({\sin(\varphi)\sqrt{s^2-t_0^2}\over 2}
\right)^{n-2}{s^2-t_0^2\cos^2(\varphi)\over 4\sqrt{s^2-t_0^2}}ds d\varphi.
\label{L3}
\end{equation}
From \eqref{L3} and the estimate $\sin(\varphi)\sqrt{s^2-t_0^2}\le \sqrt{\tau^2-t_0^2}$ for $s\in (t_0, \tau)$, we obtain
\begin{eqnarray}
{\rm Vol}_n({\cal E}_{1,n}(\tau,t_0e_1,0))&\le& {\rm Vol}_{n-2}(\S^{n-2})\left({\sqrt{\tau^2-t_0^2}\over 2}\right)^{n-2}
\int_{t_0}^{\tau}\int_0^\pi{s^2-t_0^2\cos^2(\phi)\over 4\sqrt{s^2-t_0^2}}ds d\varphi\nonumber\\
&\le& {1\over 2}{\rm Vol}_{n-2}(\S^{n-2})\left({\sqrt{\tau^2-t_0^2}\over 2}\right)^{n-2}{\rm Vol}({\cal E}_{1,2}(\tau, t_0e_1,0)).
\label{L4}
\end{eqnarray}
We remind that ${\rm Vol}({\cal E}_{1,2}(\tau, t_0e_1,0))={\pi(t_0+\tau)\sqrt{\tau^2-t_0^2}\over 4}$. Therefore \eqref{lA1b} follows from \eqref{L4}. Lemma \ref{lem:vol} is proved.
\end{proof}

\begin{proof}[Proof of Lemma \ref{lem:n=3nulaubord}]
Let $(\mu,z,z')\in (0,T)\times\pa X\times\R^3$ be such that $\mu>|z-z'|>0$. 
Using the change of variables $y={t_0\over 2}(1,0)+(s\cos(\varphi),{\sqrt{s^2-t_0^2}\over 2}\sin(\varphi)\omega)$ (see \eqref{P2}), 
$\varphi\in  (0,\pi)$, $s\in (t_0,\mu)$, $\omega\in \S^1$, we obtain 
\begin{equation}
B(\mu,z,z')={\pi \over 4}\int_{t_0}^\mu\int_0^{\pi}B_1(\mu,s,\varphi)d\varphi ds,\label{lemp2}
\end{equation}
where
\begin{equation}
B_1(\mu,s,\varphi)=(s^2-t_0^2\cos^2(\varphi))\sin(\varphi)\ln\left({\mu-t_0\cos(\varphi)\over \mu-s}\right),\label{lemp3a}
\end{equation}
for $\varphi\in (0,2\pi)$ and $s\in (t_0,\mu)$.
Using \eqref{lemp3a} and the estimates $\ln\left({\mu-t_0\cos(\varphi)\over \mu-s}\right)\le\ln\left({\mu+t_0\over \mu-s}\right)$,  
$s^2-t_0^2\cos^2(\varphi)\le \mu^2$, we obtain
\begin{equation}
\int_0^\pi B_1(\mu,s,\varphi)d\varphi\le \mu^2\int_0^\pi\sin(\varphi)d\varphi\ln\left({\mu+t_0\over \mu-s}\right),
\label{lemp4b}
\end{equation}
for $s\in (t_0,\mu)$.
Combining \eqref{lemp2}, \eqref{lemp4b} and \eqref{P5} we obtain
\begin{equation}
B(\mu,z,z')\le
{\mu^2\pi\over 2}(\mu-t_0)\left(\ln\left({\mu+t_0\over \mu-t_0}\right)+1\right),\label{lemp5}
\end{equation}
which proves \eqref{lemp0}.
\end{proof}

\section{The distributional kernel of the operators $H_m$ and the proof of Proposition \ref{prop:kernelA}}
\label{proof_kernelA}
Before we prove Proposition \ref{prop:kernelA} we shall introduce and prove Proposition \ref{prop:kernelH} given below, which gives the distributional kernel of the operators $H_m$ defined by \eqref{E1a}.

Let $\bar E$ denotes the nonnegative mesurable function from $\R^n\times\R^n$ to $\R$ defined by
\begin{equation}
\bar E(x_1,x_2)=e^{-\int_0^{|x_1-x_2|}\sigma(x_1-s{x_1-x_2\over|x_1-x_2|},{x_1-x_2\over |x_1-x_2|})ds}\Theta(x_1,x_2),\textrm{ for a.e. }(x_1,x_2)\in \R^n\times\R^n,\label{N1}
\end{equation} 
where $\Theta$ is defined by \eqref{B4.0}.
For $m\ge 3$, we define recursively the nonnegative measurable real function $\bar E(x_1,\ldots,x_m)$ by the formula
\begin{equation} 
\bar E(x_1,\ldots, x_m)=\bar E(x_1,\ldots,x_{m-1})\bar E(x_{m-1},x_m),\label{N2}
\end{equation}
for $(x_1,\ldots,x_m)\in (\R^n)^m$.

Concerning the distributional kernel of the $H_m$, $m\ge 2$, we have the following result.
\vskip 2mm
\begin{proposition}
\label{prop:kernelH}
We have
\begin{equation}
H_m(t)\phi(x,v)=\int_{X\times \S^{n-1}}\beta_m(t,x,v,x',v')\phi(x',v')dx'dv',\label{E3}
\end{equation}
for $t\in (0,T)$ and a.e. $(x,v)\in X\times \S^{n-1}$ and for $m\ge 2$, where
\begin{eqnarray}
\beta_2(t,x,v,x',v')&=&\int_0^t\chi_{(0,t-s_2)}(|x'-(x-s_2v')|){2^{n-2}\left(t-s_2-(x-s_2v'-x')\cdot v\right)^{n-3}\over\left|x-s_2v'-x'-(t-s_2)v\right|^{2n-4}}\nonumber\\
&&\times\left[\bar E(x,x-(t-s_1-s_2)v,x'+s_2v',x')k(x-(t-s_1-s_2)v,v_1,v)\right.\nonumber\\
&&\left.\times k(x'+s_2v',v',v_1) 
\right]_{v_1={x-s_2v'-x'-(t-s_1-s_2)v\over s_1}\atop s_1={|x-s_2v'-x'-(t-s_2)v|^2\over 2(t-s_2-(x-x'-s_2v')\cdot v)}}ds_2,\label{E4}
\end{eqnarray}
for $t\in (0,T)$ and a.e. $(x,v,x',v')\in X\times\S^{n-1}\times X\times\S^{n-1}$, and where
\begin{eqnarray}
&&\!\!\!\!\!\!\!\!\!\!\!\!\!\!\!\!\!\!\!\beta_m(t,x,v,x',v')=
\int_{\left(\S^{n-1}\right)^{m-2}}\int_{s_2+\ldots+s_m\le t\atop s_i\ge 0,\ i=2\ldots m}\chi_{(0,t-s_m-\ldots-s_2)}(|x'+s_mv'+\ldots+s_2v_2-x|)\nonumber\\
&&\!\!\!\!\!\!\!\!\!\!\!\!\!\!\!\!\!\!\!\times {2^{n-2}\left(t-s_2-\ldots-s_m-(x-x'-s_2v_2-\ldots-s_{m-1}v_{m-1}-s_mv')\cdot v\right)^{n-3}\over |x-x'-s_2v_2-\ldots-s_{m-1}v_{m-1}-s_mv'-
(t-s_2-\ldots-s_m)v|^{2n-4}}\left[\bar E(x,x-(t-s_1-\ldots-s_m)v,\right.\nonumber\\
&&\!\!\!\!\!\!\!\!\!\!\!\!\!\!\!\!\!\!\! \left.x'+s_mv'+s_{m-1}v_{m-1}+\ldots+s_2v_2,x'+s_mv'+s_{m-1}v_{m-1}+\ldots+s_3v_3,\right.\nonumber\\
&&\!\!\!\!\!\!\!\!\!\!\!\!\!\!\!\!\!\!\!\ldots, x'+s_mv',x') k(x-(t-s_1-\ldots-s_m)v,v_1,v)k(x'+s_mv'+s_{m-1}v_{m-1}+\ldots+s_2v_2,v_2,v_1)\nonumber\\
&&\!\!\!\!\!\!\!\!\!\!\!\!\!\!\!\!\!\!\!\ldots k(x'+s_mv'+s_{m-1}v_{m-1}+\ldots+s_{i+1}v_{i+1},v_{i+1},v_i)\ldots \nonumber\\
&&\!\!\!\!\!\!\!\!\!\!\!\!\!\!\!\!\!\!\!\left.k(x'+s_mv',v',v_{m-1})
\right]_{v_1={x-x'-s_2v_2-\ldots-s_{m-1}v_{m-1}-s_mv'-(t-s_1-\ldots-s_m)v\over s_1}
\atop s_1={|x-x'-s_2v_2-\ldots-s_{m-1}v_{m-1}-s_mv'-(t-s_2-\ldots-s_m)v|^2\over 2(t-s_2-\ldots-s_m-(x-x'-s_2v_2-\ldots s_{m-1}v_{m-1}-s_mv')\cdot v)}}
ds_2\ldots ds_mdv_2\ldots dv_{m-1},\label{E7b}
\end{eqnarray}
for $t\in (0,T)$ and a.e. $(x,v,x',v')\in X\times\S^{n-1}\times X\times\S^{n-1}$, $m\ge 3$.
\end{proposition}

\begin{proof}[Proof of Proposition \ref{prop:kernelH}]
Note that
\begin{eqnarray*}
H_2(t)\phi(x,v)&=&\left(\int_0^t\int_0^{t-s_1}U_1(t-s_1-s_2)A_2U_1(s_1)A_2 U_1(s_2)\phi ds_2ds_1\right)(x,v)\\
&=&\left(\int_0^t\left(\int_0^{t-s_2} U_1(t-s_1-s_2)A_2U_1(s_1)A_2ds_1\right)U_1(s_2)\phi ds_2\right)(x,v)\\
&=&\int_0^t\int_0^{t-s_2}\bar E(x,x-(t-s_1-s_2)v)\int_{\S^{n-1}}k(x-(t-s_1-s_2)v,v_1,v)\\
&&\times\bar E(x-(t-s_1-s_2)v,x-(t-s_1-s_2)v-s_1v_1)\\
&&\times\int_{\S^{n-1}}k(x-(t-s_2-s_1)v-s_1v_1,v_2,v_1)\\
&&\times\bar E(x-(t-s_1-s_2)v-s_1v_1,x-(t-s_1-s_2)v-s_1v_1-s_2v_2)\\
&&\times\phi(x-(t-s_1-s_2)v-s_1v_1-s_2v_2,v_2)dv_2dv_1ds_1ds_2,
\end{eqnarray*}
for $t\in (0,T)$ and $(x,v)\in X\times\S^{n-1}$, where functions $\bar E$ are defined by \eqref{N1}--\eqref{N2}.

Using the change of variables ``$y(s_1,v_1)=(t-s_2-s_1)v+s_1v_1$'' we obtain
\begin{eqnarray*}
H_2(t)\phi(x,v)&=&\int_0^t\int_{\S^{n-1}}\left[\bar E(x,x-(t-s_1-s_2)v,x-y,x-y-s_2v_2)k(x-(t-s_1-s_2)v,v_1,v)\right.\\
&&\times \left.k(x-y,v_2,v_1)
\right]_{v_1={y-(t-s_1-s_2)v\over s_1}\atop s_1={|y-(t-s_2)v|^2\over 2(t-s_2-y\cdot v)}}\\
&&\times {2^{n-2}\left((t-s_2)-y\cdot v\right)^{n-3}\over\left|y-(t-s_2)v\right|^{2n-4}}\phi(x-y-s_2v_2,v_2)dy dv_2ds_2.
\end{eqnarray*}
Hence we obtain \eqref{E3}.
Note that
\begin{eqnarray*}
(H_3(t)\phi)(x,v)&=&\int_0^tH_2(t-s_3)A_2U_1(s_3)\phi ds_3\\
&=&\int_0^t\int_{X\times \S^{n-1}}\beta_2(t-s_3,x,v,x_2,v_2)(A_2U_1(s_3))\phi(x_2,v_2)dx_2dv_2ds_3\\
&=&\int_{X\times \S^{n-1}}\int_0^t\beta_2(t-s_3,x,v,x_2,v_2)\int_{\S^{n-1}}k(x_2,v',v_2) \bar E(x_2,x_2-s_3v')\\
&&\times \phi(x_2-s_3v',v')dv'ds_3dx_2dv_2.
\end{eqnarray*}
Hence
\begin{equation}
(H_3(t)\phi)(x,v)=\int_{X\times \S^{n-1}}\beta_3(t,x,v,x',v')\phi(x',v')dx'dv',\label{E5}
\end{equation}
where
\begin{eqnarray}
\beta_3(t,x,v,x',v')&=&\int_{\S^{n-1}}\int_0^t\int_0^{t-s_3}\chi_{(0,t-s_3-s_2)}(|x'+s_3v'-x+s_2v_2|)\nonumber\\
&&\times {2^{n-2}\left(t-s_2-s_3-(x-s_2v_2-x'-s_3v')\cdot v\right)^{n-3}\over |x-x'-s_2v_2-s_3v'-
(t-s_2-s_3)v|^{2n-4}}\nonumber\\
&&\times\left[\bar E(x,x-(t-s_1-s_2-s_3)v,x'+s_2v_2+s_3v',x'+s_3v',x')\right.\nonumber\\
&&\times k(x-(t-s_3-s_2-s_1)v,v_1,v)k(x'+s_2v_2+s_3v',v_2,v_1)\nonumber\\
&&\times\left.k(x'+s_3v',v',v_2)
\right]_{v_1={x-x'-s_2v_2-s_3v'-(t-s_1-s_2-s_3)v\over s_1}\atop s_1={|x-x'-s_2v_2-s_3v'-(t-s_2-s_3)v|^2\over 2(t-s_2-s_3-(x-x'-s_2v_2-s_3v')\cdot v)}}
ds_2ds_3dv_2.\label{E6}
\end{eqnarray}
The proof of \eqref{E7b} follows by induction from \eqref{E5} and \eqref{E1b}.
\end{proof}

\begin{proof}[Proof of \eqref{E8b}--\eqref{E8d}]
We recall that
\begin{equation}
\left(A_2G_-(s)\phi_S\right)(z,w)=\int_{\pa
    X}\left[k(z,v',w)S(x',v')|\nu(x')\cdot v'| \right]_{v'={z-x'\over
      |z-x'|}}{E(z,x')\over |z-x'|^{n-1}}\phi(s-|z-x'|,x')d\mu(x'),\label{E9}
\end{equation}
for a.e. $(z,w)\in X\times \S^{n-1}$ and $\phi\in L^1((0,\eta)\times\pa X)$ (see the derivation of \eqref{E8.0} and \eqref{E8a} given in Section \ref{sec:albedoaveraged}).

Let $m=2$. Then from \eqref{E1a} and \eqref{E1b} it follows that
\begin{equation*}
\begin{array}{l}
\displaystyle A_{2,S,W}(\phi)(t,x)=\int_{\S^{n-1}_{x,+}}(\nu(x)\cdot v)W(x,v)\int_{-\infty}^t\int_0^{t-s}\int_{\S^{n-1}} \int_{\pa X}\left[k(x-(t-s-s_1)v,v_1,v)\right.\\
\displaystyle\times \left.k(x-(t-s-s_1)v-s_1v_1,v',v_1)S(x',v')|\nu(x')\cdot v'|\right]_{v'={x-(t-s-s_1)v-s_1v_1-x'\over |x-(t-s-s_1)v-s_1v_1-x'|}}E(x,x-(t-s-s_1)v,\\
\displaystyle x-(t-s-s_1)v-s_1v_1,x'){\phi(s-|x-(t-s-s_1)v-s_1v_1-x'|,x')\over |x-(t-s-s_1)v-s_1v_1-x'|^{n-1}}d\mu(x')dv_1ds_1dsdv.
\end{array}
\end{equation*}
Performing the change of variables $y(s_1,v_1)=(t-s-s_1)v+s_1v_1$, we obtain
\begin{eqnarray}
&&A_{2,S,W}(\phi)(t,x)=\int_{\S^{n-1}_{x,+}\times \pa X\times \R^n}(\nu(x)\cdot v)W(x,v)\int_{-\infty}^t\chi_{(0,t-s)}(|y|)\nonumber\\
&&\!\!\!\!\!\!\!\!\!\!\!\!\!\!\!\!\!\!\!\!\!\!\!\!\!\times \left[E(x,x-(t-s-s_1)v,x-y,x')k(x-(t-s-s_1)v,v_1,v)k(x-y,v',v_1)S(x',v')\right.\nonumber\\
&&\!\!\!\!\!\!\!\!\!\!\!\!\!\!\!\!\!\!\!\!\!\!\!\!\!\times\left.|\nu(x')\cdot v'|\right]_{{s_1={|(t-s)v-y|^2\over 2(t-s-y\cdot v)}\atop v_1={y-(t-s-s_1)v\over s_1}}\atop v'={x-y-x'\over |x-y-x'|}}
{2^{n-2}(t-s-y\cdot v)^{n-3}\phi(s-|x-y-x'|,x')\over |(t-s)v-y|^{2n-4}|x-y-x'|^{n-1}}ds dy d\mu(x')dv.\label{E12}
\end{eqnarray}
Performing the change of variables ``$y$''$=x-x'-y$ and $t'=s-|y|$ we obtain \eqref{E8b}.

Let $m=3$. Then from \eqref{E2b}, \eqref{E3} and \eqref{E9} it follows that
\begin{eqnarray}
&&A_{3,S,W}(\phi)(t,x)=\int_{\S^{n-1}_{x,+}}(\nu(x)\cdot v)W(x,v)\int_{-\infty}^t\int_{X\times \S^{n-1}}\beta_2(t-s,x,v,x_2,v_2)\label{E13}\\
&&\!\!\!\!\!\!\!\!\!\!\!\!\!\!\!\!\int_{\pa X}\left[k(x_2,v',v_2)S(x',v')|\nu(x')\cdot v'|\right]_{v'={x_2-x'\over |x_2-x'|}}{E(x_2,x')\over |x_2-x'|^{n-1}}\phi(s-|x_2-x'|,x')d\mu(x')
dx_2dv_2dsdv,\nonumber
\end{eqnarray}
for $t\in (0,T)$ and $x\in \pa X$. From \eqref{E13} and \eqref{E4} we obtain 
\begin{eqnarray}
&&\!\!\!\!\!\!\!\!\!\!\!\!\!\!\!\!\!\!\!\!A_{3,S,W}(\phi)(t,x)=\int_{\S^{n-1}_{x,+}}(\nu(x)\cdot v)W(x,v)\int_{X\times \S^{n-1}\times \pa X}\int_{-\infty}^t\int_0^{t-s}\chi_{(0,t-s-s_2)}(|x_2-(x-s_2v_2)|)\nonumber\\
&&\!\!\!\!\!\!\!\!\!\!\!\!\!\!\!\!\!\!\!\!{2^{n-2}\left(t-s-s_2-(x-s_2v_2-x_2)\cdot v\right)^{n-3}\over|x_2-x'|^{n-1}\left|x-s_2v_2-x_2-(t-s-s_2)v\right|^{2n-4}}\nonumber\\
&&\!\!\!\!\!\!\!\!\!\!\!\!\!\!\!\!\!\!\!\!\times\left[E(x,x-(t-s_1-s_2)v,x_2+s_2v_2,x_2,x')k(x-(t-s_1-s_2)v,v_1,v)k(x_2+s_2v_2,v_2,v_1)\right.\nonumber\\
&&\!\!\!\!\!\!\!\!\!\!\!\!\!\!\!\!\!\!\!\!\left.k(x_2,v',v_2)S(x',v')|\nu(x')\cdot v'|
\right]_{{v_1={x-s_2v_2-x_2-(t-s-s_1-s_2)v\over s_1}\atop s_1={|x-s_2v_2-x_2-(t-s-s_2)v|^2\over 2(t-s-s_2-(x-x_2-s_2v_2)\cdot v)}}\atop v'={x_2-x'\over |x_2-x'|}}\phi(s-|x_2-x'|,x')ds_2dsdx_2dv_2d\mu(x')dv.
\nonumber
\end{eqnarray}
Performing the change of variables $y_2=s_2v_2$ and $y_3=x_2-x'$ we obtain \eqref{E8d} for ``$m=3$''.

Let $m\ge 3$.
From \eqref{E2b}, \eqref{E3}, \eqref{E7b} and \eqref{E9} it follows that
\begin{eqnarray}
A_{m+1,S,W}(\phi)(t,x)&=&\int_{\S^{n-1}_{x,+}}\int_{\pa X}(\nu(x)\cdot v)W(x,v)\int_X\int_{(-\infty,t-|x_m-x'|)\times \S^{n-1}}\!\!\!\!\!\!\!\!\!\!\!\!
\!\!\!\!\!\!\!\!\!\!\!\!\beta_m(t-t'-|x_m-x'|,x,v,x_m,v_m)\nonumber\\
&&\!\!\!\!\!\!\!\!\!\!\!\!\!\!\!\!\!\!\!\!\!\!\!\!\left[k(x_m,v',v_m)S(x',v')|\nu(x')\cdot v'|\right]_{v'={x_m-x'\over |x_m-x'|}}{E(x_m,x')\over |x_m-x'|^{n-1}}\phi(t',x')d\mu(x')
dt'dx_mdv_mdv\nonumber\\
&&\!\!\!\!\!\!\!\!\!\!\!\!\!\!\!\!\!\!\!\!\!\!\!\!=\int_{(0,\eta)\times \pa X}\gamma_{m+1}(t-t',x,x')\phi(t',x')dt'd\mu(x'),\label{E10}
\end{eqnarray}
where
\begin{eqnarray}
&&\gamma_{m+1}(\tau,x,x'):=
\int_{\S^{n-1}_{x,+}}(\nu(x)\cdot v)W(x,v)\int_{X\times \S^{n-1}}\chi_{(0,+\infty)}(\tau-|x_m-x'|)\nonumber\\
&&\int_{\left(\S^{n-1}\right)^{m-2}}\int_{s_2+\ldots+s_m\le \tau-|x_m-x'|\atop s_i\ge 0,\ i=2\ldots m}
\chi_{(0,\tau-|x_m-x'|-s_m-\ldots-s_2)}(|x_m+s_mv_m+\ldots+s_2v_2-x|)\nonumber\\
&&\times {2^{n-2}\left(\tau-|x_m-x'|-s_2-\ldots-s_m-(x-x_m-s_2v_2-\ldots-s_mv_m)\cdot v\right)^{n-3}\over |x_m-x'|^{n-1}|x-x_m-s_2v_2-\ldots-s_mv_m-
(\tau-|x_m-x'|-s_2-\ldots-s_m)v|^{2n-4}}\nonumber\\
&&\times \left[E(x,x-(\tau-|x_m-x'|-s_1-\ldots-s_m)v,x_m+s_mv_m\ldots+s_2v_2,x_m+s_mv_m+\ldots+s_3v_3,
\ldots,\right.\nonumber\\
&&x_m+s_mv_m,x_m,x') k(x-(\tau-|x_m-x'|-s_1-\ldots-s_m)v,v_1,v)k(x_m+s_mv_m+\ldots+s_2v_2,v_2,v_1)\nonumber\\
&&\ldots k(x_m+s_mv_m+\ldots+s_{i+1}v_{i+1},v_{i+1},v_i)\ldots \nonumber\\
&&\left.k(x_m+s_mv_m,v_m,v_{m-1})k(x_m,v',v_m)S(x',v')|\nu(x')\cdot v'|
\right]_{{v_1={x-x_m-s_2v_2-\ldots-s_mv_m-(\tau-|x_m-x'|-s_1-\ldots-s_m)v\over s_1}\atop s_1={|x-x_m-s_2v_2-\ldots-s_mv_m-(\tau-|x_m-x'|-s_2-\ldots-s_m)v|^2\over 2(t-s_2-\ldots-s_m-(x-x'-s_2v_2-\ldots s_mv_m)\cdot v)}}\atop v'={x_m-x'\over |x_m-x'|}}\nonumber\\
&&ds_2\ldots ds_mdv_2\ldots dv_{m-1}dx_m dv_m dv.\label{E11}
\end{eqnarray}
Performing the change of variables $y_i=s_iv_i$, $i=2\ldots m$, and $y_{m+1}=x_{m+1}-x'$, we obtain \eqref{E8d} for ``$m\ge 4$''.
\end{proof}

\section*{Acknowledgment}
This paper was funded in part by grant NSF DMS-0554097. The authors
would like to thank Ian Langmore and Fran\c cois Monard for stimulating 
discussions on the inverse transport problem.

\bibliographystyle{plain}
\bibliography{../../bibliography}

\end{document}